\documentclass[fleqn,usenatbib,useAMS]{mnras}

\usepackage{etoolbox}
\makeatletter
\patchcmd\@combinedblfloats{\box\@outputbox}{\unvbox\@outputbox}{}{%
   \errmessage{\noexpand\@combinedblfloats could not be patched}%
}%
 \makeatother

\usepackage{graphicx}	% Including figure files
\usepackage{amsmath}	% Advanced maths commands
\usepackage{amssymb}	% Extra maths symbols
\usepackage{multicol}        % Multi-column entries in tables
\usepackage{bm}		% Bold maths symbols, including upright Greek
\usepackage{pdflscape}	% Landscape pages
\usepackage{booktabs,caption}
\usepackage[flushleft]{threeparttable}
\newcommand{\EBnumber}{35}
\newcommand{\PMnumber}{45}

\newcommand{\gdor}{$\gamma$\,Dor}
\newcommand{\cpd}{\mathrm{d^{-1}}}

\usepackage[T1]{fontenc}
\usepackage{ae,aecompl}

\usepackage{newtxtext,newtxmath}
\title[Tidal effect of $\gamma$\,Dor stars in binaries]{The effect of tides on near-core rotation: analysis of \EBnumber\ \textit{Kepler} $\gamma$\,Doradus stars in eclipsing and spectroscopic binaries}

\author[G. Li]{Gang Li$^{1,2}$, Zhao Guo$^{3,4}$, Jim Fuller$^{5}$, Timothy R. Bedding$^{1,2}$, Simon J. Murphy$^{1,2}$, 
\newauthor{Isabel L. Colman$^{1,2}$, Daniel R. Hey$^{1,2}$}\\
$^{1}$Sydney Institute for Astronomy (SIfA), School of Physics, University of Sydney, NSW 2006, Australia\\
$^{2}$Stellar Astrophysics Centre, Department of Physics and Astronomy, Aarhus University, Ny Munkegade 120, DK-8000 Aarhus C, Denmark\\
$^{3}$Department of Astronomy \& Astrophysics, 525 Davey Laboratory, The Pennsylvania State University, University Park, PA, 16802, USA.\\
$^{4}$Center for Exoplanets and Habitable Worlds, 525 Davey Laboratory, The Pennsylvania State University, University Park, PA 16802, USA \\
$^{5}$TAPIR, Mailcode 350-17, California Institute of Technology, Pasadena, CA 91125, USA}

\date{Last updated 2015 May 22; in original form 2013 September 5}

\pubyear{2015}

\begin{document}
\label{firstpage}
\pagerange{\pageref{firstpage}--\pageref{lastpage}}
\maketitle

\begin{abstract}
We systematically searched for gravity- and Rossby-mode period spacing patterns in \textit{Kepler} eclipsing binaries with $\gamma$\,Doradus pulsators. These stars provide an excellent opportunity to test the theory of tidal synchronisation and angular momentum transport in F- and A-type stars. We discovered \EBnumber\ systems that show clear patterns, including the spectroscopic binary KIC\,10080943. Combined with \PMnumber\ non-eclipsing binaries with \gdor\ components that have been found using pulsation timing, we measured their near-core rotation rates and asymptotic period spacings. 
We find that many stars are tidally locked if the orbital periods are shorter than 10\,days, in which the near-core rotation periods given by the traditional approximation of rotation (TAR) are consistent with the orbital period. Compared to the single stars, $\gamma$\,Dor stars in binaries tend to have slower near-core rotation rates, likely a consequence of tidal spin-down. We also find three stars that have extremely slow near-core rotation rates. To explain these, we hypothesise that unstable tidally excited oscillations can transfer angular momentum from the star to the orbit, and slow the star below synchronism, a process we refer to as `inverse tides'.
\end{abstract}


%%%%%%%%%%%%%%%%%%%%%%%%%%%%%%%%%%%%%%%%%%%%%%%%%%

%%%%%%%%%%%%%%%%% BODY OF PAPER %%%%%%%%%%%%%%%%%%

% The MNRAS class isn't designed to include a table of contents, but for this document one is useful.
% I therefore have to do some kludging to make it work without masses of blank space.

\section{Introduction}
%Gang Li: g mode

%Zhao Guo: binary and synchronisation
Binary and multiple systems are common, probably outnumbering single stars throughout much of the HR diagram \citep{Duchene_2013, Moe_2017, Guszejnov_2017, Murphy_2018_review}. Eclipsing binaries provide important tests for calibrating stellar parameters, giving stellar masses and radii to an accuracy of $\sim 2 \%$ \citep{Andersen_1991} and, in ideal cases, to 0.2\% \citep{Maxted_2020}. This also improves the accuracy of the parameters of orbiting planets \citep[e.g.][]{Doyle_2011}. Well-measured stellar parameters in binary systems allow us to test stellar evolution models or study mass transfer \citep[e.g.][]{Torres_2002, del_Burgo_2018, Faulkner_1971}, and determine distances \citep[e.g.][]{Hilditch_2005, North_2010}.

A critical question about binarity is how tides affect the system. In a binary system with a radiative-envelope main-sequence star, the periodic gravitational disturbance by its companion star can excite oscillations and lead to circularization and synchronization \citep{Zahn_1975, Zahn_1977}. Radiative damping of low-frequency tidal oscillations can lead to angular momentum exchange between orbital motion and stellar rotation, or orbital energy dissipation into thermal energy, or angular momentum transfer by the tidally excited oscillations \citep[e.g.][]{Goldreich_1989_a,Goldreich_1989_b, Lee_1993}.

In this work, we report observations of the near-core rotation rates of \EBnumber\ binaries, which are important for understanding tidal effects in binary systems. We use self-excited gravity modes to measure the near-core rotation rates in $\gamma$\,Doradus stars, which are main-sequence stars with masses between $1.4$ and $2.0$ solar masses \citep{Balona_1994, Kaye_1999}. The gravity modes carry information about the near-core regions, such as the chemical composition gradient \citep{Miglio_2008}, and the near-core rotation rate \citep{Bouabid_2013, Ouazzani_2017}, hence allow us to infer conditions in the stellar interiors. Thanks to the high-precision continuous photometry from the \textit{Kepler} space mission \citep{Borucki_2010, Koch_2010}, we can resolve the gravity modes in over 600 $\gamma$\,Dor stars \citep{Li_2019_r_modes, Li_2019_all_gdor}. Most show dipole sectoral g modes with decreasing period spacings, which are the period differences between two consecutive modes \citep[e.g.][]{VanReeth_2015, Li_2019_r_modes}. The typical near-core rotation rate is around $1\,\mathrm{d^{-1}}$ \citep{Li_2019_all_gdor}, measured by the traditional approximation of rotation (TAR) \citep[e.g.][]{Eckart_1960, Lee_1997_rotating, Townsend_2003, VanReeth_2016_TAR, Li_2019_all_gdor}. Somewhat surprisingly, there are also many slowly-rotating $\gamma$\,Dor stars, showing nearly-identical period spacings or rotational splittings \citep[e.g.][]{Kurtz_2014, Saio_2015, Murphy_2016, Keen_2015, Li_2018, Li_2019_all_gdor}. 

In addition to g modes, we also see Rossby modes (r modes) in about one fifth of the \gdor{} stars \citep{VanReeth_2016_TAR, Li_2019_r_modes, Li_2019_all_gdor}. These modes are restored primarily by the Coriolis force \citep[e.g.][]{Rossby_1939, Papaloizou_1978_first_r_mode, Lee_1997_rotating, Saio_2018_Rossby_mode}, have pulsation periods longer than the rotation period in the observer's frame, and show increasing period spacing patterns as a function of period \citep{Provost_1981_eigenfreq_of_Rossby_mode, Saio_1982_Rossby_mode}. We can use g and r modes together to measure the near-core rotation rates. When a star rotates rigidly, both g and r modes are sensitive to the inner regions. However, if the core rotates more rapidly than the surface, which is consistent with the theory and observation \citep[e.g.]{Rieutord_2006, Hypolite_2014, Li_2019_all_gdor}, r modes will be more sensitive to the outer layers, whereas the g modes are more sensitive to the near-core regions. This situation is rare in the observations but provides a new approach to detect radial differential rotation \citep{VanReeth_2018}. Rossby modes with small period spacings are still unresolved by the 4-yr \textit{Kepler} data, hence they show an amplitude hump near the rotation frequency \citep[e.g.][]{Saio_2018_Rossby_mode}. These humps offer a new opportunity to measure the rotation rates of stars, including stars in eclipsing binaries \citep{Saio_2019_r_mode_binaries}.  

In this work, we report our study of \gdor{} stars in eclipsing binaries that show clear period spacing patterns. In Section~\ref{sec: data reduction}, we describe the data reduction method, including the eclipse removal, detection of period spacing patterns, and the TAR fitting. We show the results and discussions in Section~\ref{sec: results}. Finally, we make conclusions in Section~\ref{sec: conclusions}.

%\citep[e.g.][]{Huber_2013, Berger_2018}. 

\section{Data reduction}\label{sec: data reduction}

%Gang Li write this
\subsection{Eclipse removal}
We downloaded 4-yr \textit{Kepler} long-cadence (LC; 29.45-min sampling) Simple Aperture Photometry (SAP) light curves using the \textsc{python} package \textsc{Lightkurve} \citep{Light_curve_2018}. We applied a high-pass filter to the light curves to remove any slow trend, during which the flux was divided by the median in a ten-day moving window.

We measured the times of primary eclipses and calculated the differences from a linear ephemeris to determine the best orbital periods  \citep[e.g.][]{Sterken_2005}. The procedure involved the following steps: Given an initial time of the first primary light minimum $T_{0, \mathrm{ini}}$ and initial orbital period $P_\mathrm{orb, ini}$, we searched for the eclipses around the calculated times of light minima $C=T_{0, \mathrm{ini}}+EP_\mathrm{orb, ini}$, where $E$ is the integer cycle number. We fitted a parabola to the local light curves to get the observed times of light minima (`$O$'). The slope of the `$O$ minus $C$' vs cycle number was used to corrected the initial orbital period. 
The light curves were folded by the orbital periods to obtain the phased light curves, and the latter were rebinned to generate the binned light curves. The bin sizes were small enough to assure that the eclipses were well-sampled. The binned light curves were subtracted from the original light curves to remove the eclipses, and the residuals were used for forward seismic analysis.  

\begin{figure}
    \centering
    \includegraphics[width=1\linewidth]{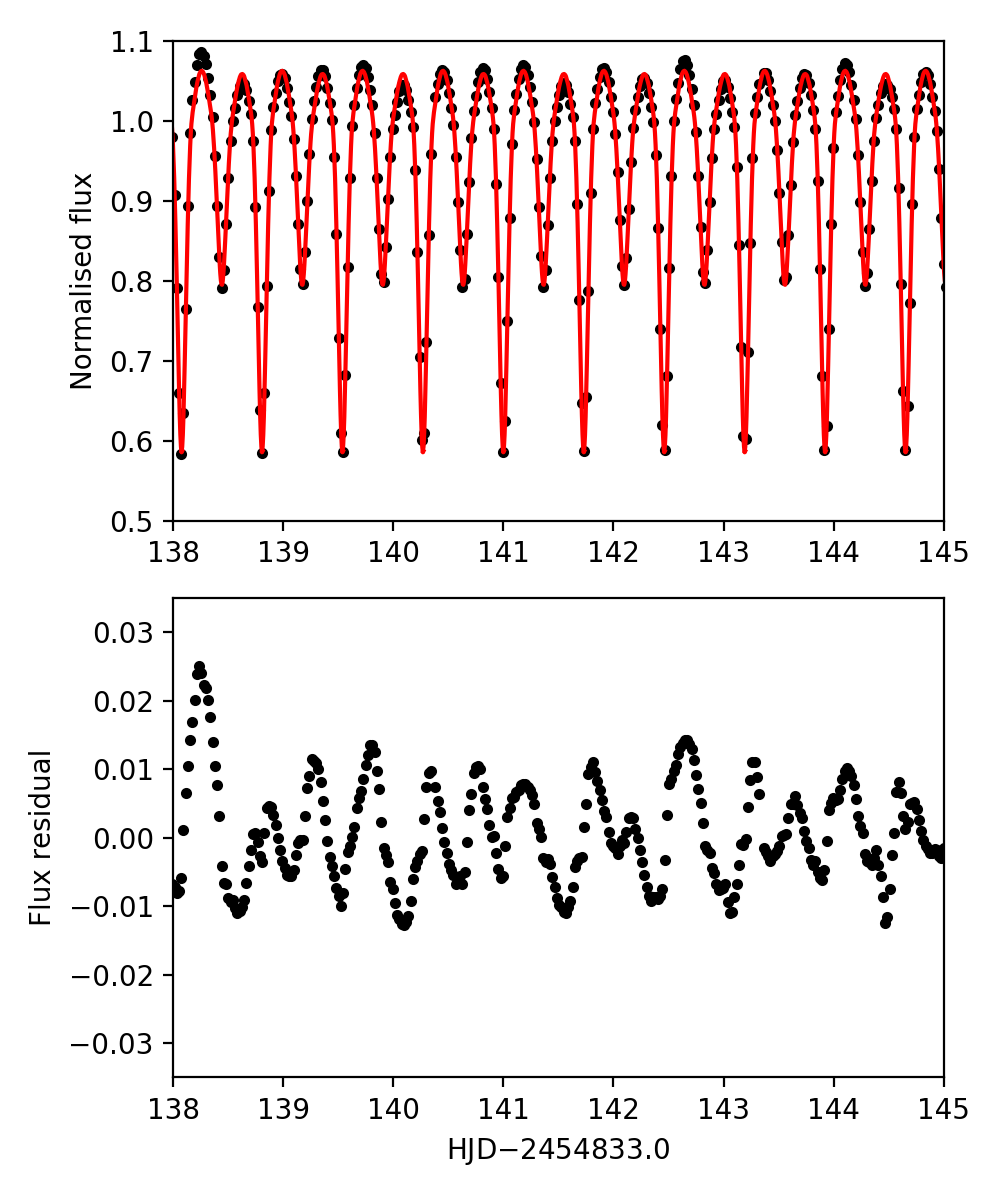}
    \caption{The light curve of KIC\,3228863. Top: the light curve before removing the eclipses. The x-axis is the Kepler Barycentric Julian Day (BKJD), which is a Julian day minus 2454833.0. The y-axis is the normalised flux. The black dots are the raw flux data, while the red line is the binned light curve. Bottom: the light curve residual. }
    \label{fig:example_eclipse_removal}
\end{figure}

Figure\,\ref{fig:example_eclipse_removal} shows the result of the eclipses removal, using KIC\,3228863 as an example. The binary system and its third component were studied using the \textit{Kepler} photometry and ground-based spectroscopy by \cite{Lee_2019_3228863}. Our calculated period is $0.73094425\pm0.00000004\,\mathrm{d}$, consistent with their measurement. In the top panel, we display seven days of the \textit{Kepler} data after the normalisation and detrending, where the primary eclipses block about 40\% of the total flux. We also plot the binned light curve. The residuals between the raw data and the binned light curve are shown in the bottom panel, where the \gdor{}-type pulsations are revealed with a total amplitude around 1\%.

Some eclipsing binaries have eccentric orbits. For those whose phases of secondary light minima are around 0.5, we assumed that their eccentricities are equal to zero. This assumption is reasonable for short-period binaries since they typically show circular orbits \citep{Shporer_2016}. For the long-period eccentric binaries, the light minima at phase near 0.5 only appear when the longitude of periastron is near $90^\circ$, which is not likely to happen \citep[see equation 18 in][]{Matson_2016}. For those with obvious eccentricity, we checked the previous literature and collected their eccentricities. 

In an eccentric orbit, true synchronous rotation is impossible, so a `pseudo-synchronous' period $P_\mathrm{ps}$ is calculated instead. For this, we used equation 42 in \cite{Hut_1981}:
\begin{equation}
    P_\mathrm{ps}=P_\mathrm{orb}\frac{\left(1+3e^2+\frac{3}{8}e^4\right)\left(1-e^2\right)^{3/2}}{1+\frac{15}{2}e^2+\frac{45}{8}e^4+\frac{5}{16}e^6},\label{equ:p_ps}
\end{equation}
where $e$ is the eccentricity and $P_\mathrm{orb}$ is the orbital period. Pseudo-synchronous spin can lead to a zero net torque during each orbit, hence the spin will not evolve \citep[e.g.][]{Hut_1981, Welsh_2011_KOI-54}.

\subsection{Oscillation signal extraction}
We calculated the amplitude spectra of the residual light curves where the eclipses have been removed. The frequencies were obtained by the prewhitening algorithm described by \cite{Li_2018}, with the criterion that the signal-to-noise ratio (S/N) should be larger than 3.5. Although the well-accepted criterion is $\text{S/N}>4$ \citep{Breger_1993}, we still adopted the peaks which have slightly smaller S/N but follow the patterns. 

\begin{figure*}
    \centering
    \includegraphics[width=1\linewidth]{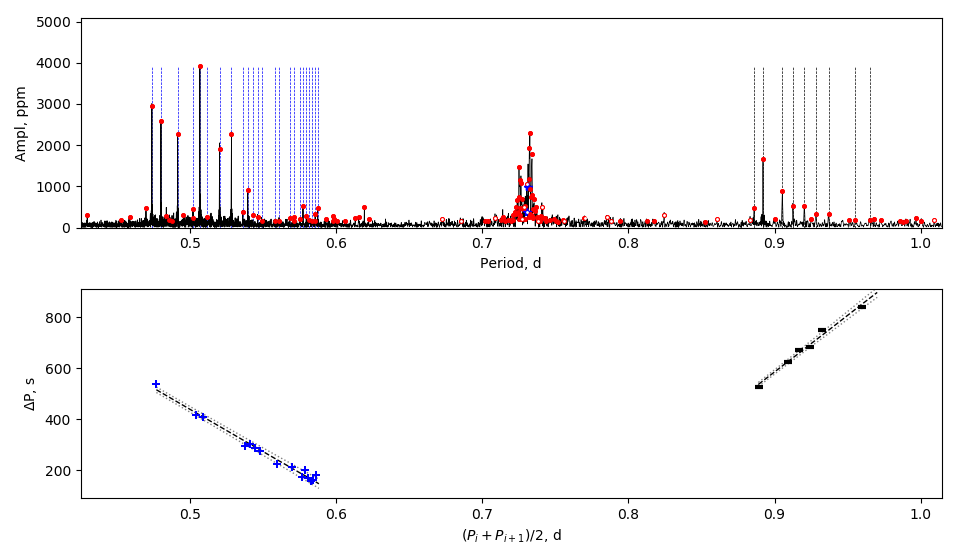}
    \caption{Top: the amplitude spectrum of KIC\,3228863 after removing the eclipses. The blue vertical dashed lines on the left show the peaks of $l=1, m=1$ g~modes. The black vertical dashed lines on the right mark the peaks of $k=-2, m=-1$ r modes. The hump at $\sim0.73$\,d might be the surface modulation signal. Bottom: the period spacings of g and r modes. The blue crosses are the pattern of $l=1, m=1$ g modes and the black symbols are those of $k=-2, m=-1$ r~modes. The dashed lines show the best-fitting linear relation, whose uncertainties are shown by the grey dotted lines surrounding them. }
    \label{fig:KIC3228863}
\end{figure*}

The g- and r-mode patterns were identified by the algorithm described by \cite{Li_2018}. Mode identification was made based on the slope--period diagram founded on the 611 \gdor{} stars by \cite{Li_2019_all_gdor}, where the patterns clustered into different groups due to different kinds of modes. 
Figure~\ref{fig:KIC3228863} displays the amplitude spectrum and oscillation patterns of KIC\,3228863. 
Having removed the eclipses, we find clear oscillation patterns of g and r modes. We also find a hump at the orbital period $\sim0.73\,\mathrm{d}$, which could be caused by variations of the eclipses that prevented them from being totally removed. Mechanisms that can cause such eclipse variations include: apsidal motion \citep[e.g.][]{Hambleton_2013}, spot evolution and motion \citep[e.g.][]{Sriram_2017, Czesla_2019}, or instrumental effects.

We fitted the period spacings in each star using the traditional approximation of rotation (TAR) to obtain the near-core rotation rates \citep{VanReeth_2016_TAR, Li_2019_r_modes, Li_2019_all_gdor}. The pulsation periods in the co-rotating frame are given by 
\begin{equation}
P^\mathrm{TAR}_{nlm, \mathrm{co}}=\frac{\Pi_0}{\sqrt{\lambda_{l,m,s}}}\left(n+ \varepsilon_g \right),  \label{equ:TAR_P}
\end{equation}
where $\Pi_0=2\pi^2  \left( \int \frac{N}{r} \mathrm{d}r\right)^{-1}$ is the asymptotic period spacing, $N$ is the buoyancy frequency, $n$ is the radial order, and the phase term $\varepsilon_g$ was fixed as $0.5$. The phase term does not change the result since the radial orders are typically high ($\sim50$), and we fit the period spacings rather than individual periods. 
The symbol $\lambda_{l,m,s}$ is the eigenvalue of the Laplace tidal equation, which is specified by the angular degree $l$ for g modes or the value $k$ for r modes, the azimuthal order $m$, and the spin parameter~$s$ \citep[][]{Eckart_1960, Lee_1997_rotating, Townsend_2003, Saio_2018_Rossby_mode, Saio_2018}. The value $k$ is equal to $l-|m|$ in g modes. For r modes, $k$ is used because the angular degree $l$ is undefined \citep{Lee_1997_rotating}, and an even (odd) $|k|$ corresponds to temperature perturbations symmetric (antisymmetric) to the equator. We used the convention that positive values of $m$ denote prograde modes. Most \gdor{} stars rotate rapidly, and in this situation for prograde sectoral ($l=m$) g modes we have
\begin{equation}
    \lambda_{l=m} \approx m^2,
\end{equation}
and for $k\leq 2$ r modes,
\begin{equation}
    \lambda_{k\leq2} \approx m^2\left(2|k|-1\right)^{-2}, 
\end{equation}
which are the most commonly observed modes \citep{Berthomieu_1978, Townsend_2003, Saio_2018, Takata_2020}. Accurate values of $\lambda$ can be obtained from the stellar oscillation code \textsc{gyre} \citep{Townsend_2013, Townsend_2018_gyre_improve}. 
The spin parameter is defined as
\begin{equation}
s\equiv\frac{2f_\mathrm{rot}}{f_\mathrm{co}},\label{equ:spin_parameter}
\end{equation}
where $f_\mathrm{rot}$ is the rotation frequency and $f_\mathrm{co}$ is the pulsation frequency in the co-rotating frame. After calculating the pulsation periods in the co-rotating frame by eq.~\ref{equ:TAR_P}, the TAR frequency in the inertial reference frame is given by 
\begin{equation}
   f^\mathrm{TAR}_{nlm, \mathrm{in}} =1/P^\mathrm{TAR}_{nlm, \mathrm{co}}+m f_\mathrm{rot},
\end{equation}
and the period in the observer's reference frame is the inverse of the frequency. 
Finally, we can obtain the near-core rotation rates, the asymptotic spacings, and the radial orders by comparing the observed and calculated period spacing patterns.

For example, we applied our TAR fitting algorithm to KIC\,3228863 and found that its near-core rotation rate is $1.3716\pm0.0013\,\mathrm{d^{-1}}$. The ratio between the near-core rotation period and the orbital period is $0.9975\pm0.0009$, showing that the near-core region of KIC\,3228863 rotates synchronously with the orbital motion.

\section{Results}\label{sec: results}

\subsection{Sample selection}

\cite{Gaulme_2019} reported 115 \gdor~stars by conducting a systematic search for stellar pulsators in all the \textit{Kepler} eclipsing binaries from the Villanova EB database\footnote{\url{http://keplerebs.villanova.edu}} \citep[e.g.][]{Prsa_2011, Slawson_2011, Matijevic_2012, Conroy_2014, Kirk_2016}. We searched for g- and r-mode patterns in these 115 \gdor~stars and also in the \gdor~binaries reported in previous literature.

The known \gdor~binaries whose period spacings were not previously measured are: KIC\,3228863 \citep{Lee_2019_3228863, Lee_2014_3228863}; KIC\,4150611 \citep{Helminiak_2017_4150611}; KIC\,6048106 \citep{Lee_2016_6048106, Samadi_Ghadim_2018_6048106}; KIC\,6206751 \citep{Lee_2018_6206751}; KIC\,8569819 \citep{Kurtz_2015_8569819}; KIC\,9236858 \citep{Kjurkchieva_2016_9236858}; and KIC\,9851944 \citep{Guo_2016_9851944}. These works focused mainly on the dynamics of the binary systems, hence they did not measure the period spacings of g modes.

The known \gdor~binaries whose period spacings have already been measured are: KIC\,3341457 \citep{Li_2019_r_modes, Li_2019_all_gdor}; KIC\,4142768 \citep{Guo_2019_4142768}; KIC\,7385478 \citep{Guo_2019_7385478, Ozdarcan_2017_7385478}; KIC\,8197406 \citep{Li_2019_all_gdor}; KIC\,8197761 \citep{Li_2018}; KIC\,9592855 \citep{Guo2017}; KIC\,10080943 in which both components pulsate \citep{Keen_2015, Schmid_2016, Schmid_2015_10080943}; and KIC\,10486425 \citep{Zhang_2018}. 

We also included the eclipsing binary KIC\,2438249 which is located in the open cluster NGC\,6791. 
We produced light curves for all 17 available quarters of KIC\,2438249 using image subtraction photometry (Colman et al., in prep), which involves the fine resampling and centroid realignment of every frame of image data, followed by the subtraction of an average to remove non-variable background flux. 

In total, we found \EBnumber{} binaries with clear \gdor\,--\,type patterns, comprising 15 stars from previous literature and 20 new stars. We list our results in Table~\ref{tab:rot_Pi0_table}. For the binary systems, we list the orbital periods, eccentricities, and pseudo-synchronous periods if eccentric. For the seismic analysis, we list the mode identifications ($l$, $k$, $m$), asymptotic spacings, near-core rotation frequencies, the ranges of the radial orders, and the spin parameters. We also provide the ratio between the near-core and orbital periods, as well as the references for the stars that have been investigated. The amplitude spectra and period spacing patterns for all \EBnumber{} binaries are shown in Appendix~\ref{appendix}. Since two components in KIC\,10080943 and 12470041 pulsate, there are 37 figures in the appendix.

\begin{landscape}
\begin{table} 
\centering
\caption{Results for \EBnumber{} \gdor{} stars in eclipsing binariees. We list KIC numbers, \textit{Kepler} magnitudes $K_p$, orbital periods $P_\mathrm{orb}$, eccentricities $e$, pseudo-synchronous periods $P_\mathrm{ps}$, mode identifications (for g modes, we give the angular degrees $l$ and the azimuthal orders $m$, while for r modes, we list the value $k$ and $m$), asymptotic spacings $\Pi_0$, near-core rotation rates $f_\mathrm{rot}$, the ranges of radial orders $n$, ranges of spin parameters $s$, the ratios between the near-core and pseudo-synchrnous periods $P_\mathrm{rot}/P_\mathrm{ps}$, and references of the binary stars and the well-studied star KIC\,10080943. }\label{tab:rot_Pi0_table} 
\begin{tabular}{rrrrrrrrrrrrrrrr} 
\hline
KIC & $K_p$ & \multicolumn{1}{c}{$P_\mathrm{orb}$}  & \multicolumn{1}{c}{$e$} &  \multicolumn{1}{c}{$P_\mathrm{ps}$}  & $l$ & $k$ & $m$  & \multicolumn{1}{c}{$\Pi_0$}  & \multicolumn{1}{c}{$f_\mathrm{rot}$} & \multicolumn{2}{c}{$n$} & \multicolumn{2}{c}{$s$} & $P_\mathrm{rot}/P_\mathrm{ps}$ & References \\ 
   &   &  \multicolumn{1}{c}{days}  &  & \multicolumn{1}{c}{days} &  &  &   & \multicolumn{1}{c}{seconds} & \multicolumn{1}{c}{$\mathrm{d^{-1}}$}  &   min   &  max   & min          &     max  \\ 
\hline
1295531 & $11.94$& $1.686441(5)$ & \phantom{$0.0000(0)$}$0$& \phantom{$3.694(2)$}$-$& $1$ &       & \phantom{$-$}$1$ & \phantom{$-$}$4155\pm 27$\phantom{$0$}\phantom{$0$} & $0.5670(24)$ & $\phantom{0}20$ & $\phantom{0}50$ & $\phantom{0}0.8$ & $\phantom{0}2.4$& $1.046\pm0.005$\phantom{0}\phantom{0}&          \\
2438249 & $15.64$& $1.4463477(11)$ & \phantom{$0.0000(0)$}$0$& \phantom{$3.694(2)$}$-$& $1$ &       & \phantom{$-$}$1$ & \phantom{$-$}$5430\pm 100$\phantom{$0$} & $0.6682(25)$ & $\phantom{0}33$ & $\phantom{0}89$ & $\phantom{0}2.5$ & $\phantom{0}7.5$& $1.035\pm0.004$\phantom{0}\phantom{0}&          \\
3228863 & $11.81$& $0.73094425(4)$ & \phantom{$0.0000(0)$}$0$& \phantom{$3.694(2)$}$-$& $1$ &       & \phantom{$-$}$1$ & \phantom{$-$}$4321\pm 23$\phantom{$0$}\phantom{$0$} & $1.3716(13)$ & $\phantom{0}28$ & $\phantom{0}61$ & $\phantom{0}3.7$ & $\phantom{0}8.2$& $0.9975\pm0.0009$\phantom{0}& a, b \\
        &        &            &         &        &       & $-2$ & $-1$ &         &         & $\phantom{0}15$ & $\phantom{0}25$ & $\phantom{0}8.1$ & $11.2$&         &          \\
3341457 & $13.87$& $0.528179(5)$ & \phantom{$0.0000(0)$}$0$& \phantom{$3.694(2)$}$-$& $1$ &       & \phantom{$-$}$1$ & \phantom{$-$}$3840\pm 40$\phantom{$0$}\phantom{$0$} & $1.8592(14)$ & $\phantom{0}41$ & $\phantom{0}79$ & $\phantom{0}6.5$ & $12.7$& $1.0183\pm0.0008$\phantom{0}& c, d \\
        &        &            &         &        &       & $-2$ & $-1$ &         &         & $\phantom{0}34$ & $\phantom{0}60$ & $17.2$ & $30.2$&         &          \\
3867593 & $13.55$& $73.33776(5)$ & $-$& \phantom{$3.694(2)$}$-$& $1$ &       & \phantom{$-$}$1$ & \phantom{$-$}$4300\pm 160$\phantom{$0$} & $1.166(14)$ & $\phantom{0}22$ & $\phantom{0}37$ & $\phantom{0}2.3$ & $\phantom{0}4.1$& $0.01169\pm0.00014$&          \\
3869825 & $13.32$& $4.800663(8)$ & \phantom{$0.0000(0)$}$0$& \phantom{$3.694(2)$}$-$& $1$ &       & \phantom{$-$}$1$ & \phantom{$-$}$4300\pm 400$\phantom{$0$} & $0.851(25)$ & $\phantom{0}36$ & $\phantom{0}47$ & $\phantom{0}2.8$ & $\phantom{0}3.7$& $0.245\pm0.007$\phantom{0}\phantom{0}&          \\
4142768 & $12.12$& $13.99583(5)$ & $0.5820(20)$& $3.69(3)$& $2$ &       & \phantom{$-$}$2$ & \phantom{$-$}$4262\pm 4$\phantom{$0$}\phantom{$0$}\phantom{$0$} & $0.00037(12)$ & $\phantom{0}33$ & $\phantom{0}55$ & $\phantom{0}0.0$ & $\phantom{0}0.0$& $730\pm240$\phantom{0}\phantom{0}\phantom{0}\phantom{.}& e \\
        &        &            &         &        & $1$ &       & \phantom{$-$}$1$ &         &         & $\phantom{0}27$ & $\phantom{0}59$ & $\phantom{0}0.0$ & $\phantom{0}0.0$&         &          \\
4150611 & $7.90$& $8.6530978(22)$ & $0.374(7)$& $4.6(10)$& $1$ &       & \phantom{$-$}$1$ & \phantom{$-$}$4050\pm 80$\phantom{$0$}\phantom{$0$} & $1.58(10)$ & $\phantom{0}16$ & $\phantom{0}37$ & $\phantom{0}2.2$ & $\phantom{0}5.3$& $0.139\pm0.003$\phantom{0}\phantom{0}& f \\
4932691 & $13.62$& $18.112094(3)$ & $0.379(5)$& $9.41(14)$& $1$ &       & \phantom{$-$}$1$ & \phantom{$-$}$4720\pm 70$\phantom{$0$}\phantom{$0$} & $0.331(13)$ & $\phantom{0}16$ & $\phantom{0}21$ & $\phantom{0}0.4$ & $\phantom{0}0.6$& $0.321\pm0.014$\phantom{0}\phantom{0}& g \\
4947528 & $13.91$& $0.49698822(4)$ & \phantom{$0.0000(0)$}$0$& \phantom{$3.694(2)$}$-$& $1$ &       & \phantom{$-$}$1$ & \phantom{$-$}$4150\pm 260$\phantom{$0$} & $2.017(20)$ & $\phantom{0}23$ & $\phantom{0}44$ & $\phantom{0}4.2$ & $\phantom{0}8.2$& $1.00\pm0.010$\phantom{0}\phantom{0}&          \\
5565486 & $14.96$& $2.8250483(4)$ & \phantom{$0.0000(0)$}$0$& \phantom{$3.694(2)$}$-$& $1$ &       & \phantom{$-$}$1$ & \phantom{$-$}$4810\pm 70$\phantom{$0$}\phantom{$0$} & $0.458(8)$ & $\phantom{0}19$ & $\phantom{0}28$ & $\phantom{0}0.8$ & $\phantom{0}1.2$& $0.772\pm0.014$\phantom{0}\phantom{0}&          \\
5809827 & $13.53$& $1.22211021(15)$ & \phantom{$0.0000(0)$}$0$& \phantom{$3.694(2)$}$-$& $1$ &       & \phantom{$-$}$1$ & \phantom{$-$}$9700\pm 800$\phantom{$0$} & $1.017(16)$ & $\phantom{0}19$ & $\phantom{0}26$ & $\phantom{0}4.1$ & $\phantom{0}5.7$& $0.804\pm0.013$\phantom{0}\phantom{0}&          \\
6048106 & $14.09$& $1.559360(4)$ & \phantom{$0.0000(0)$}$0$& \phantom{$3.694(2)$}$-$& $1$ &       & \phantom{$-$}$1$ & \phantom{$-$}$3290\pm 40$\phantom{$0$}\phantom{$0$} & $0.570(15)$ & $\phantom{0}13$ & $\phantom{0}24$ & $\phantom{0}0.4$ & $\phantom{0}0.8$& $1.12\pm0.03$\phantom{0}\phantom{0}\phantom{0}& h, i \\
6206751 & $12.14$& $1.24534226(13)$ & \phantom{$0.0000(0)$}$0$& \phantom{$3.694(2)$}$-$& $1$ &       & \phantom{$-$}$1$ & \phantom{$-$}$4180\pm 60$\phantom{$0$}\phantom{$0$} & $0.856(3)$ & $\phantom{0}35$ & $\phantom{0}79$ & $\phantom{0}2.6$ & $\phantom{0}6.3$& $0.939\pm0.003$\phantom{0}\phantom{0}& j \\
6290382 & $13.18$& $6.128161(5)$ & \phantom{$0.0000(0)$}$0$& \phantom{$3.694(2)$}$-$& $1$ &       & \phantom{$-$}$1$ & \phantom{$-$}$4070\pm 60$\phantom{$0$}\phantom{$0$} & $1.002(8)$ & $\phantom{0}16$ & $\phantom{0}29$ & $\phantom{0}1.3$ & $\phantom{0}2.5$& $0.1629\pm0.0014$\phantom{0}&          \\
6292398 & $9.83$& $9.239875(5)$ & \phantom{$0.0000(0)$}$0$& \phantom{$3.694(2)$}$-$& $4$ &       & \phantom{$-$}$4$ & \phantom{$-$}$3937\pm 21$\phantom{$0$}\phantom{$0$} & $1.3333(11)$ & $\phantom{0}36$ & $\phantom{0}66$ & $\phantom{0}1.0$ & $\phantom{0}1.9$& $0.08117\pm0.00007$&          \\
        &        &            &         &        & $3$ &       & \phantom{$-$}$3$ &         &         & $\phantom{0}33$ & $\phantom{0}62$ & $\phantom{0}1.3$ & $\phantom{0}2.4$&         &          \\
        &        &            &         &        & $2$ &       & \phantom{$-$}$2$ &         &         & $\phantom{0}26$ & $\phantom{0}67$ & $\phantom{0}1.5$ & $\phantom{0}4.0$&         &          \\
        &        &            &         &        & $1$ &       & \phantom{$-$}$1$ &         &         & $\phantom{0}26$ & $\phantom{0}74$ & $\phantom{0}3.0$ & $\phantom{0}8.8$&         &          \\
        &        &            &         &        &       & $-2$ & $-1$ &         &         & $\phantom{0}41$ & $\phantom{0}52$ & $15.4$ & $19.3$&         &          \\
7385478 & $11.47$& $1.65547262(4)$ & \phantom{$0.0000(0)$}$0$& \phantom{$3.694(2)$}$-$& $1$ &       & \phantom{$-$}$1$ & \phantom{$-$}$4330\pm 70$\phantom{$0$}\phantom{$0$} & $0.659(14)$ & $\phantom{0}14$ & $\phantom{0}20$ & $\phantom{0}0.7$ & $\phantom{0}1.1$& $0.917\pm0.019$\phantom{0}\phantom{0}& k, l \\
7515679 & $12.25$& $5.546081(4)$ & $-$& \phantom{$3.694(2)$}$-$& $2$ &       & \phantom{$-$}$2$ & \phantom{$-$}$7200\pm 900$\phantom{$0$} & $0.562(12)$ & $\phantom{0}59$ & $\phantom{0}81$ & $\phantom{0}2.6$ & $\phantom{0}3.6$& $0.321\pm0.007$\phantom{0}\phantom{0}&          \\
8197406 & $12.57$& $2.513006(4)$ & \phantom{$0.0000(0)$}$0$& \phantom{$3.694(2)$}$-$& $2$ &       & \phantom{$-$}$2$ & \phantom{$-$}$4940\pm 120$\phantom{$0$} & $0.391(3)$ & $\phantom{0}57$ & $107$ & $\phantom{0}1.1$ & $\phantom{0}2.2$& $1.018\pm0.008$\phantom{0}\phantom{0}& d \\
8197761 & $10.65$& $9.8686667(27)$ & \phantom{$0.0000(0)$}$0$& \phantom{$3.694(2)$}$-$& $1$ &       & \phantom{$-$}$1$ & \phantom{$-$}$3862\pm 5$\phantom{$0$}\phantom{$0$}\phantom{$0$} & $0.00336(13)$ & $\phantom{0}27$ & $\phantom{0}50$ & $\phantom{0}0.0$ & $\phantom{0}0.0$& $30\pm1$\phantom{0}\phantom{0}\phantom{0}\phantom{0}\phantom{0}\phantom{.}& m, n \\
8330092 & $13.48$& $0.32172416(4)$ & \phantom{$0.0000(0)$}$0$& \phantom{$3.694(2)$}$-$& $1$ &       & \phantom{$-$}$1$ & \phantom{$-$}$3640\pm 200$\phantom{$0$} & $1.36(4)$ & $\phantom{0}16$ & $\phantom{0}23$ & $\phantom{0}1.6$ & $\phantom{0}2.4$& $2.29\pm0.06$\phantom{0}\phantom{0}\phantom{0}&          \\
8429450 & $13.10$& $2.70515386(25)$ & \phantom{$0.0000(0)$}$0$& \phantom{$3.694(2)$}$-$& $2$ &       & \phantom{$-$}$2$ & \phantom{$-$}$5790\pm 200$\phantom{$0$} & $0.026(20)$ & $\phantom{0}17$ & $\phantom{0}20$ & $\phantom{0}0.0$ & $\phantom{0}0.0$& $14\pm11$\phantom{0}\phantom{0}\phantom{0}\phantom{0}\phantom{.}&          \\
8548416 & $13.34$& $1.16364475(7)$ & \phantom{$0.0000(0)$}$0$& \phantom{$3.694(2)$}$-$& $2$ &       & \phantom{$-$}$2$ & \phantom{$-$}$5710\pm 250$\phantom{$0$} & $0.90(10)$ & $\phantom{0}51$ & $\phantom{0}56$ & $\phantom{0}2.9$ & $\phantom{0}3.2$& $0.96\pm0.010$\phantom{0}\phantom{0}&          \\
        &        &            &         &        & $1$ &       & \phantom{$-$}$1$ &         &         & $\phantom{0}32$ & $\phantom{0}41$ & $\phantom{0}3.5$ & $\phantom{0}4.6$&         &          \\
8569819 & $13.04$& $20.849926(9)$ & \phantom{$0.0000(0)$}$0$& \phantom{$3.694(2)$}$-$& $1$ &       & \phantom{$-$}$1$ & \phantom{$-$}$4061\pm 16$\phantom{$0$}\phantom{$0$} & $0.615(4)$ & $\phantom{0}\phantom{0}6$ & $\phantom{0}25$ & $\phantom{0}0.2$ & $\phantom{0}1.2$& $0.0780\pm0.0005$\phantom{0}& o \\
9108579 & $11.56$& $1.16962726(5)$ & \phantom{$0.0000(0)$}$0$& \phantom{$3.694(2)$}$-$& $2$ &       & \phantom{$-$}$2$ & \phantom{$-$}$4210\pm 230$\phantom{$0$} & $0.846(13)$ & $\phantom{0}51$ & $\phantom{0}62$ & $\phantom{0}1.9$ & $\phantom{0}2.4$& $1.011\pm0.016$\phantom{0}\phantom{0}&          \\
        &        &            &         &        & $1$ &       & \phantom{$-$}$1$ &         &         & $\phantom{0}41$ & $\phantom{0}48$ & $\phantom{0}3.1$ & $\phantom{0}3.7$&         &          \\
9236858 & $13.04$& $2.53708215(15)$ & \phantom{$0.0000(0)$}$0$& \phantom{$3.694(2)$}$-$& $1$ &       & \phantom{$-$}$1$ & \phantom{$-$}$3990\pm 50$\phantom{$0$}\phantom{$0$} & $0.390(14)$ & $\phantom{0}14$ & $\phantom{0}29$ & $\phantom{0}0.4$ & $\phantom{0}0.8$& $1.01\pm0.04$\phantom{0}\phantom{0}\phantom{0}& p \\
9592855 & $12.21$& $1.2193249(10)$ & \phantom{$0.0000(0)$}$0$& \phantom{$3.694(2)$}$-$& $1$ &       & \phantom{$-$}$1$ & \phantom{$-$}$3960\pm 120$\phantom{$0$} & $0.829(27)$ & $\phantom{0}13$ & $\phantom{0}22$ & $\phantom{0}0.8$ & $\phantom{0}1.4$& $0.99\pm0.03$\phantom{0}\phantom{0}\phantom{0}& r \\
9850387 & $13.54$& $2.74849909(20)$ & \phantom{$0.0000(0)$}$0$& \phantom{$3.694(2)$}$-$& $2$ &       & \phantom{$-$}$2$ & \phantom{$-$}$3894\pm 7$\phantom{$0$}\phantom{$0$}\phantom{$0$} & $0.0053(15)$ & $\phantom{0}38$ & $\phantom{0}47$ & $\phantom{0}0.0$ & $\phantom{0}0.0$& $69\pm19$\phantom{0}\phantom{0}\phantom{0}\phantom{0}\phantom{.}& q \\
        &        &            &         &        & $1$ &       & \phantom{$-$}$1$ &         &         & $\phantom{0}21$ & $\phantom{0}40$ & $\phantom{0}0.0$ & $\phantom{0}0.0$&         &          \\
9851944 & $11.24$& $2.16390187(14)$ & \phantom{$0.0000(0)$}$0$& \phantom{$3.694(2)$}$-$& $2$ &       & \phantom{$-$}$2$ & \phantom{$-$}$3500\pm 500$\phantom{$0$} & $0.41(5)$ & $\phantom{0}37$ & $\phantom{0}42$ & $\phantom{0}0.5$ & $\phantom{0}0.6$& $1.13\pm0.15$\phantom{0}\phantom{0}\phantom{0}& s \\
\hline
 \end{tabular} 
\begin{tablenotes}
\item \emph{Note}: (a)  \cite{Lee_2019_3228863}; (b)  \cite{Lee_2014_3228863}; (c)  \cite{Li_2019_r_modes}; (d)  \cite{Li_2019_all_gdor}; (e)  \cite{Guo_2019_4142768}; (f)  \cite{Helminiak_2017_4150611}; (g)  \cite{Kjurkchieva_2017}; (h)  \cite{Lee_2016_6048106}; (i)  \cite{Samadi_Ghadim_2018_6048106}; (j)  \cite{Lee_2018_6206751}; (k)  \cite{Guo_2019_7385478}; (l)  \cite{Ozdarcan_2017_7385478}; (m)  \cite{Li_2018}; (n)  \cite{Sowicka_2017}; (o)  \cite{Kurtz_2015_8569819}; (p)  \cite{Kjurkchieva_2016_9236858}; (q)  \cite{Zhang_2020}; (r)  \cite{Guo2017}; (s)  \cite{Guo_2016_9851944}; (t)  \cite{Keen_2015}; (u)  \cite{Schmid_2016}; (v)  \cite{Schmid_2015_10080943}; (w)  \cite{Zhang_2018}. 
 \end{tablenotes}
 \end{table} 
\end{landscape}

\setcounter{table}{0}

\begin{landscape}
\begin{table} 
\centering
\caption{continued.}\label{tab:rot_Pi0_table_2} 
\begin{tabular}{rrrrrrrrrrrrrrrr} 
\hline
KIC & $K_p$ & \multicolumn{1}{c}{$P_\mathrm{orb}$}  & \multicolumn{1}{c}{$e$} &  \multicolumn{1}{c}{$P_\mathrm{ps}$}  & $l$ & $k$ & $m$  & \multicolumn{1}{c}{$\Pi_0$}  & \multicolumn{1}{c}{$f_\mathrm{rot}$} & \multicolumn{2}{c}{$n$} & \multicolumn{2}{c}{$s$} & $P_\mathrm{rot}/P_\mathrm{ps}$ & References \\ 
   &   &  \multicolumn{1}{c}{days}  &  & \multicolumn{1}{c}{days} &  &  &   & \multicolumn{1}{c}{seconds} & \multicolumn{1}{c}{$\mathrm{d^{-1}}$}  &   min   &  max   & min          &     max  \\ 
\hline
10080943A & $11.81$& $15.3364(3)$ & $0.449(5)$& $6.5(10)$& $1$ &       & \phantom{$-$}$1$ & \phantom{$-$}$4090\pm 3$\phantom{$0$}\phantom{$0$}\phantom{$0$} & $0.0884(7)$ & $\phantom{0}24$ & $\phantom{0}38$ & $\phantom{0}0.1$ & $\phantom{0}0.2$& $1.75\pm0.03$\phantom{0}\phantom{0}\phantom{0}& t, u, v \\
        &        &            &         &        & $1$ &       & $-1$ &         &         & $\phantom{0}30$ & $\phantom{0}43$ & $\phantom{0}0.1$ & $\phantom{0}0.2$&         &          \\
10080943B & $-$& $-$ & $-$& $-$& $1$ &       & \phantom{$-$}$1$ & \phantom{$-$}$4072\pm 3$\phantom{$0$}\phantom{$0$}\phantom{$0$} & $0.1361(12)$ & $\phantom{0}21$ & $\phantom{0}28$ & $\phantom{0}0.1$ & $\phantom{0}0.2$& $1.138\pm0.022$\phantom{0}\phantom{0}& t, u, v \\
        &        &            &         &        & $1$ &       & \phantom{$-$}$0$ &         &         & $\phantom{0}24$ & $\phantom{0}28$ & $\phantom{0}0.2$ & $\phantom{0}0.2$&         &          \\
        &        &            &         &        & $1$ &       & $-1$ &         &         & $\phantom{0}22$ & $\phantom{0}33$ & $\phantom{0}0.1$ & $\phantom{0}0.2$&         &          \\
10486425 & $12.46$& $5.27482053(26)$ & \phantom{$0.0000(0)$}$0$& \phantom{$3.694(2)$}$-$& $2$ &       & \phantom{$-$}$2$ & \phantom{$-$}$4740\pm 130$\phantom{$0$} & $0.233(16)$ & $\phantom{0}18$ & $\phantom{0}25$ & $\phantom{0}0.2$ & $\phantom{0}0.2$& $0.81\pm0.06$\phantom{0}\phantom{0}\phantom{0}& w \\
11820830 & $12.08$& $12.731948(12)$ & $-$& \phantom{$3.694(2)$}$-$& $1$ &       & \phantom{$-$}$1$ & \phantom{$-$}$2390\pm 140$\phantom{$0$} & $0.411(16)$ & $\phantom{0}75$ & $\phantom{0}94$ & $\phantom{0}1.4$ & $\phantom{0}1.9$& $0.191\pm0.007$\phantom{0}\phantom{0}&          \\
11973705 & $9.12$& $6.772142(5)$ & \phantom{$0.0000(0)$}$0$& \phantom{$3.694(2)$}$-$& $1$ &       & \phantom{$-$}$1$ & \phantom{$-$}$4040\pm 250$\phantom{$0$} & $1.46(4)$ & $\phantom{0}15$ & $\phantom{0}19$ & $\phantom{0}1.8$ & $\phantom{0}2.4$& $0.101\pm0.003$\phantom{0}\phantom{0}&          \\
12470041A & $13.41$& $14.667697(4)$ & $-$& \phantom{$3.694(2)$}$-$& $2$ &       & \phantom{$-$}$2$ & \phantom{$-$}$3720\pm 80$\phantom{$0$}\phantom{$0$} & $0.295(6)$ & $\phantom{0}42$ & $\phantom{0}68$ & $\phantom{0}0.4$ & $\phantom{0}0.7$& $0.231\pm0.005$\phantom{0}\phantom{0}&          \\
12470041B & $-$& $-$ & $-$& \phantom{$3.694(2)$}$-$& $1$ &       & \phantom{$-$}$1$ & \phantom{$-$}$6090\pm 90$\phantom{$0$}\phantom{$0$} & $0.460(4)$ & $\phantom{0}25$ & $\phantom{0}34$ & $\phantom{0}1.4$ & $\phantom{0}1.9$& $0.1483\pm0.0014$\phantom{0}&          \\
12785282 & $13.51$& $0.78874991(4)$ & \phantom{$0.0000(0)$}$0$& \phantom{$3.694(2)$}$-$& $1$ &       & \phantom{$-$}$1$ & \phantom{$-$}$4100\pm 1000$ & $1.25(9)$ & $\phantom{0}31$ & $\phantom{0}35$ & $\phantom{0}3.3$ & $\phantom{0}3.8$& $1.01\pm0.07$\phantom{0}\phantom{0}\phantom{0}&          \\
\hline
 \end{tabular} 
\begin{tablenotes}
\item \emph{Note}: (a)  \cite{Lee_2019_3228863}; (b)  \cite{Lee_2014_3228863}; (c)  \cite{Li_2019_r_modes}; (d)  \cite{Li_2019_all_gdor}; (e)  \cite{Guo_2019_4142768}; (f)  \cite{Helminiak_2017_4150611}; (g)  \cite{Kjurkchieva_2017}; (h)  \cite{Lee_2016_6048106}; (i)  \cite{Samadi_Ghadim_2018_6048106}; (j)  \cite{Lee_2018_6206751}; (k)  \cite{Guo_2019_7385478}; (l)  \cite{Ozdarcan_2017_7385478}; (m)  \cite{Li_2018}; (n)  \cite{Sowicka_2017}; (o)  \cite{Kurtz_2015_8569819}; (p)  \cite{Kjurkchieva_2016_9236858}; (q)  \cite{Zhang_2020}; (r)  \cite{Guo2017}; (s)  \cite{Guo_2016_9851944}; (t)  \cite{Keen_2015}; (u)  \cite{Schmid_2016}; (v)  \cite{Schmid_2015_10080943}; (w)  \cite{Zhang_2018}. 
 \end{tablenotes}
 \end{table} 
\end{landscape}

\subsection{Near-core rotation rates and asymptotic spacings} \label{subsec:rotation_Pi0}
\begin{figure}
    \centering
    \includegraphics[width=1\linewidth]{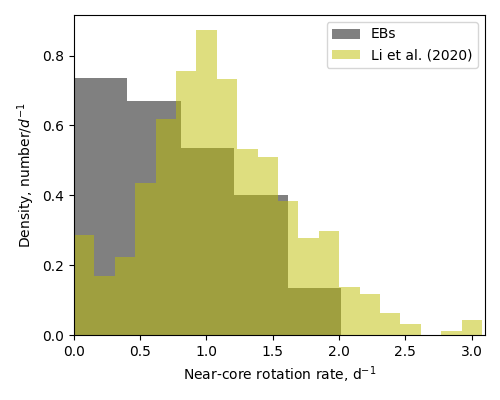}
    \caption{The distributions of the near-core rotation rates. The y-axis is normalised density so that the area under each histogram will sum to 1. The grey histogram shows the distribution of the \EBnumber{} binaries studied in this work. The yellow histogram contains 611 \gdor{} stars by \protect\cite{Li_2019_all_gdor}. }
    \label{fig:near-core-rotation_histogram}
\end{figure}

\begin{figure}
    \centering
    \includegraphics[width=1\linewidth]{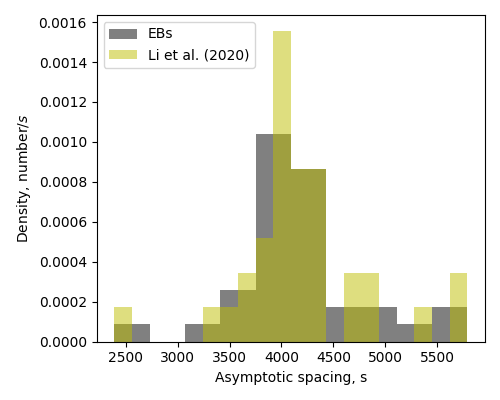}
    \caption{The distributions of the asymptotic spacings. The colours have the same meanings as Fig.~\ref{fig:near-core-rotation_histogram}.}
    \label{fig:Pi0_hist}
\end{figure}

\begin{figure*}
    \centering
    \includegraphics[width=1\linewidth]{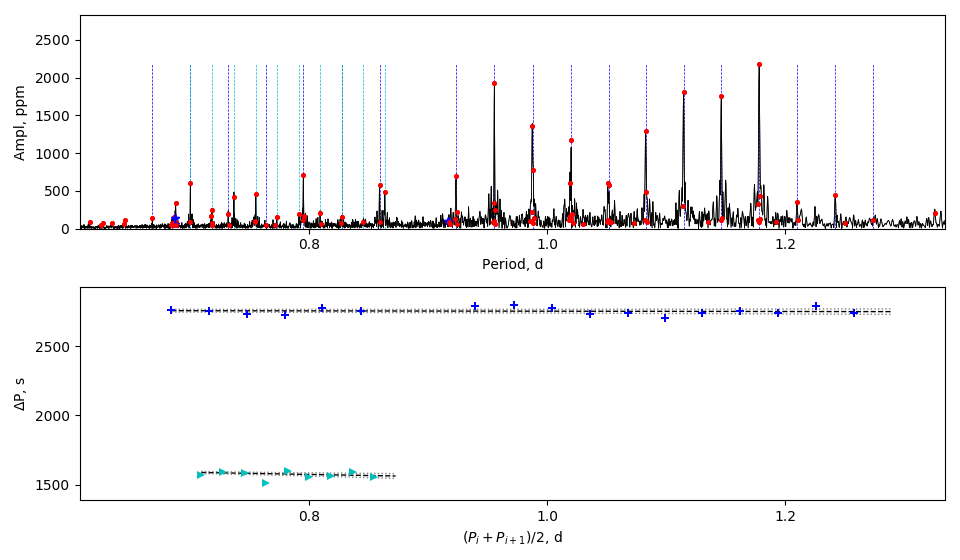}
    \caption{Top: the amplitude spectrum of KIC\,9850387 after removing the eclipses from the light curve. The blue vertical dashed lines show the peaks of $l=1$ g modes, and the green vertical dashed lines mark the $l=2$ g modes. The blue stars show the locations of orbital period harmonics. Bottom: the period spacings of $l=1$ and $2$ g modes. The blue crosses are the pattern of $l=1$ g modes and the green symbols are those of $l=2$ g modes. The dashed lines show the best-fitting linear relation, whose uncertainties are shown by the grey dotted lines surrounding them.}
    \label{fig:KIC9850387}
\end{figure*}

We show the distribution of the near-core rotation rates of the \EBnumber{} binaries in grey in Fig.~\ref{fig:near-core-rotation_histogram}. As a comparison, the near-core rotation rates of 611 \gdor{} stars by \cite{Li_2019_all_gdor} are also plotted as the yellow histogram. 

The near-core rotation rates of the eclipsing binary sample are significantly lower than those from the 611-star sample by \cite{Li_2019_all_gdor}. For the latter, \cite{Li_2019_all_gdor} reported that most stars rotate around $1\,\cpd$ while some stars rotate slowly, forming an excess at the left end of the distribution. Conversely, the near-core rotations of the eclipsing binaries reported in this work are significantly slower, where the number decreases with increasing rotation rate. In a binary system where stars initially rotate faster than the orbital frequency, the tidal force decelerates the surface of the components and gradually slows down the near-core regions. \textcolor{black}{In the 611-star sample by \cite{Li_2019_all_gdor}, there are still some binaries, such as the extremely shallow eclipsing binaries KIC\,3341457 and KIC\,8197761, the double-pulsator binary KIC\,10080943, the phase-modulation binaries by \cite{Murphy_2018}, and the spectroscopic binary KIC\,4480321 by \cite{Lampens_2018}. All the known \gdor{} stars with short orbital periods ($P_\mathrm{orb}<20\,\mathrm{d}$) show the near-core rotation rates slower than the peak value of the distribution from \cite{Li_2019_all_gdor} in Fig.~\ref{fig:near-core-rotation_histogram}. Therefore, we speculate that the slow-rotator excess found by \cite{Li_2019_all_gdor} might arise from non-eclipsing binaries with small semi-major axes. }

The most rapid rotator in the eclipsing binary sample is KIC\,4947528, whose near-core rotation rate is $2.017\pm0.020\,\cpd$. The orbital period of KIC\,4947528 is almost equal to the near-core rotation period, showing that the system has been tidally locked. 

We also find three extremely slowly-rotating stars. 
For these stars, the measurements of the near-core rotation rates have considerable uncertainties, because the slopes are typically small and are dominated by the fluctuations caused by the chemical abundance gradients, rather than the rotational effect \citep[e.g.][]{Miglio_2008}. However, we still confirm that these stars are rotating at extremely low rates. 
The slowest one is KIC\,4142768, whose period spacing patterns have an average gradient of zero but show fluctuations and were reported by \cite{Guo_2019_4142768}. The near-core rotation rate of KIC\,4142768 is $0.00037\pm0.00012\,\cpd$, which is much lower than the orbital frequency ($\sim0.071\,\mathrm{d^{-1}}$). \cite{Aerts_2019arXiv191212300A} derived a near-core rotation rate $0.0060\pm0.0026\,\mathrm{d^{-1}}$ for KIC\,4142768, which is faster than our value but is still very slow. KIC\,9850387 is another extremely slow rotator, whose amplitude spectrum and period spacing patterns are shown in Fig.~\ref{fig:KIC9850387}. The period spacings are almost identical, implying an extremely slow rotation rate, and we obtained a near-core rotation rate of $0.0053\pm0.0015\,\cpd$ \cite[see a recent study in][]{Zhang_2020}. The last one is KIC\,8197761, whose near-core rotation rate is $0.00336\pm0.00013\,\cpd$, which was measured by rotational splittings by \cite{Sowicka_2017} and \cite{Li_2018}, hence shows a smaller uncertainty. \cite{Sowicka_2017} made a spectroscopic observation of KIC\,8197761 and found that the stellar surface is synchronised with the orbit, but the stellar core rotates $\sim 30$ times slower. The strong differential rotation is unexpected \citep[e.g.][]{Jermyn_2020}, and we will give a discussion in Section~\ref{subsec:Anti-tide-hypothesis}. 

Figure~\ref{fig:Pi0_hist} displays the distributions of the asymptotic spacings, both for the eclipsing binary sample (grey) and the 611-star sample by \cite{Li_2019_all_gdor} (yellow). The distributions are similar, with the mean values both being around $4000\,\mathrm{s}$. KIC\,11820830 has the smallest asymptotic spacing (2390\,s), and its period spacing pattern shows a significant fluctuation, hence we infer that the star has evolved to the end of the main sequence. KIC\,5809827 has the largest asymptotic spacing in our sample (9727\,s) and the star has nearly equal periods between the near-core region and the orbital motion. This star might be a very young star, since the asymptotic spacing decreases with stellar evolution. We considered whether it might be a Slowly Pulsating B (SPB) star since it shows a higher asymptotic spacing than \gdor{} stars \citep[e.g.][]{Papics_2017}. However, the effective temperature given by \cite{Mathur_2017} is $6369 \pm 226\,\mathrm{K}$, implying that it is not likely to be a B-type star. 

\subsection{The relation between the near-core and orbital periods}\label{subsec:near_core_orbital_periods}

\begin{figure*}
    \centering
    \includegraphics[width=\linewidth]{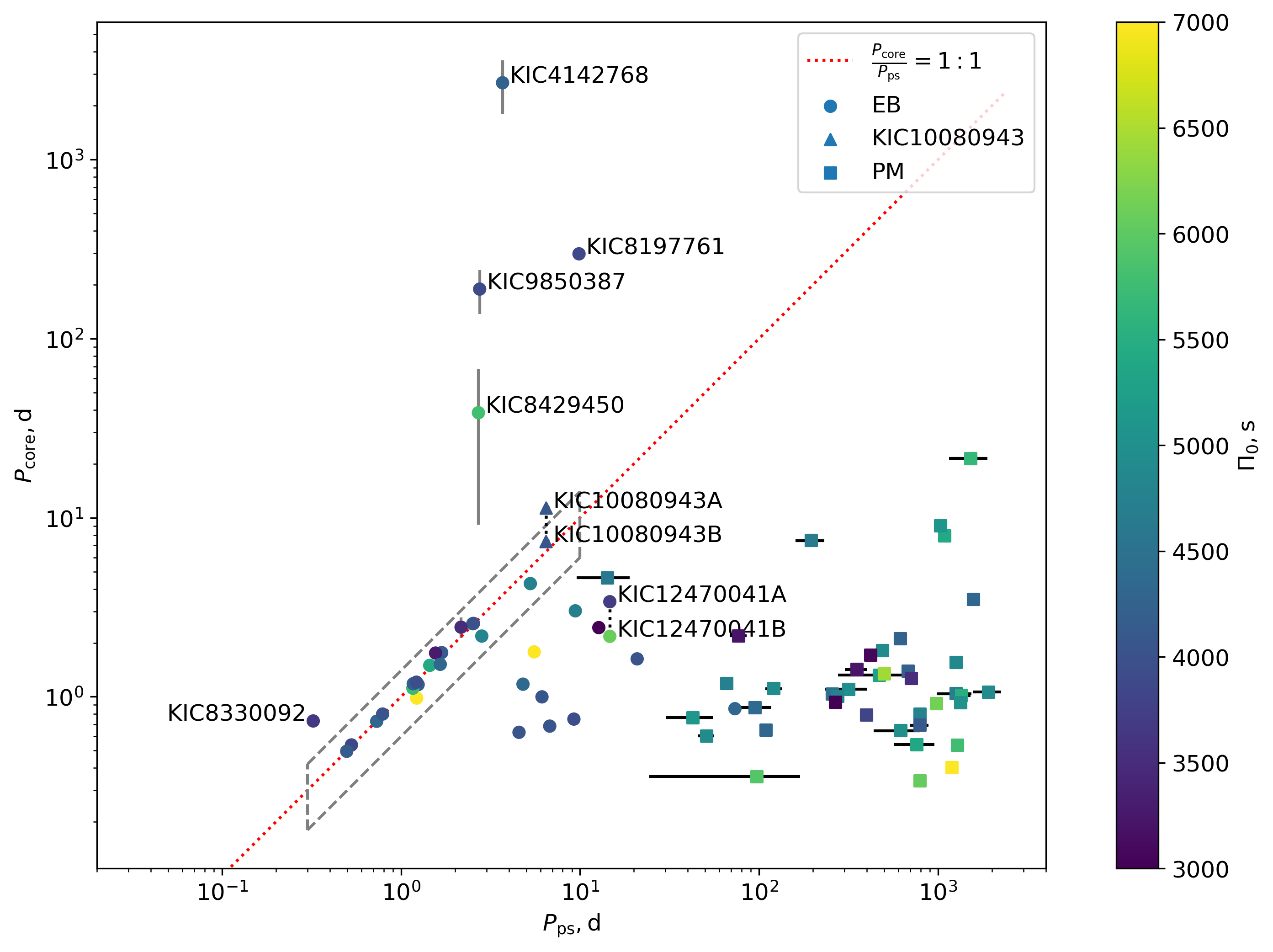}
    \caption{The relation between the near-core rotation periods $P_\mathrm{core}$ and pseudo-synchronous periods $P_\mathrm{ps}$, coloured by their asymptotic spacings. For those whose eccentricities are zero or unknown, we use the orbital period instead of $P_\mathrm{ps}$. The circles are the eclipsing binaries reported by this work, while the squares are the phase-modulation (PM) binaries reported by \protect \cite{Murphy_2018}. The dotted red line displays the location where the near-core rotation period is equal to the pseudo-synchronous period. The grey dashed lines enclose the tidally locked regions, whose $P_\mathrm{core}/P_\mathrm{ps}$ is between 0.7 and 1.3. Note that the boundaries of the colour bar are set between 3000\,s and 7000\,s, hence the stars marked by light yellow might have asymptotic spacings larger than 7000\,s.}
    \label{fig:core-orbital-relation}
\end{figure*}

\begin{figure}
    \centering
    \includegraphics[width=\linewidth]{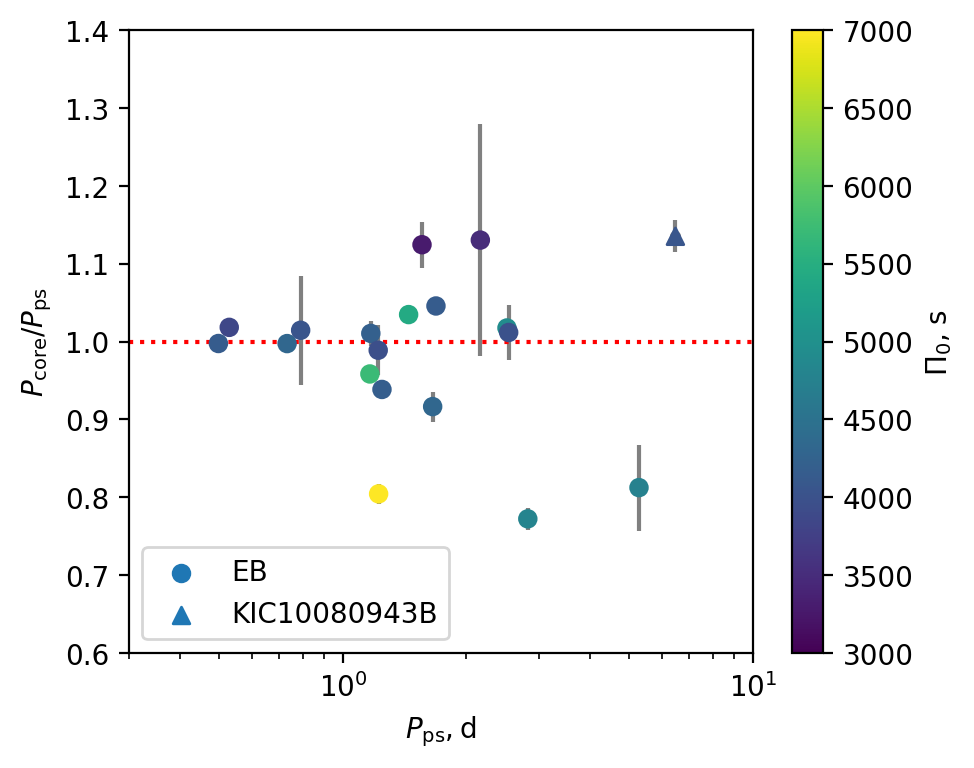}
    \caption{The ratios between the near-core and pseudo-synchronous periods of the tidally locked binaries shown in the grey box in Fig.~\ref{fig:core-orbital-relation}, coloured by their asymptotic spacings. The boundaries of the colourbar are set between 3000\,s and 7000\,s. }
    \label{fig:core_orbital_period_ratio}
\end{figure}

Figure~\ref{fig:core-orbital-relation} shows the relation between the near-core rotation periods and the orbital periods. For eccentric systems, we use the pseudo-synchronous periods $P_\mathrm{ps}$ (Eq.~\ref{equ:p_ps}). We plot the \EBnumber{} binaries (circles) by this work, and \PMnumber{} phase-modulation (PM) binaries (squares) reported by \cite{Murphy_2018} whose near-core rotations were measured using the g-mode patterns by \cite{Li_2019_all_gdor}. Due to the detection limit, the PM binaries typically have longer orbital periods from tens to thousands of days. The \gdor{} components of those systems are rotating freely since the tidal forces are weak, and they have near-core rotation rates around $1\,\cpd$. 

We classified the eclipsing binaries into three groups based on the ratio between the near-core and orbital periods $P_\mathrm{core}/P_\mathrm{orb}$. 

\begin{itemize}
    \item $P_\mathrm{core}/P_\mathrm{orb}<1$: with the orbital periods around $10\,\mathrm{d}$, we see some eclipsing binaries whose near-core rotation periods are smaller than the orbital periods (below the red dotted line). These stars lie at the transition between the freely-rotating and the tidally-locked binaries, hence they have slightly higher near-core rotation periods than the PM binaries around $10^3\,\mathrm{d}$, but not significantly. %Their near-core rotation rates are slightly higher than those of PM binaries because the tidal forces are dragging the rotations of the near-core regions. 
    
    \item $P_\mathrm{core}/P_\mathrm{orb} \approx 1$: many stars in eclipsing binaries have almost the same orbital and near-core rotation periods, falling along the $1:1$ straight line in the diagram (the red dotted line). We identify a tidally-locked star if the rate of its near-core and orbital periods is between $0.7$ and $1.3$, which is indicated by the grey dashed box in Fig.~\ref{fig:core-orbital-relation}. Figure~\ref{fig:core_orbital_period_ratio} displays the period ratios of the tidally-locked stars. We see that there is a smaller spread at the short-period side, presumably because the tidal force is stronger. The tidally-locked star with the shortest orbital period is KIC\,4947528. Binary offers an independent test of the TAR, and it proves that the TAR works effectively for dipole sectoral g modes with near-core rotation period of $0.5\,\mathrm{d}$, or with spin parameter up to $8$.
    
    \item $P_\mathrm{core}/P_\mathrm{orb} > 1$: apart from three extremely slowly rotating stars introduced in Section~\ref{subsec:rotation_Pi0}, there are three other stars that have the near-core rotation periods substantially longer than their orbital periods. One is the well-studied star KIC\,10080943, which is a non-eclipsing spectroscopic binary in which both components are \gdor{} pulsators \citep[e.g.][]{Keen_2015, Schmid_2015_10080943, Schmid_2016}. For KIC\,10080943B, the ratio between the near-core and pseudo-synchronous period is $1.14$, hence it is tidally locked. However, the ratio is $1.75$ for KIC\,10080943A, so this component is not pseudo-synchronous. The second one is KIC\,8429450. This star only shows four g-mode peaks, hence there is a large uncertainty on the near-core rotation rate. However, it is not equal to the orbital period within the uncertainty range. The third star is KIC\,8330092. This star has the shortest orbital period in the sample but its \gdor{} component is not synchronous. \textcolor{black}{We will give a detailed discussion in Section~\ref{subsec:8330092}. We also find that KIC\,4480321 is a spectroscopic binary \citep{Lampens_2018}, whose orbital period is $9.16592\pm0.00006$ days and the near-core rotation rate is $121\pm4$ days \citep{Li_2018}. A hypothesis explaining the slow rotators is given in Section~\ref{subsec:Anti-tide-hypothesis}.}
\end{itemize}

\subsection{KIC\,8330092 with a pulsating third component}\label{subsec:8330092}
\begin{figure}
    \centering
    \includegraphics[width=1\linewidth]{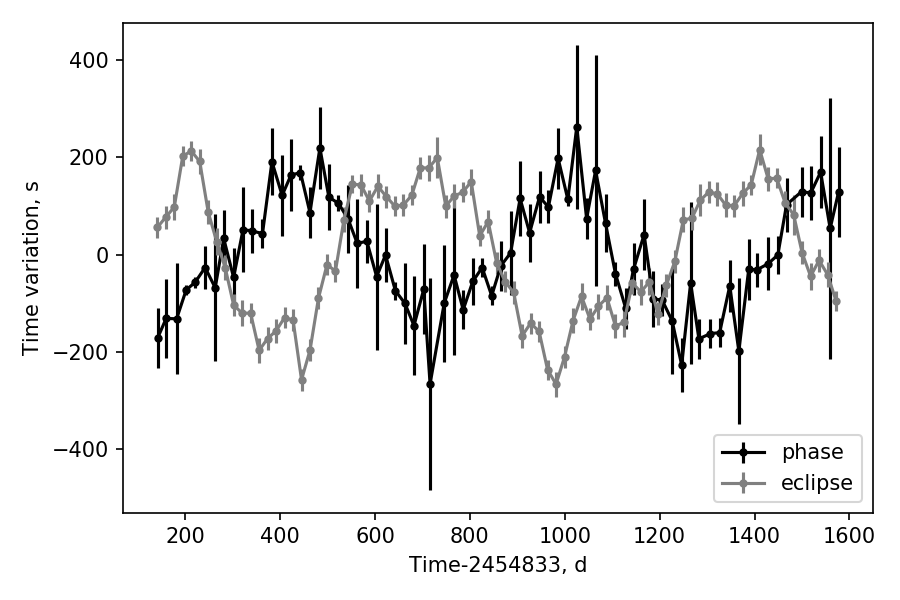}
    \caption{Time delays of the eclipses and the p-mode phases of KIC\,8330092. }
    \label{fig:time_delay_8330092}
\end{figure}

\begin{figure}
    \centering
    \includegraphics[width=\linewidth]{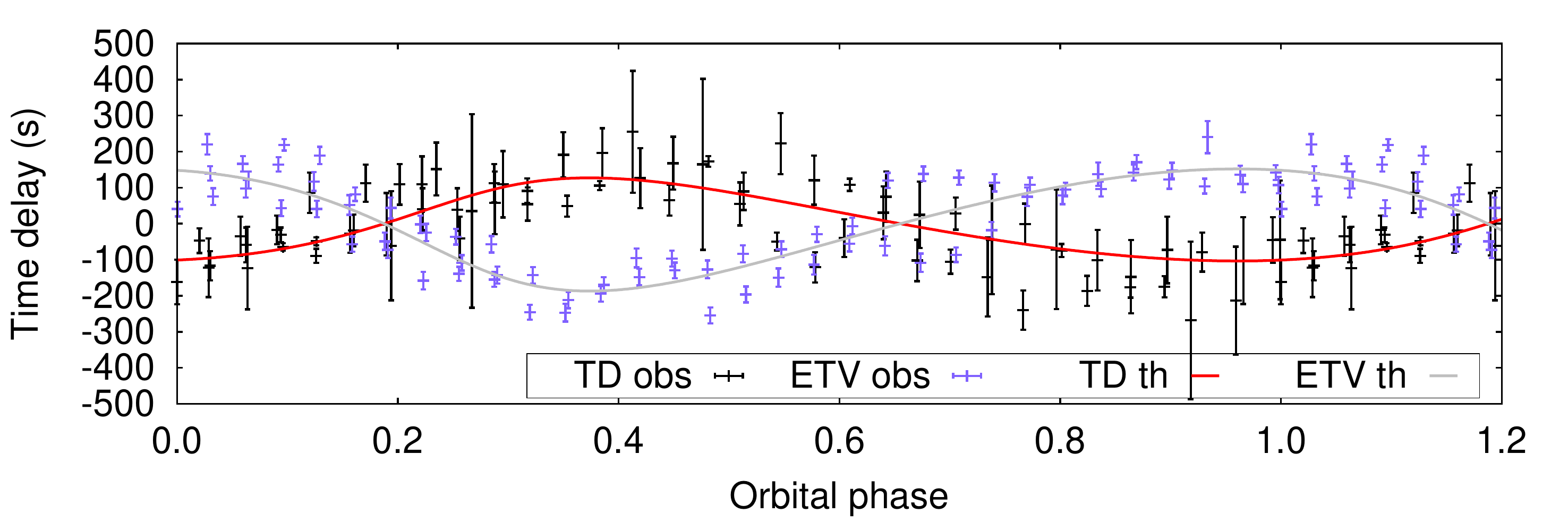}
    \caption{Phase-folded time delays of KIC\,8330092. `ETV' means the eclipsing time variation, and `TD' means the time delay of the pulsating third component. }
    \label{fig:kic8330092_best_time_delay}
\end{figure}

As mentioned in Section~\ref{subsec:near_core_orbital_periods}, the binaries with orbital periods shorter than 10\,d are likely to rotate synchronously in their near-core regions. However, KIC\,8330092 is an outlier. We checked the pixel files and did not find any contamination. 
We noticed that KIC\,8330092 shows eclipse timing variations, hence there is a third companion \cite[firstly reported by][]{Conroy_2014}. To determine which component is pulsating, we applied the phase-modulation method to four p-mode peaks of this star \citep[e.g.][]{Murphy_2015_PM, Murphy_2016_PM, Hey_2020}. Figure~\ref{fig:time_delay_8330092} displays the time delays of the eclipses and of the p-mode phases of KIC\,8330092. The variations of p-mode phases are the average of two p modes (19.674 and 18.676\,$\mathrm{d^{-1}}$), measured with the python package {\sc maelstrom} using 20-d subdivisions of the light curve. The primary eclipse timing variations were also binned every 20 days. We find that the time delays of the eclipses and pulsations are anti-correlated, which means that the p modes are from the third component, rather than the close binary.

Using the MCMC time-delay modelling code in \cite{Murphy_2016_PM}, we obtain the orbital solutions of KIC\,8330092 and its third component, as shown in Fig.~\ref{fig:kic8330092_best_time_delay}. We find that the orbital period of the third component is $625\pm3$\,d with eccentricity $0.34\pm0.03$, and the combined mass of the binary system is $0.68\pm0.04$ times the mass of the third component. Considering that the third component is a $\delta$\,Scuti star, the masses of the components in the binary are too low to be in the range of \gdor{} stars. Hence, we conclude that the g modes are also from the third component. The tidal force exerted on the third component is weak because the period of its orbit around the inner pair is long, so it is not surprising that synchronised rotation is not observed. 

\subsection{KIC\,12470041: both components have clear patterns}
\begin{figure*}
    \centering
    \includegraphics[width=1\linewidth]{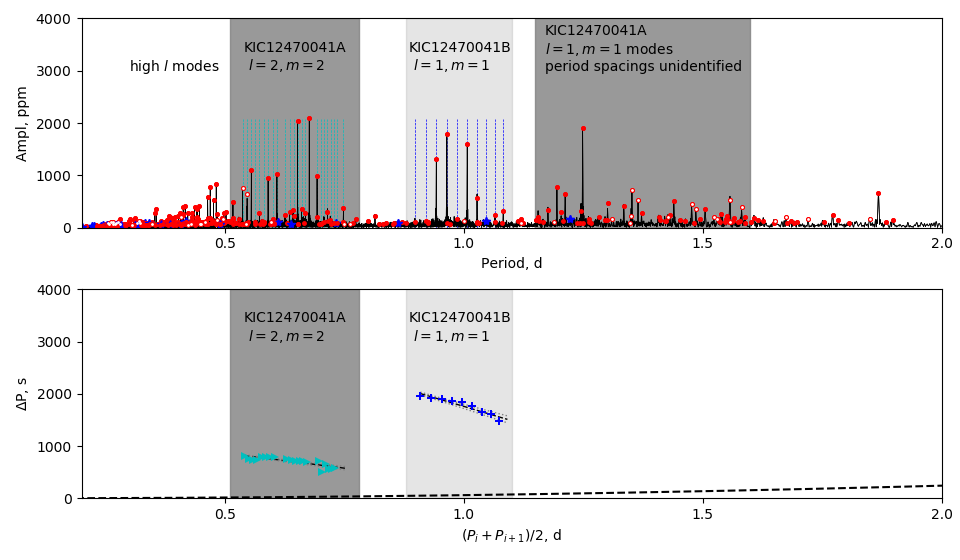}
    \caption{Amplitude spectrum and period spacing patterns of KIC\,12470041. Top: amplitude spectrum. The dark grey areas mark the spectrum of KIC\.2470041A and the light grey area shows the spectrum of KIC\,12470041B. The red points are the extracted frequencies, and the open dots are the harmonics, the blue stars mark the harmonics of the orbital period. Bottom: period spacing patterns. Note that the $l=1$ g-mode pattern of KIC\,12470041A is not identified. The dashed line shows the period resolution. }
    \label{fig:KIC012470041_mode_structure}
\end{figure*}

We detected two period spacing patterns in the spectrum of KIC\,12470041, as shown in Fig.~\ref{fig:KIC012470041_mode_structure}. We cannot fit these two patterns simultaneously under the assumption that they come from the same component. Hence, KIC\,12470041 is a system having two components that show period spacing patterns. In Fig.~\ref{fig:KIC012470041_mode_structure}, the pattern around $0.7\,\mathrm{d}$ (dark grey area on the left) is identified as the $l=2, m=2$ g modes, since it shows a smaller slope compared to dipole g modes. We name this component KIC\,12470041A. The peaks in the range $\text{from}~1.1~\text{to}~1.6\,\mathrm{d}$ (dark grey area on the right) might be the $l=1$ g modes of KIC\,12470041A, since they have a mean pulsation period twice the $l=2$ g modes, as demonstrated by \cite{Saio_2018}. However, we cannot identify any period spacings in them. The period spacings around $1\,\mathrm{d}$ (light grey area in the middle) are $l=1, m=1$ g~modes of KIC\,12470041B. 

We find that the asymptotic spacings of these two components differ greatly. The values are $3700\,\mathrm{s}$ for component A and $6090\,\mathrm{s}$ for component B. The asymptotic spacing decreases with stellar evolution, but stars with higher masses typically show higher asymptotic spacings. Since these two effects work in opposite directions, we cannot infer the evolutionary stages or masses of these two components straightforwardly in KIC 12470041. The rotation rates of these two components are also different, and neither of them is tidally locked. Forward modelling in dynamics and asteroseismology would be needed to reveal the evolutionary stages.

\subsection{Inverse tide hypothesis}\label{subsec:Anti-tide-hypothesis}

In close binaries, stellar oscillations can be tidally excited. Such modes are typically high-order stellar g modes of frequency $\sim 1\, {\rm d}^{-1}$, in the same frequency range as the unstable pulsation modes that are linearly driven in the \gdor{} instability. In the literature, it is usually assumed that the tidally excited modes are damped modes that drive the system towards the lowest energy equilibrium state, namely that of a circular orbit with the spins of each star synchronized and aligned with the orbit. However, if tidally excited modes are unstable modes, the system may not evolve toward this minimum energy state. As we show below, the tidally excited modes may instead pump the eccentricity of the orbit or drive the stellar spins away from synchronism. We refer to this process as ``inverse tides."

Following \cite{Fuller_2017} (see also \citealt{Burkart_2014}), the angular momentum transferred from a tidally excited oscillation mode to its star is
\begin{equation}
\label{jdot}
    \dot{J}_\alpha = 4 m \omega_\alpha \gamma_\alpha |a_\alpha|^2 M R^2 \, ,
\end{equation}
where $m$ is the mode's azimuthal number, $\omega_\alpha$ is the mode frequency in the star's rotating frame, and $\gamma_\alpha$ is the mode damping rate. The mode amplitude is given by 
\begin{equation}
    a_\alpha = \epsilon Q_\alpha F_{Nm} \frac{\omega_\alpha}{\sqrt{(\omega_\alpha^2 - \omega_{Nm})^2 + \gamma_\alpha^2}} \, .
\end{equation}
Here, $\epsilon$ is the tidal forcing strength, $Q_\alpha$ is the mode quadrupole moment, and $F_{Nm}$ is the temporal overlap of the forcing at an integer multiple $N$ of the orbital frequency. Definitions and conventions are described in \cite{Fuller_2017}. In these expressions, $\omega_\alpha$ is defined to be positive (i.e., the negative-frequency modes have already been accounted for). We may consider the simplest case of a circular (but not synchronized) orbit, for which tidal forcing only occurs at $N=m=2$ in the case that the star rotates slower than the orbit (i.e., prograde modes are excited), or $-N=m=-2$ in the case that the star rotates faster than the orbit (i.e., retrograde modes are excited). 

Considering equation \ref{jdot}, we see that the sign of angular momentum deposition in the star depends solely on the sign of $m \gamma_\alpha$, since all other terms are positive. In the case of a slowly rotating star with damped tidally excited oscillation modes (i.e., $m=2$ and $\gamma_\alpha > 0$), the torque is positive and the star is spun up towards synchronism. For a rapidly rotating star with damped modes (i.e., $m=-2$ and $\gamma_\alpha>0$), the torque on the star is negative and it is spun down towards synchronism. 

However, consider now the case that the tidally excited modes are linearly unstable such that the damping rate $\gamma_\alpha$ is \textit{negative}, as could easily be the case for \gdor{} stars in close binaries. In this case, the direction of angular momentum transfer will be opposite to the usual case: a slowly rotating star would be further spun down away from synchronism, and a rapidly rotating star would be further spun up away from synchronism. This inverse tidal effect could therefore cause binary \gdor{} stars to spin much slower or much faster than the orbital frequency. In both cases, the eccentricity of the orbit can be increased, and the system will move to a higher energy state, at the expense of the star's thermal energy that is pumped into the oscillation modes, and then transferred to the orbit. 

Whether this inverse tidal mechanism can operate will depend on the combined contributions of all the tidally excited oscillations from both stars. A typical star will pulsate in both stable and unstable oscillation modes, and inverse tides can only operate if the unstable modes dominate the angular momentum transfer rate. This could certainly be achieved if the unstable modes are resonantly excited, as can occur in this sample of \gdor{} binaries, where orbital frequencies and the range of unstable mode frequencies overlap. The dynamics could be somewhat complex, including resonance locking processes (\citealt{witte_1999,Fuller_2017b}). There could also exist systems where inverse tides operate in one star (acting to pump the orbit eccentricity and spin that star up or down), but normal tides operate in the other star (acting to damp the eccentricity and synchronize that star). We hope to explore such possibilities in future work. 

Another important facet of the inverse tide hypothesis is the end state of the system. In our sample, we do not observe any close binaries that rotate much faster than the orbit (with the exception of the wide binaries, for which tidal processes are likely to be negligible). However, we do find a few stars whose cores rotate much slower than the orbit \citep[there is another SPB star reported by][]{Kallinger_2017_HD_201433}, perhaps indicating the inverse tide spin-down channel is more likely to occur in real systems. However, it is still not clear why the stars would be spun down to low rotation rates, since the inverse tidal process could easily spin them down to negative rotation rates, i.e., cause the stars to spin rapidly but retrograde relative to the orbit. In order for inverse tides to explain these observations, there must be some mechanism which shuts off the process when the stellar spin frequency approaches zero.  Interestingly, the observed stars with very slow rotation cluster at orbital periods around 5 days, at the boundary between the tidally synchronized stars at shorter periods, and the longer period stars for which tides are negligible. This may indicate that inverse tides can only dominate over normal tides for these systems at intermediate orbital periods. There is much work to be done to investigate the possibility of inverse tides, but the very slowly rotating stars in this sample appear to indicate that tides can effectively spin down the cores of some \gdor{} stars.

\subsection{Where are the dipole modes?}
Five stars show only quadrupole g-mode patterns in their amplitude spectra. 
Of these five, four stars (KIC\,7515679, KIC\,8197406, KIC\,9851944, and KIC\,12470041B) still show power excesses near the expected $l=1$ periods but the spectra are too crowded to identify any pattern. 
One star (KIC\,10486425) does not have any power excess near the expected $l=1$ regions. 

The occurrence rate is $5/\EBnumber{}=14\%$. As a comparison, there are only 17 stars with missing dipole modes among the 611-star sample (2.8\%) reported by \cite{Li_2019_all_gdor}. The tidal effects of these five systems should be strong, since the longest orbital period is $\sim14.67\,\mathrm{d}$ for KIC\,12470041A and the rest are below $\sim6\,\mathrm{d}$. We infer that tidal effects are somehow encouraging quadrupole pulsations, but inhibiting dipole g modes. Further theoretical explanation and observations are needed.

\section{conclusions}\label{sec: conclusions}
Tides affect the evolution and pulsation of close binaries. We searched for \gdor{} pulsators in eclipsing binaries with \textit{Kepler} and found \EBnumber{} binary systems that have g- or r-mode patterns. These stars provide the opportunity to reveal how tidal synchronisation acts on the stellar interior in a binary system since we can measure the near-core rotation rates and asymptotic spacings by their seismic signals. 

The distribution of the near-core rotation rates of binaries is different from that of single stars. We find that the binaries tend to rotate more slowly, implying a significant tidal-locking effect on the \gdor{} components. We find that the near-core regions are more likely to be tidally locked if the orbital periods are shorter than ten days, and there are many stars with orbital periods around ten days that lie at the transition between free rotation and synchronous rotation. Assuming tidal synchronisation has occurred, the orbital period of the binary offers an independent test of the TAR, and we find that the near-core rotation periods given by the TAR are in agreement with the orbital periods down to rotation periods of $0.5\,\mathrm{d}$, or spin parameters up to 8.

We also find that three stars are rotating extremely slowly, which may signal the operation of a new mechanism we refer to as `inverse tides'. The classical tides excite damped oscillation modes, which lead to synchronisation and circularisation. However, in the `inverse tides' situation, the tidally-excited modes are unstable, allowing for angular momentum transfer that drives the system away from synchronism rather than toward it. There are still many questions that remain, such as when `inverse tides' can operate and the end state of this process. It is indeed a new aspect of tides waiting for more investigation.

\section*{Acknowledgement}

This work was supported by the Australian Research Council through DECRA DE180101104. 
Funding for the Stellar Astrophysics Centre is provided by the Danish National Research Foundation (Grant agreement no.: DNRF106). 
This work was partially supported by funding from the Center for Exoplanets and Habitable Worlds. The Center for Exoplanets and Habitable Worlds is supported by the Pennsylvania State University, the Eberly College of Science, and the Pennsylvania Space Grant Consortium. 
We appreciate Andrej Pr\v{s}a and Diana Windemuth for providing eccentricities of some EBs. We appreciate Adam Jermyn for sharing his paper draft before publication.

\section*{Data availability}
The light curves were downloaded from the public data archive at MAST\footnote{\url{https://mast.stsci.edu/}}, using the \textsc{python} package \textsc{Lightkurve} \citep{Light_curve_2018}\footnote{\url{https://docs.lightkurve.org/index.html}}. The \textit{Kepler} eclipsing binary catalogue (third revision)\footnote{\url{http://keplerebs.villanova.edu}} provided some basic parameters of some eclipsing binaries. The scientific output is available in Table~\ref{tab:rot_Pi0_table} in the article and in its online supplementary material.


\bibliographystyle{mnras}
\bibliography{ligangref} % if your bibtex file is called example.bib
\appendix

\section{Amplitude spectra and period spacing patterns}\label{appendix}
We display the amplitude spectra and period spacing patterns of the \gdor\ stars reported by this work, sorted by their descending near-core rotation rates. For each figure, the top panel shows the amplitude spectrum with x-axis of pulsation period. The extracted peaks are marked by red dots, and the likely orbital period harmonics are marked by blue stars. The identified g- or r-mode peaks are marked by the vertical lines. The bottom panel shows the period spacings, whose x-axis is the mean period. We also give the linear fits of the period spacings as a function of period, as shown by the dashed lines, whose uncertainty are given by the dotted lines surrounding it.

%\includepdf[pages=-,pagecommand={},width=2\linewidth]{gdor_binaries_appendix.pdf}
\begin{figure*}
\centering
\includegraphics[width=1\textwidth]{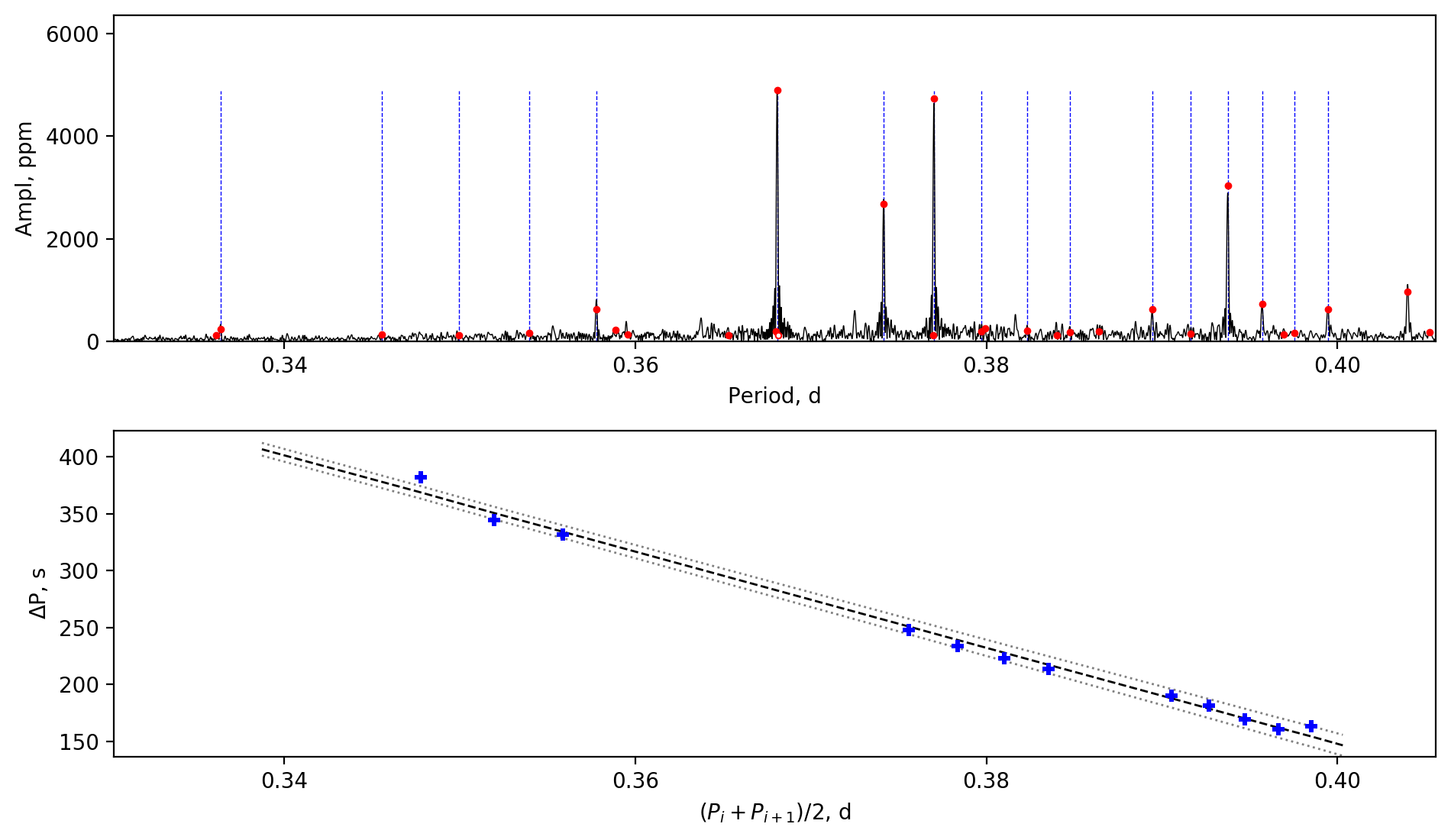}
\caption{The period spacing patterns of KIC\,4947528.}\label{fig:KIC 4947528}
\end{figure*}

\begin{figure*}
\centering
\includegraphics[width=1\textwidth]{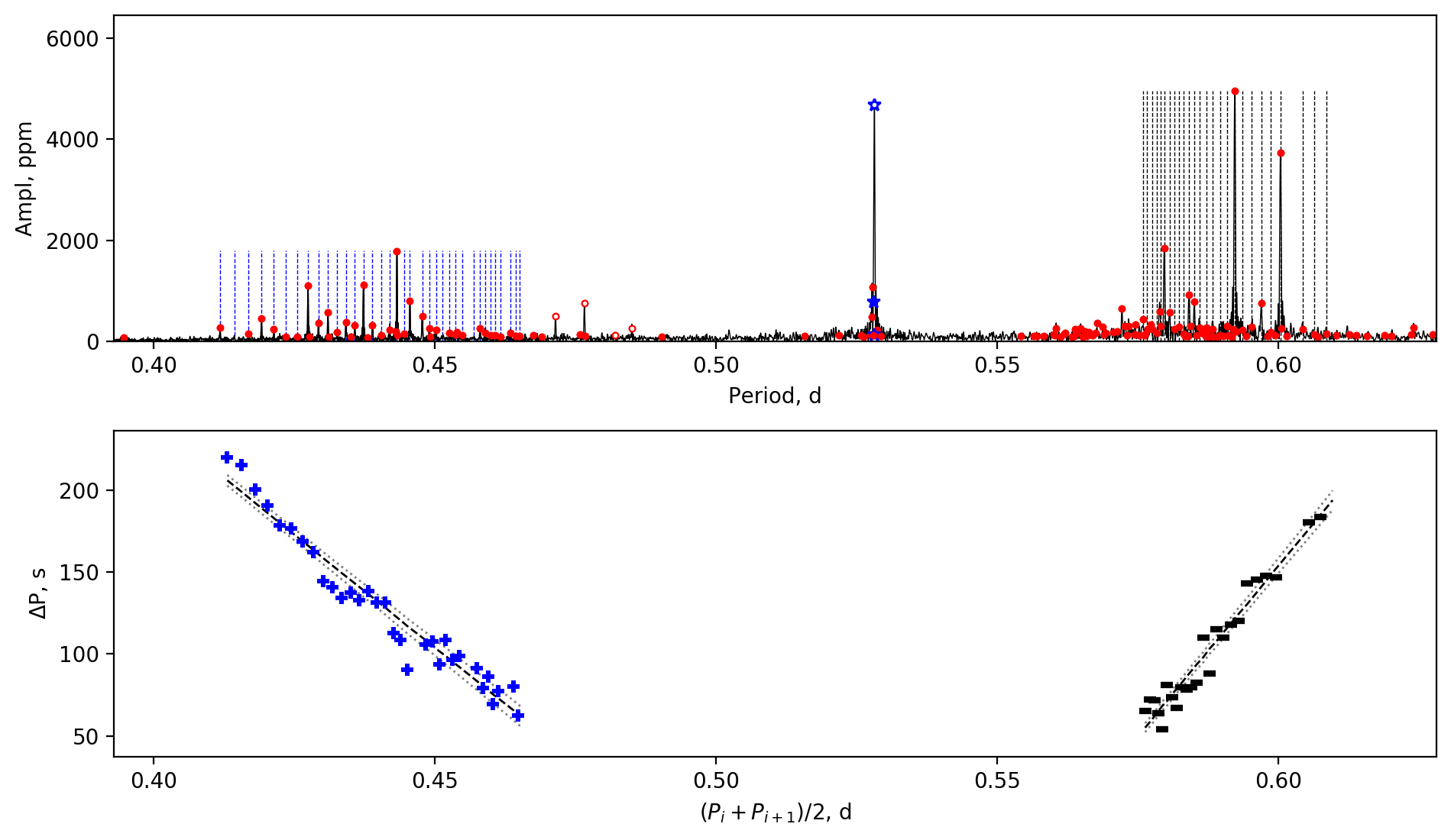}
\caption{The period spacing patterns of KIC\,3341457.}\label{fig:KIC 3341457}
\end{figure*}

\begin{figure*}
\centering
\includegraphics[width=1\textwidth]{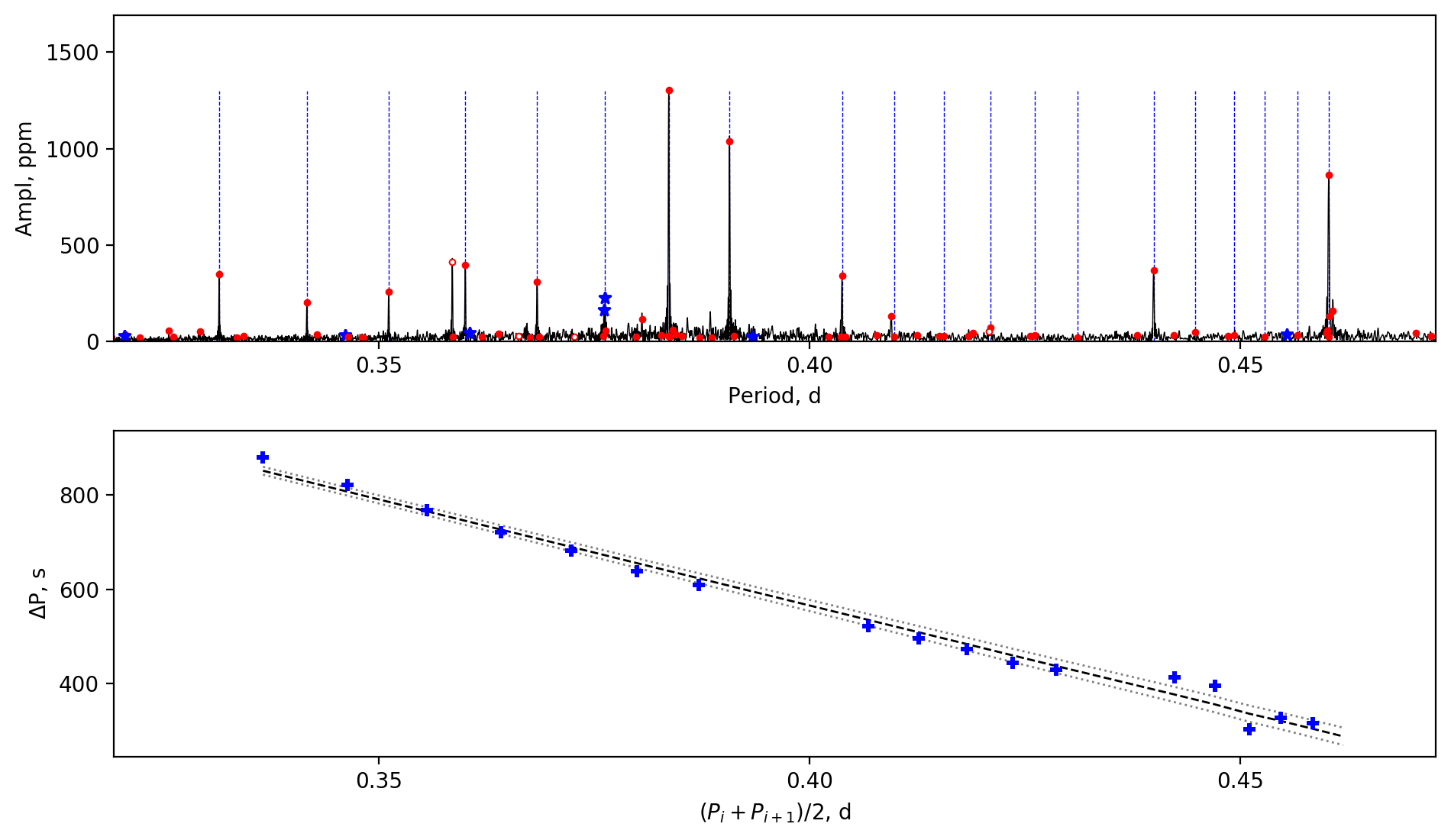}
\caption{The period spacing patterns of KIC\,4150611.}\label{fig:KIC 4150611}
\end{figure*}

\begin{figure*}
\centering
\includegraphics[width=1\textwidth]{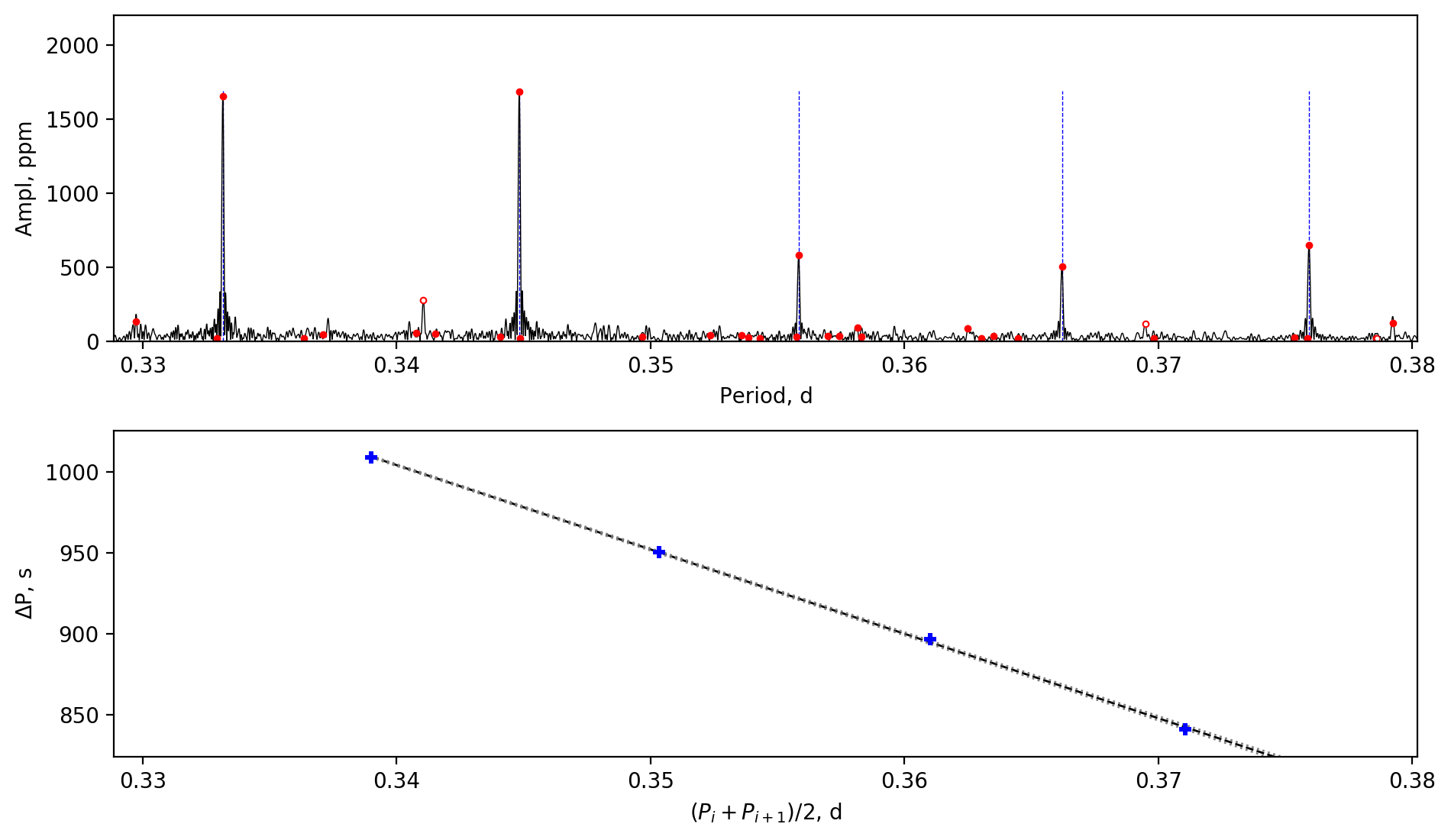}
\caption{The period spacing patterns of KIC\,11973705.}\label{fig:KIC 11973705}
\end{figure*}

\begin{figure*}
\centering
\includegraphics[width=1\textwidth]{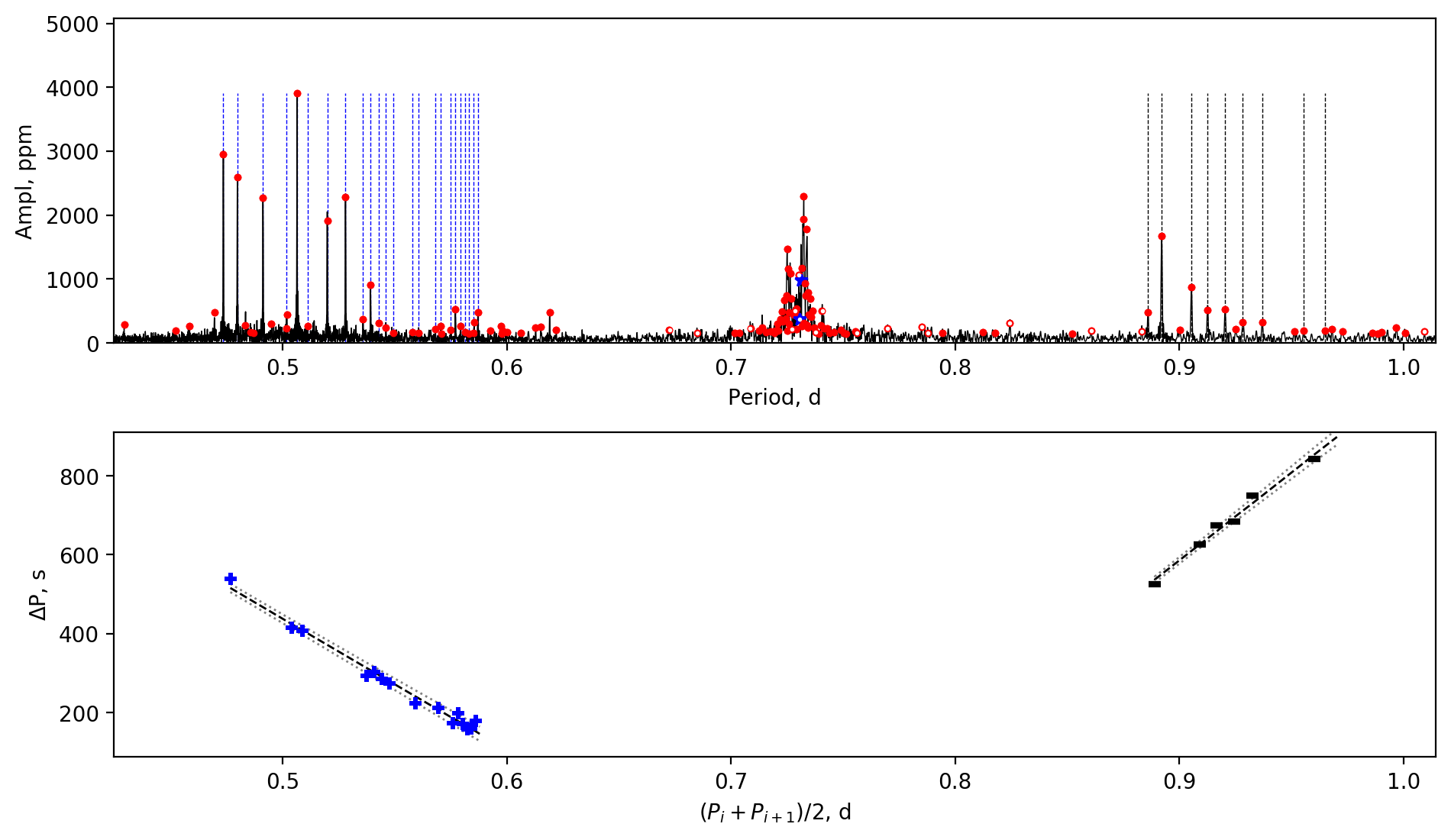}
\caption{The period spacing patterns of KIC\,3228863.}\label{fig:KIC 3228863}
\end{figure*}

\begin{figure*}
\centering
\includegraphics[width=1\textwidth]{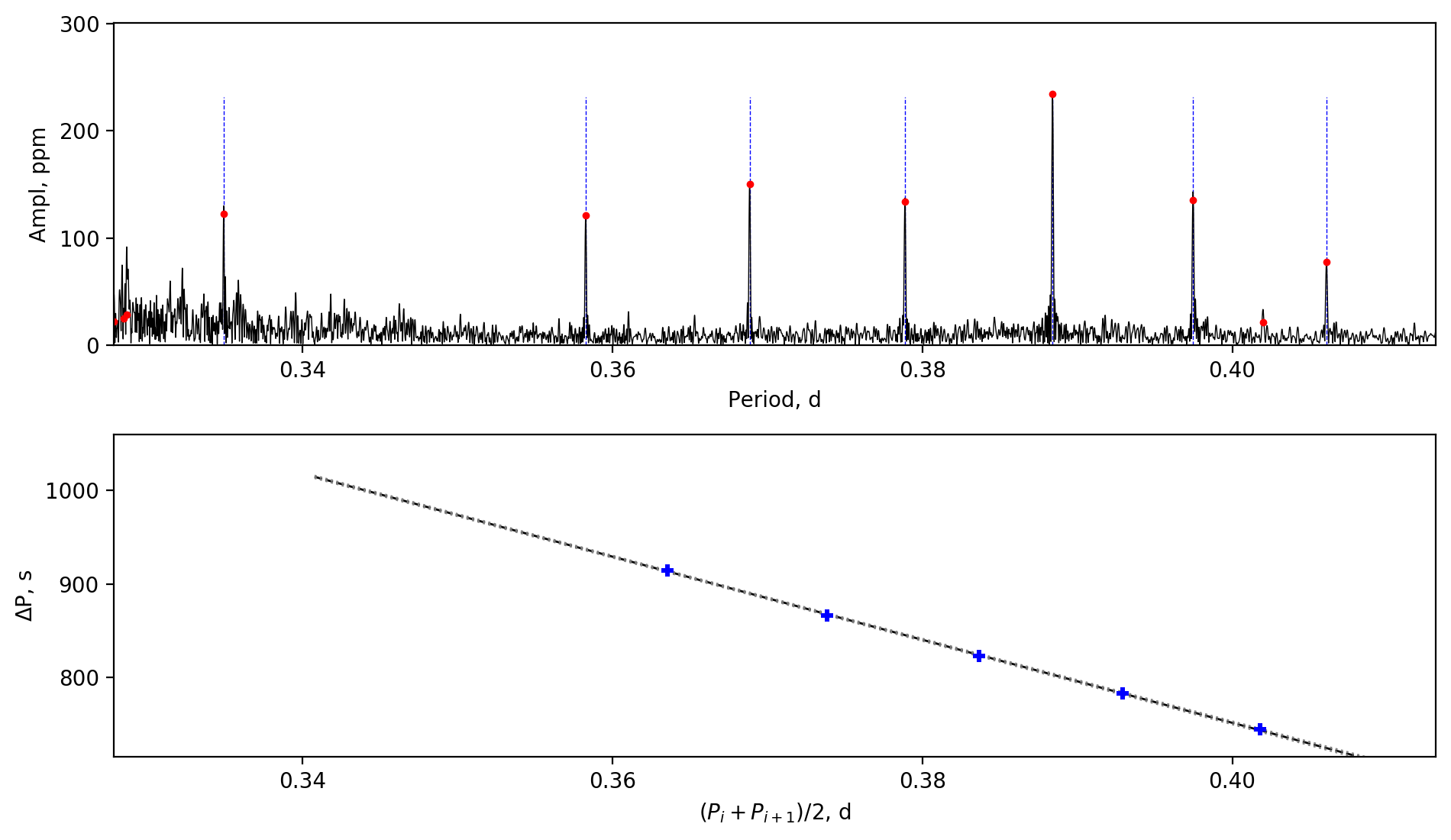}
\caption{The period spacing patterns of KIC\,8330092.}\label{fig:KIC 8330092}
\end{figure*}

\begin{figure*}
\centering
\includegraphics[width=1\textwidth]{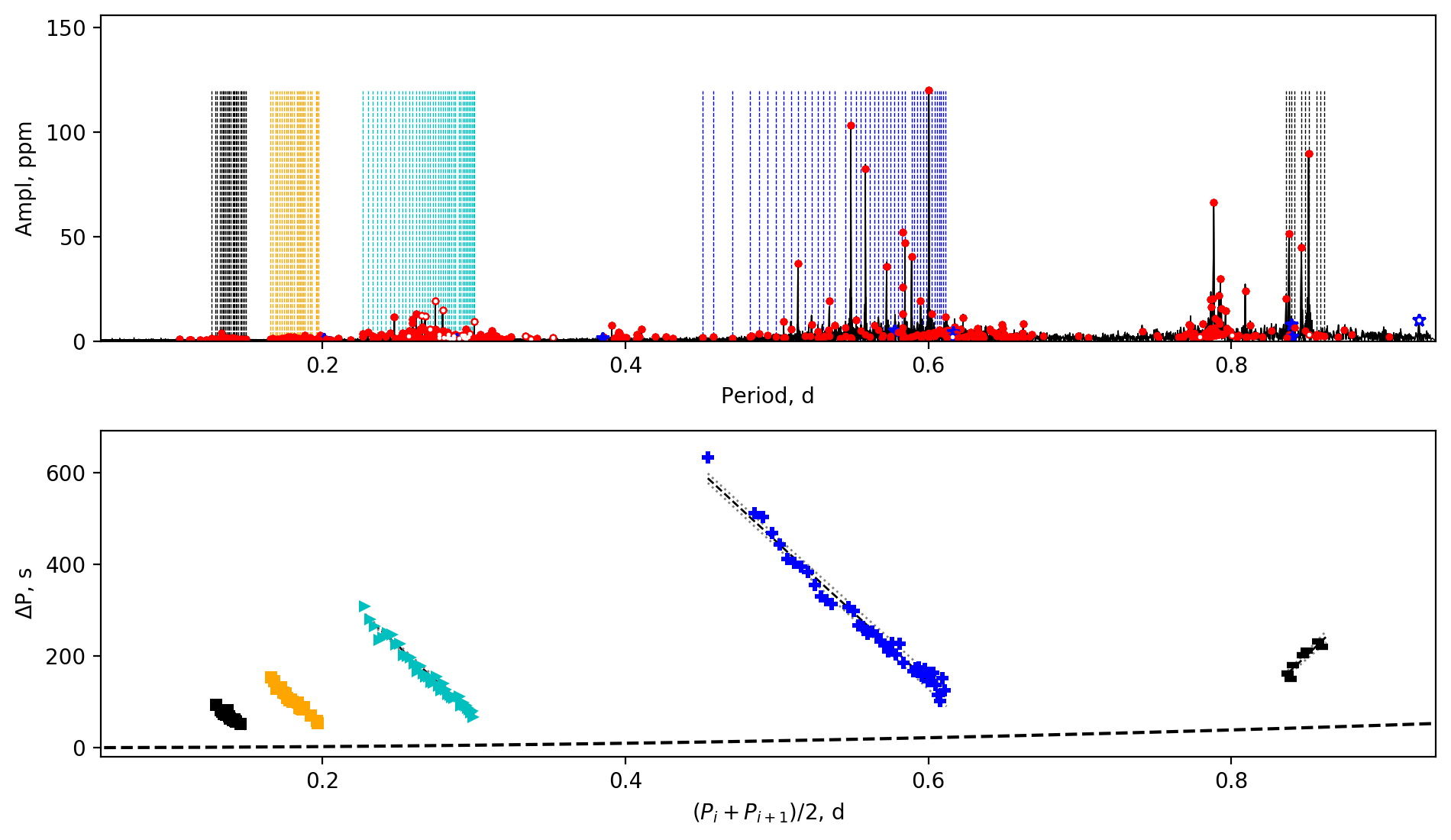}
\caption{The period spacing patterns of KIC\,6292398.}\label{fig:KIC 6292398}
\end{figure*}

\begin{figure*}
\centering
\includegraphics[width=1\textwidth]{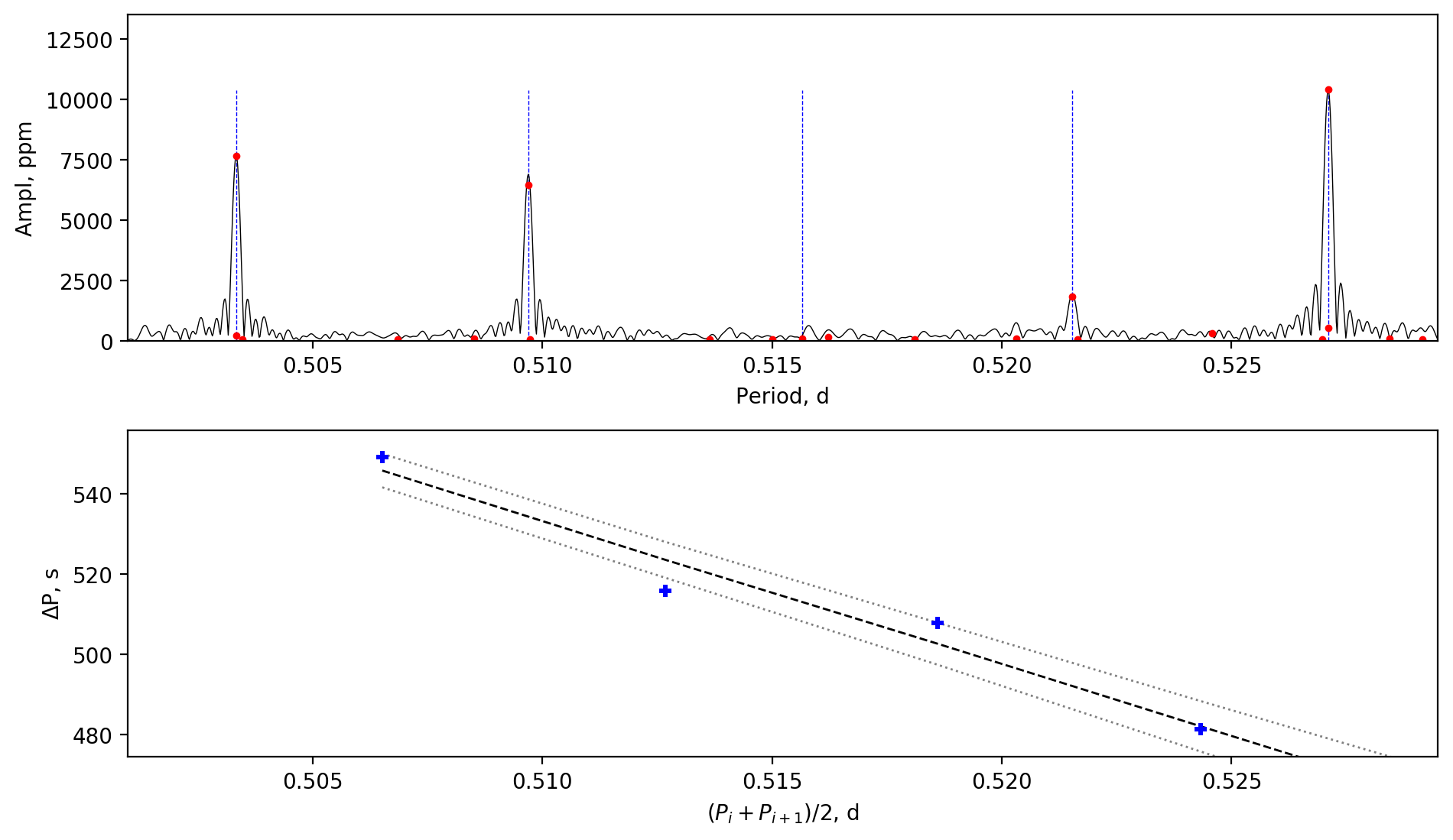}
\caption{The period spacing patterns of KIC\,12785282.}\label{fig:KIC 12785282}
\end{figure*}

\begin{figure*}
\centering
\includegraphics[width=1\textwidth]{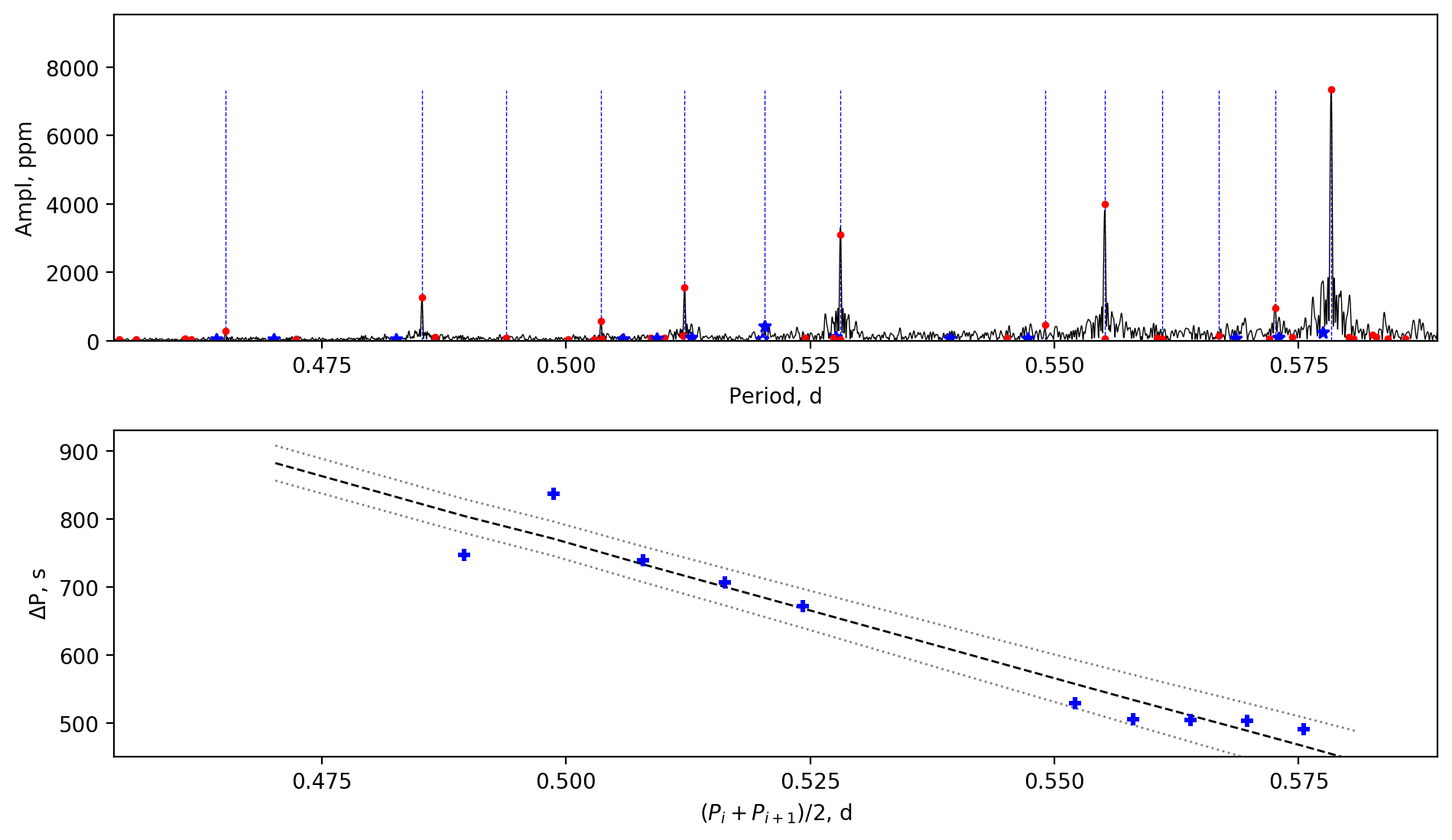}
\caption{The period spacing patterns of KIC\,3867593.}\label{fig:KIC 3867593}
\end{figure*}

\begin{figure*}
\centering
\includegraphics[width=1\textwidth]{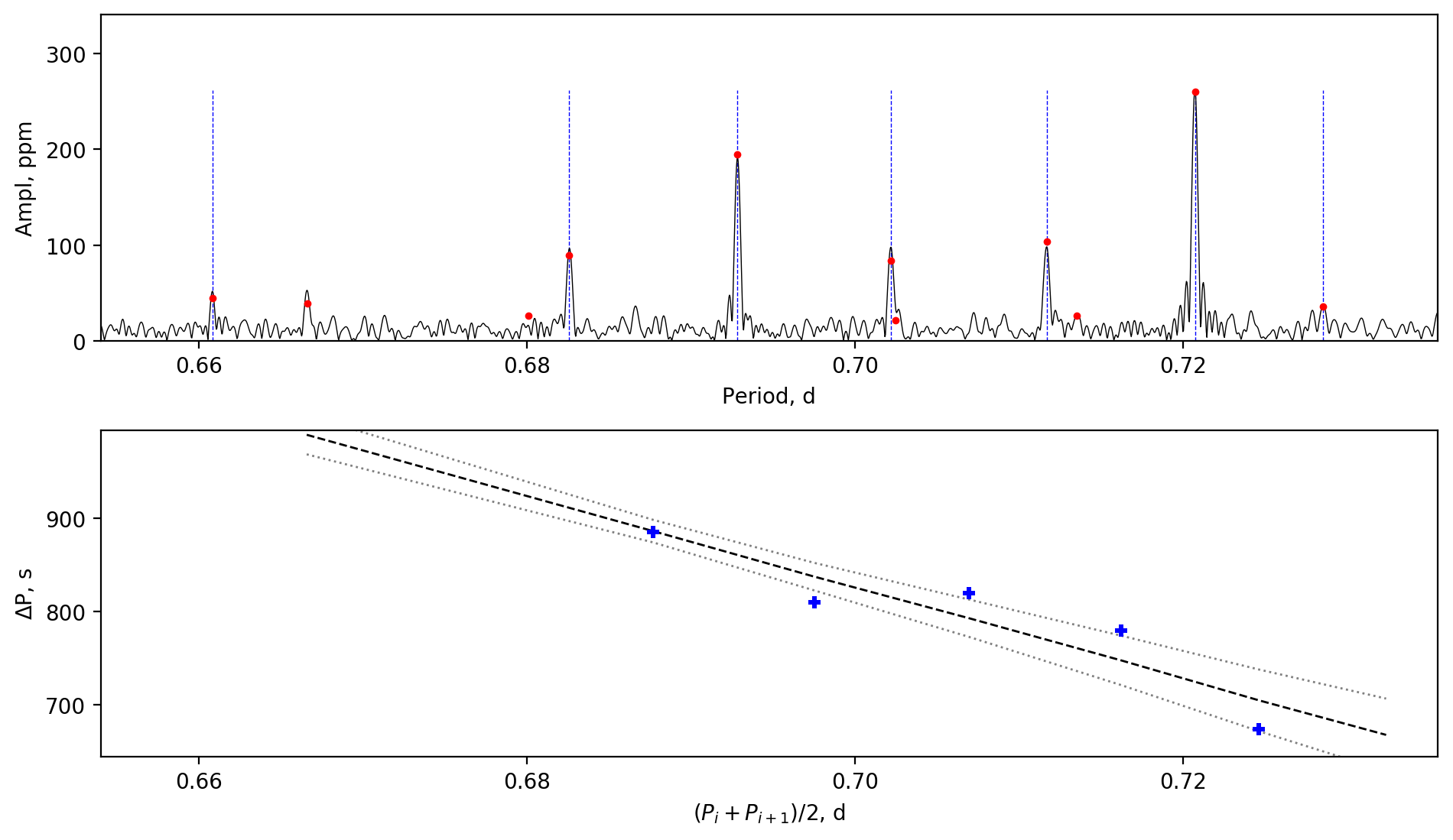}
\caption{The period spacing patterns of KIC\,5809827.}\label{fig:KIC 5809827}
\end{figure*}

\begin{figure*}
\centering
\includegraphics[width=1\textwidth]{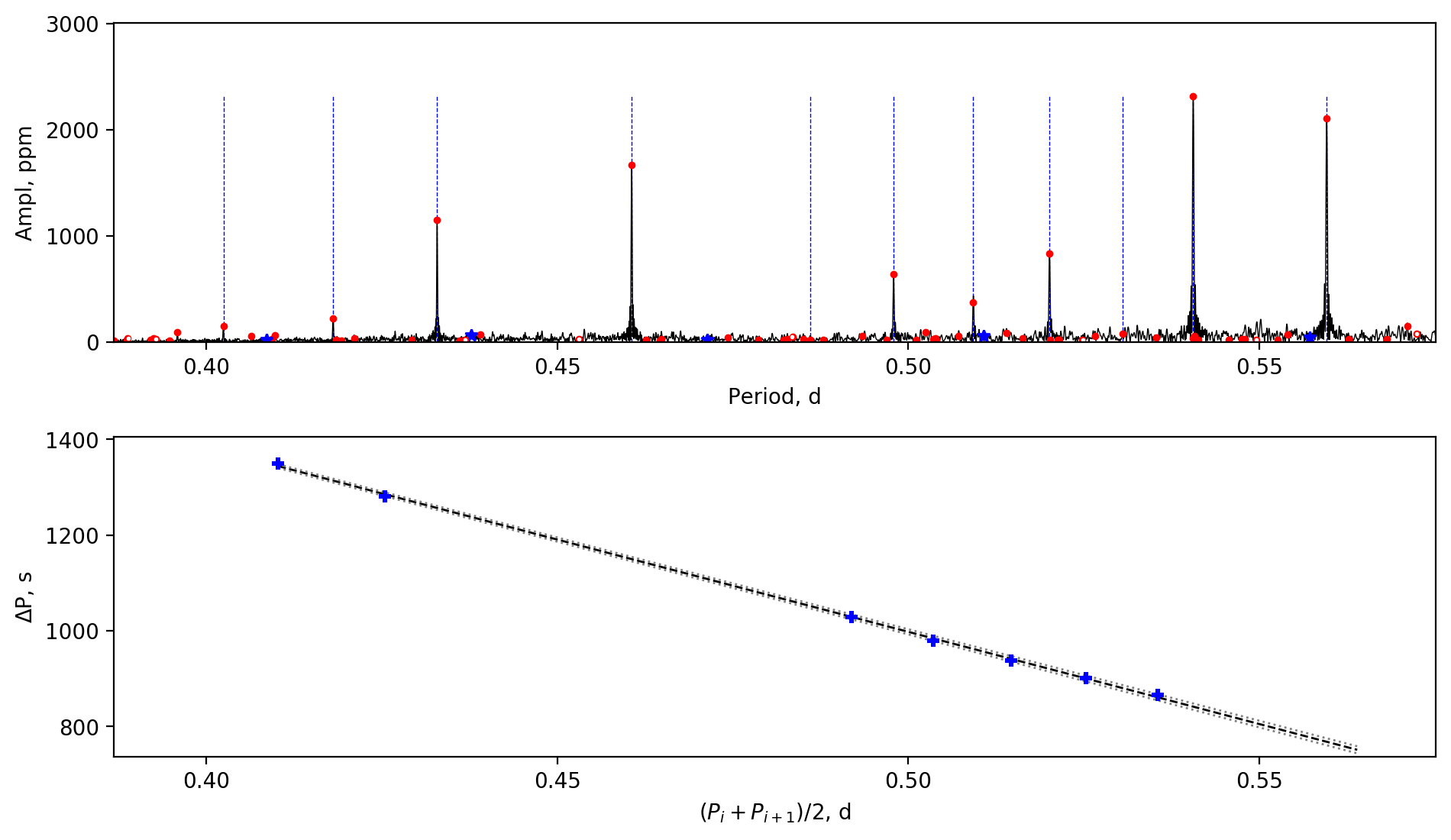}
\caption{The period spacing patterns of KIC\,6290382.}\label{fig:KIC 6290382}
\end{figure*}

\begin{figure*}
\centering
\includegraphics[width=1\textwidth]{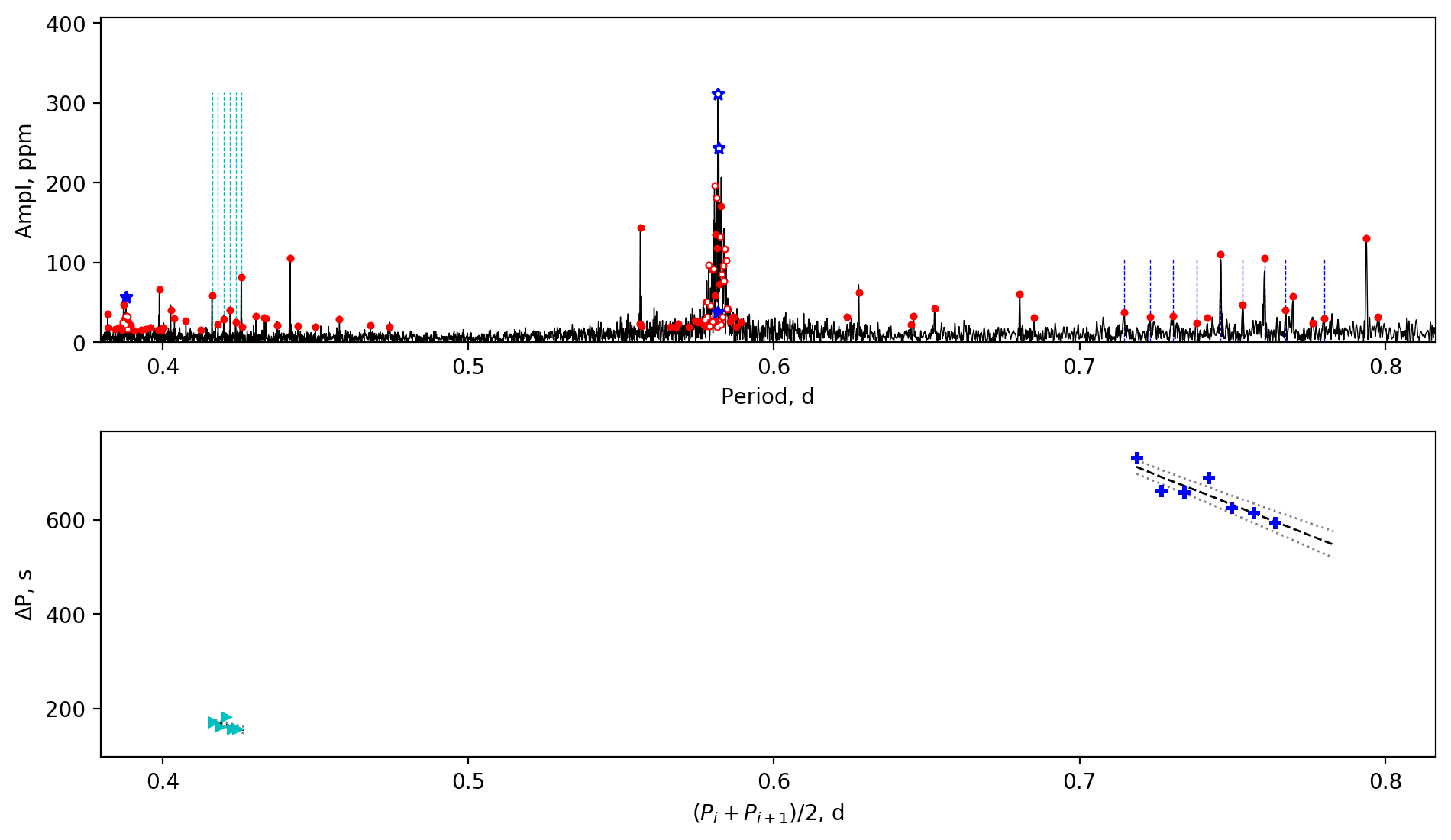}
\caption{The period spacing patterns of KIC\,8548416.}\label{fig:KIC 8548416}
\end{figure*}

\begin{figure*}
\centering
\includegraphics[width=1\textwidth]{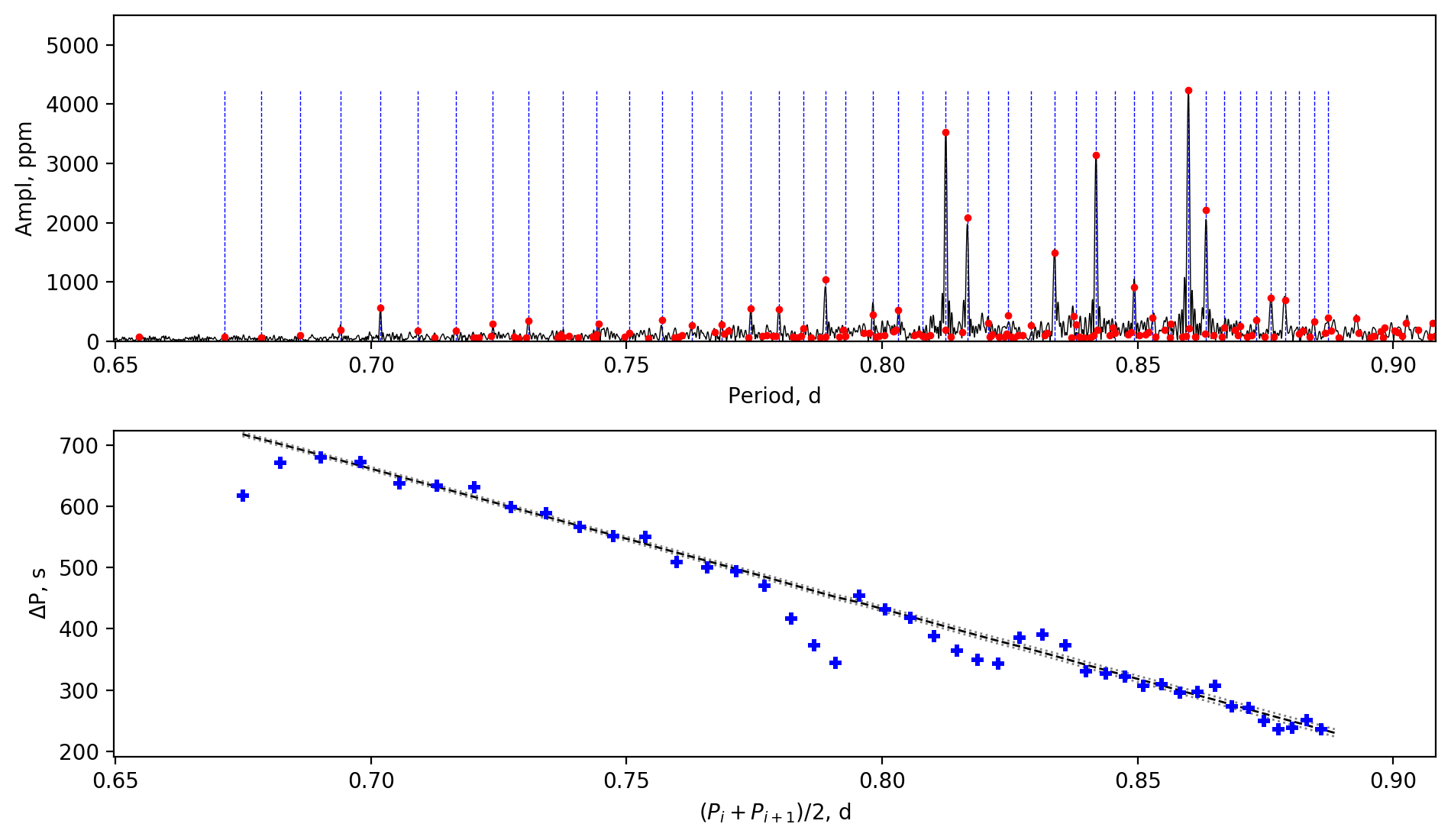}
\caption{The period spacing patterns of KIC\,6206751.}\label{fig:KIC 6206751}
\end{figure*}

\begin{figure*}
\centering
\includegraphics[width=1\textwidth]{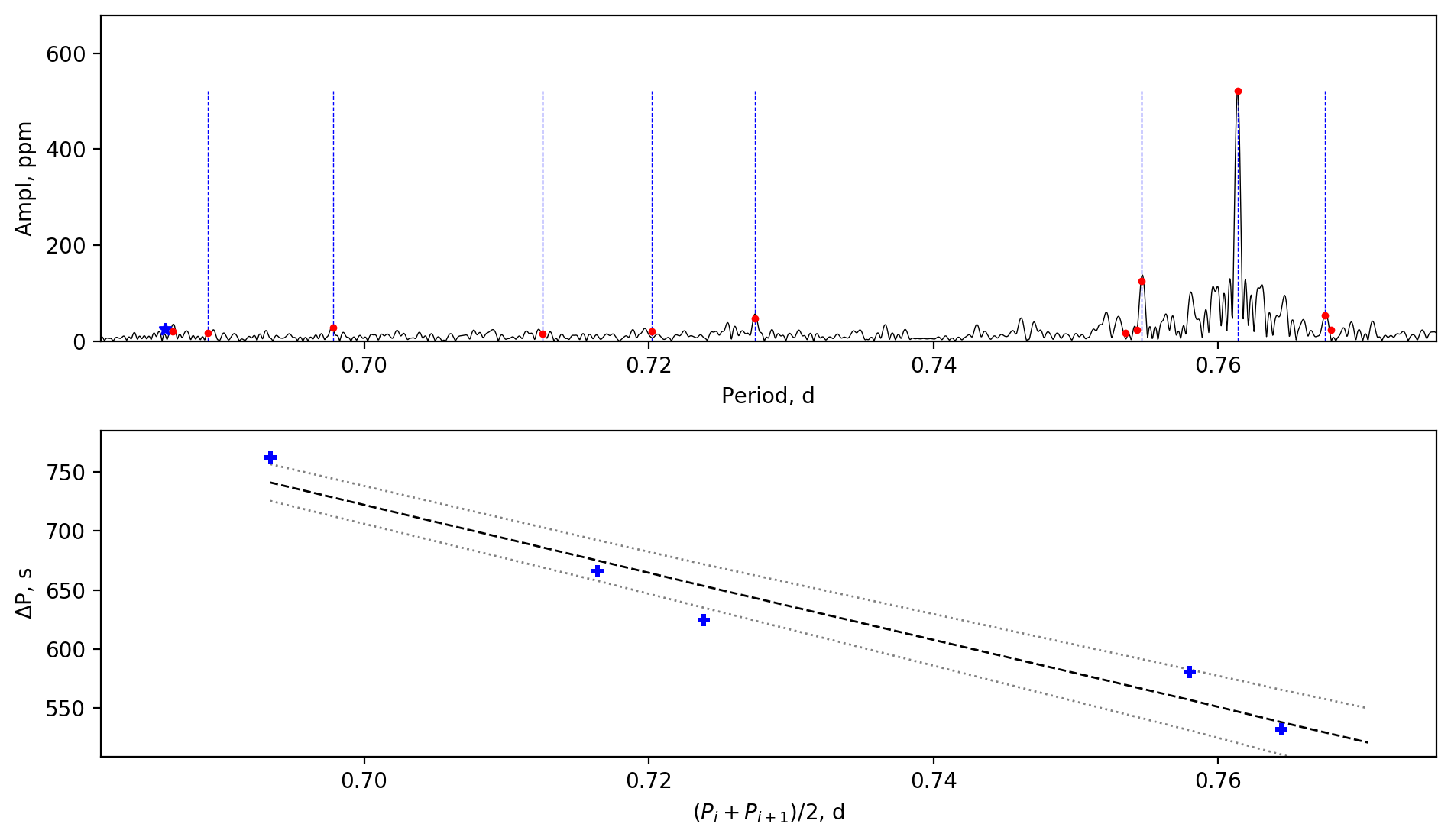}
\caption{The period spacing patterns of KIC\,3869825.}\label{fig:KIC 3869825}
\end{figure*}

\begin{figure*}
\centering
\includegraphics[width=1\textwidth]{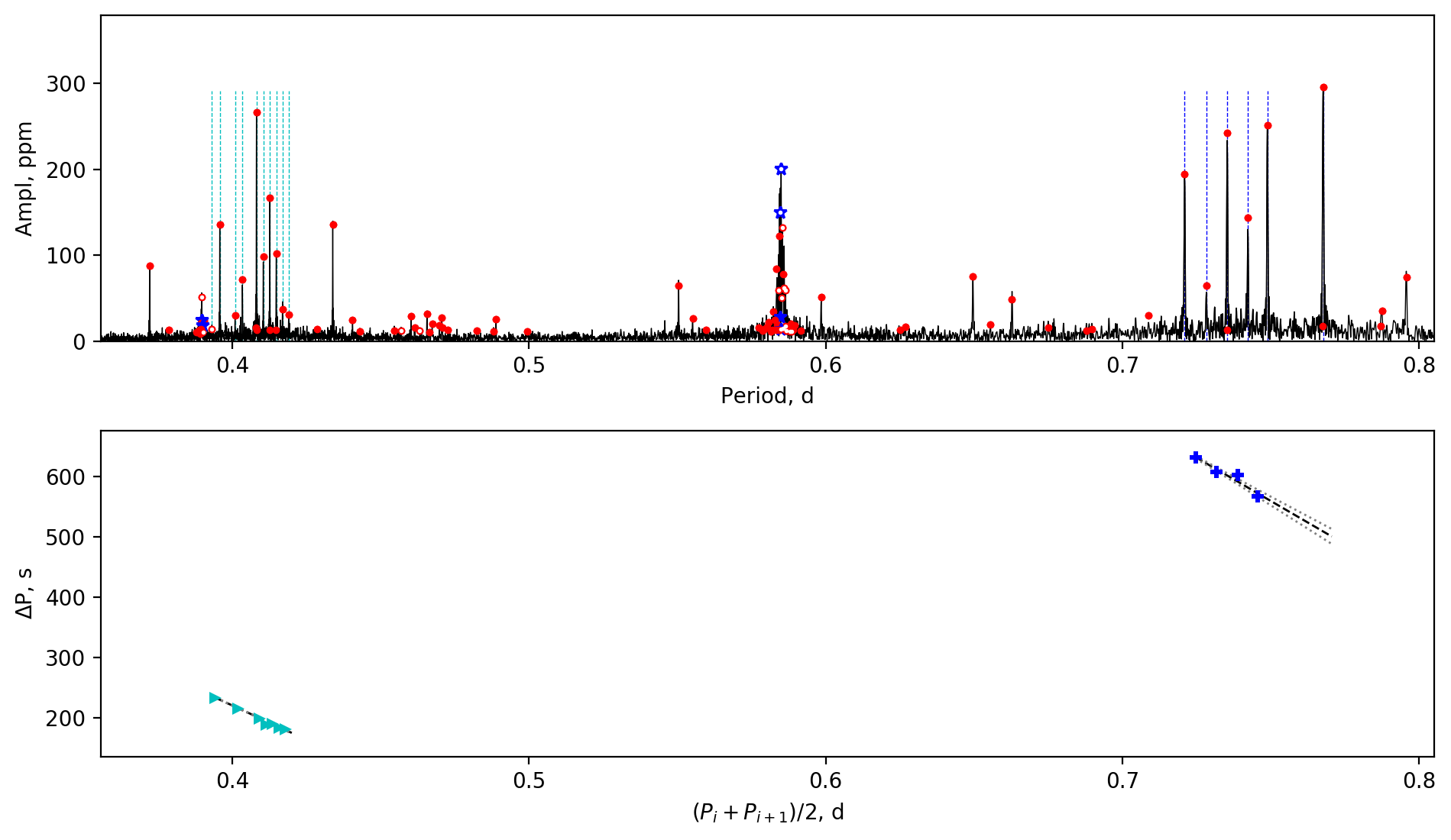}
\caption{The period spacing patterns of KIC\,9108579.}\label{fig:KIC 9108579}
\end{figure*}

\begin{figure*}
\centering
\includegraphics[width=1\textwidth]{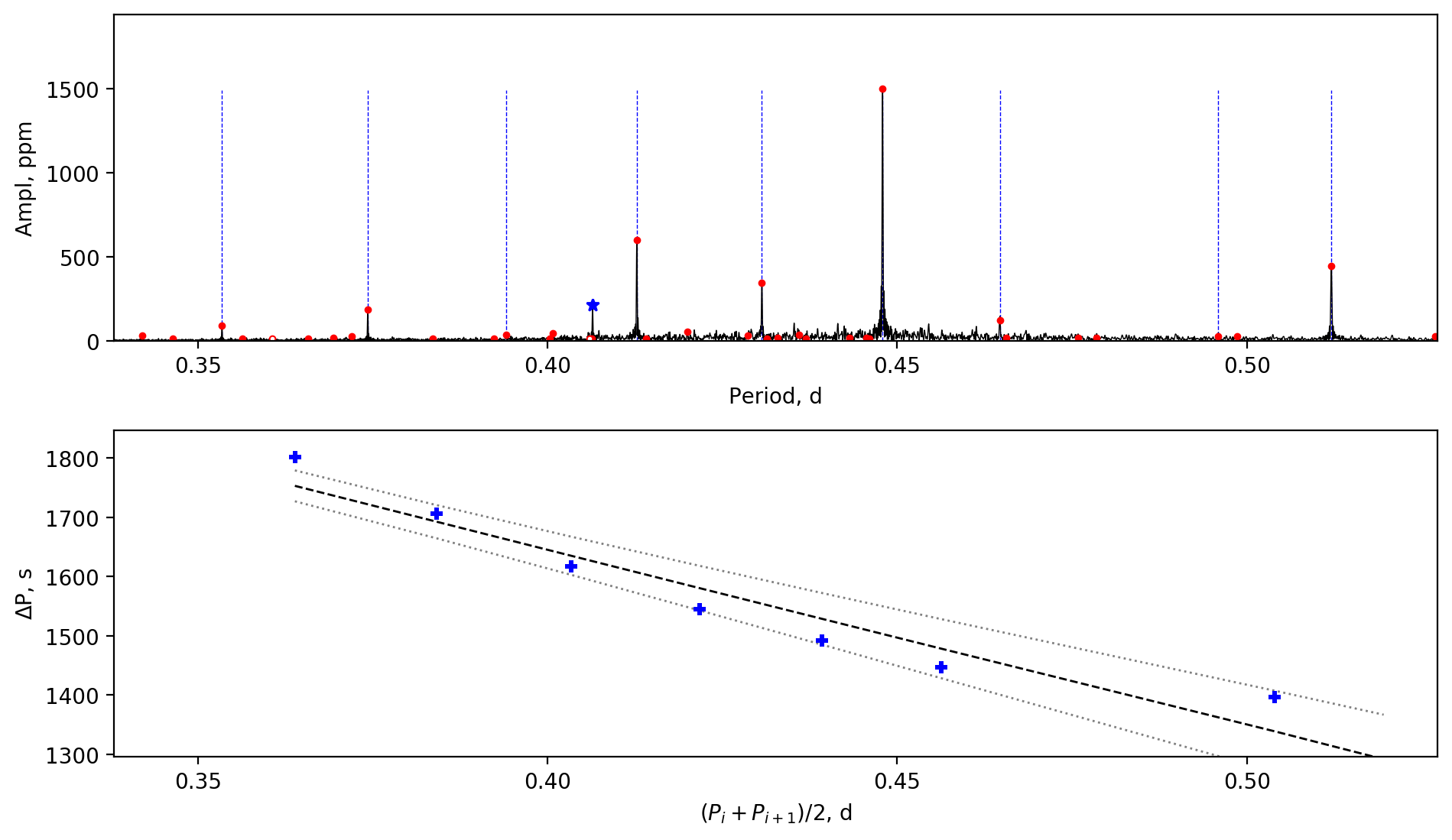}
\caption{The period spacing patterns of KIC\,9592855.}\label{fig:KIC 9592855}
\end{figure*}

\begin{figure*}
\centering
\includegraphics[width=1\textwidth]{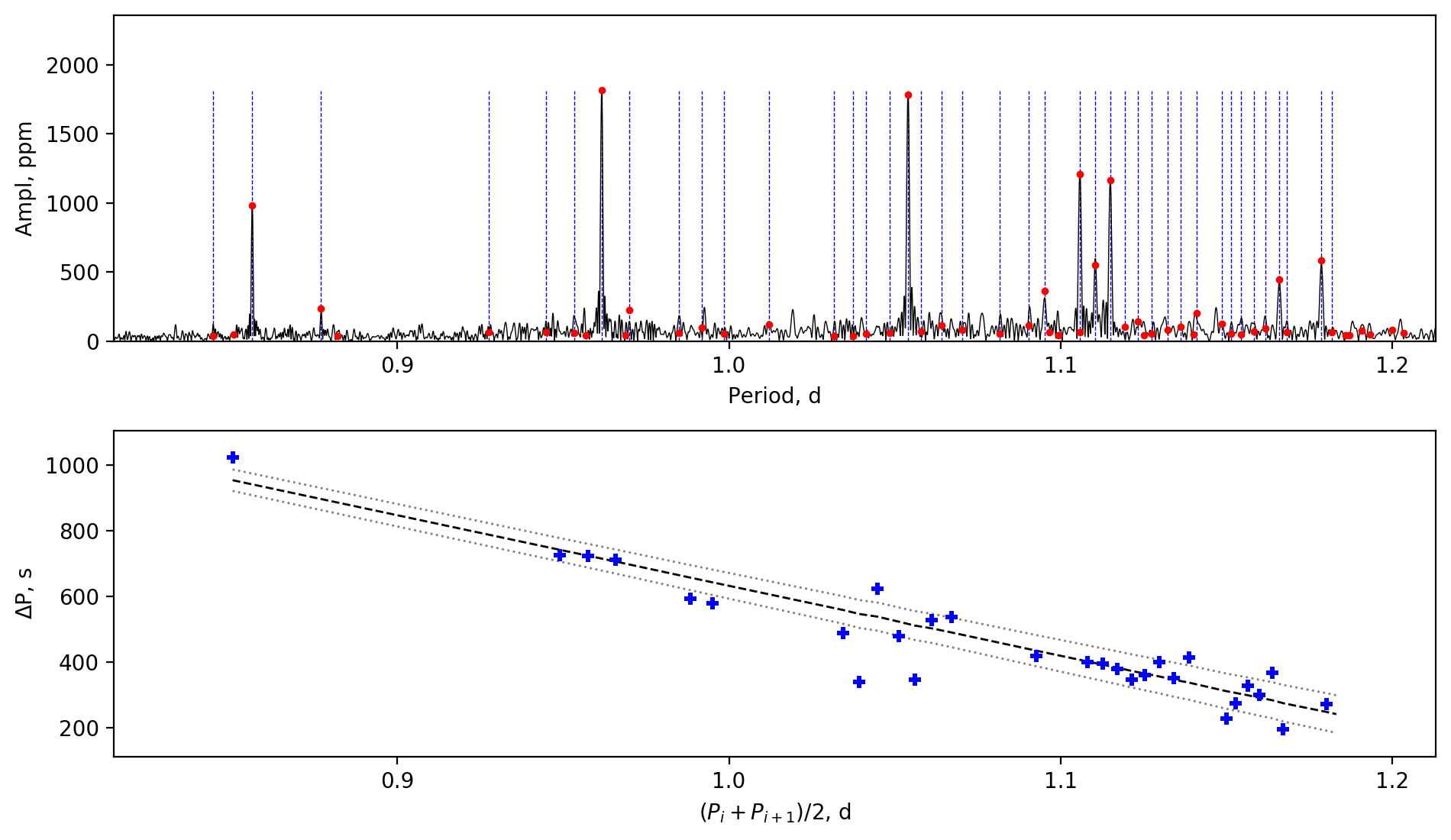}
\caption{The period spacing patterns of KIC\,2438249.}\label{fig:KIC 2438249}
\end{figure*}

\begin{figure*}
\centering
\includegraphics[width=1\textwidth]{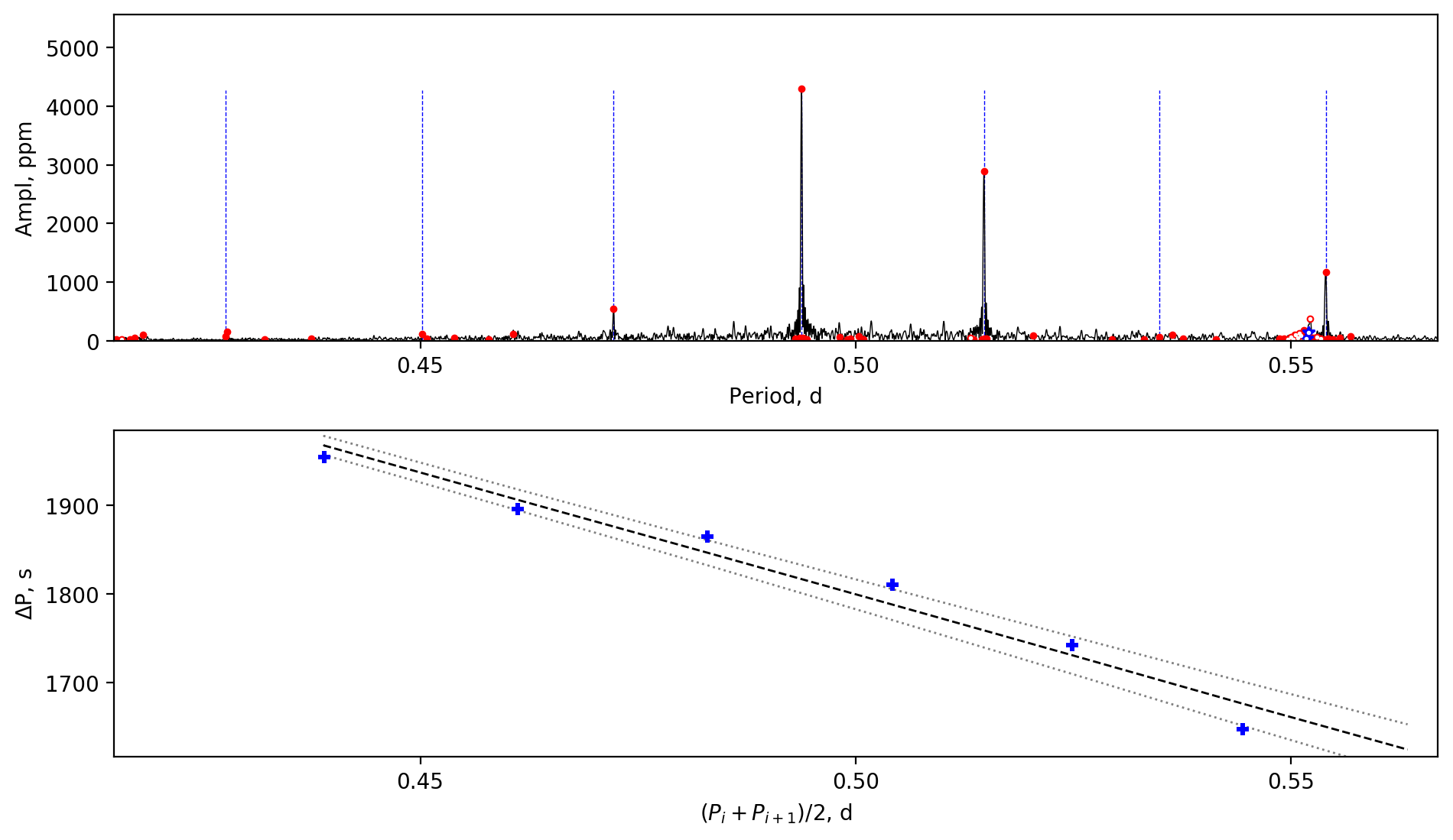}
\caption{The period spacing patterns of KIC\,7385478.}\label{fig:KIC 7385478}
\end{figure*}

\begin{figure*}
\centering
\includegraphics[width=1\textwidth]{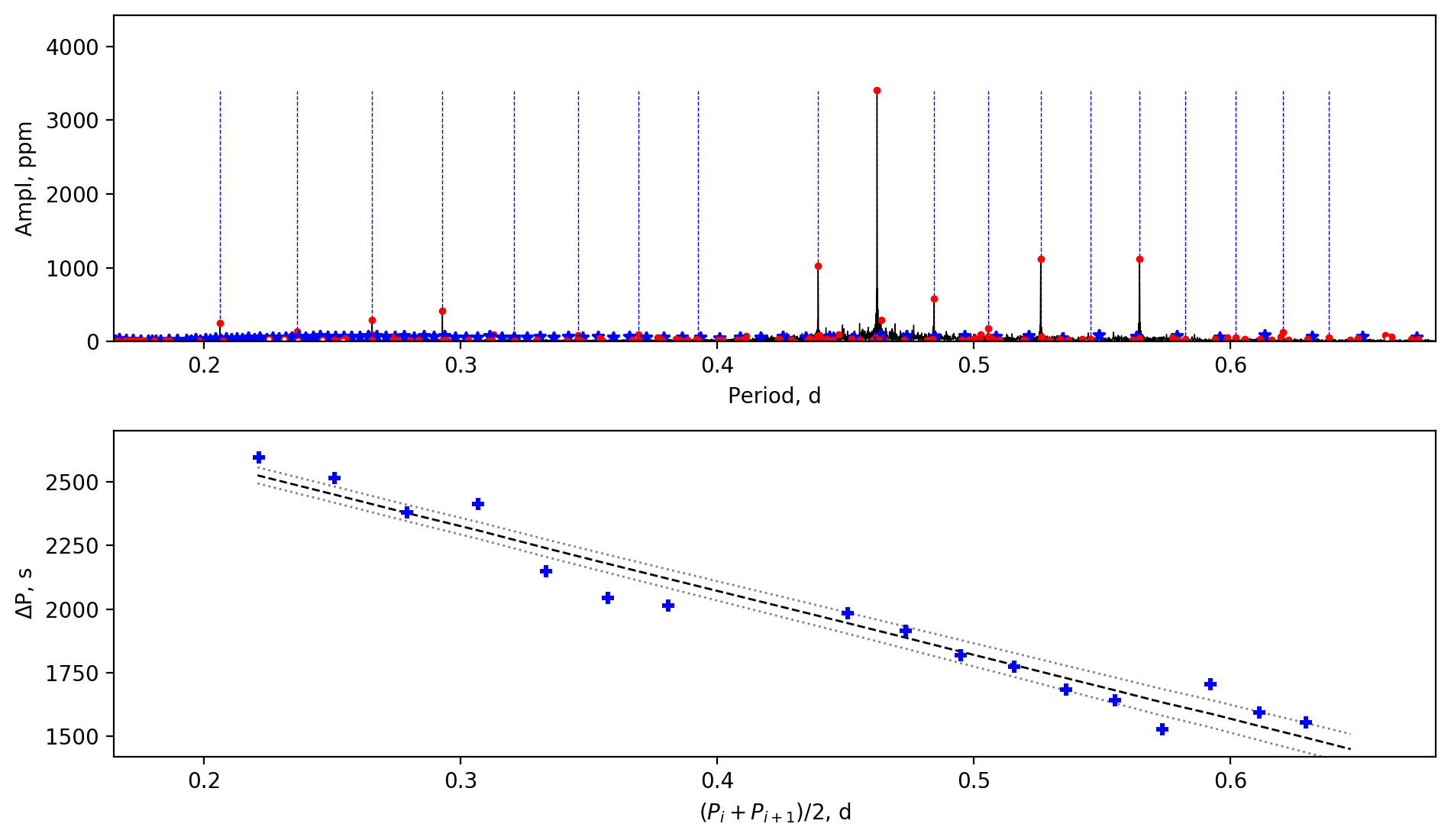}
\caption{The period spacing patterns of KIC\,8569819.}\label{fig:KIC 8569819}
\end{figure*}

\begin{figure*}
\centering
\includegraphics[width=1\textwidth]{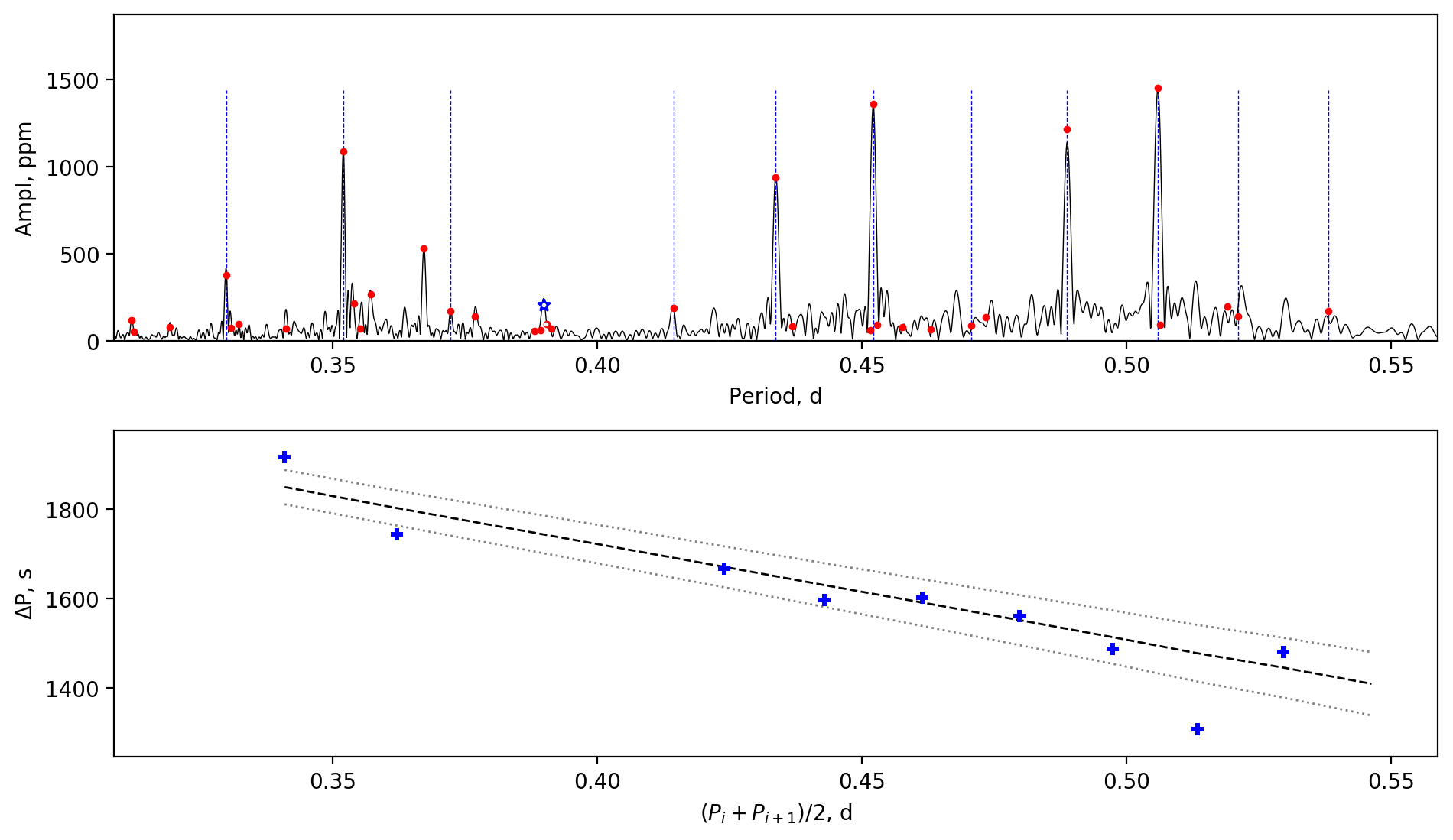}
\caption{The period spacing patterns of KIC\,6048106.}\label{fig:KIC 6048106}
\end{figure*}

\begin{figure*}
\centering
\includegraphics[width=1\textwidth]{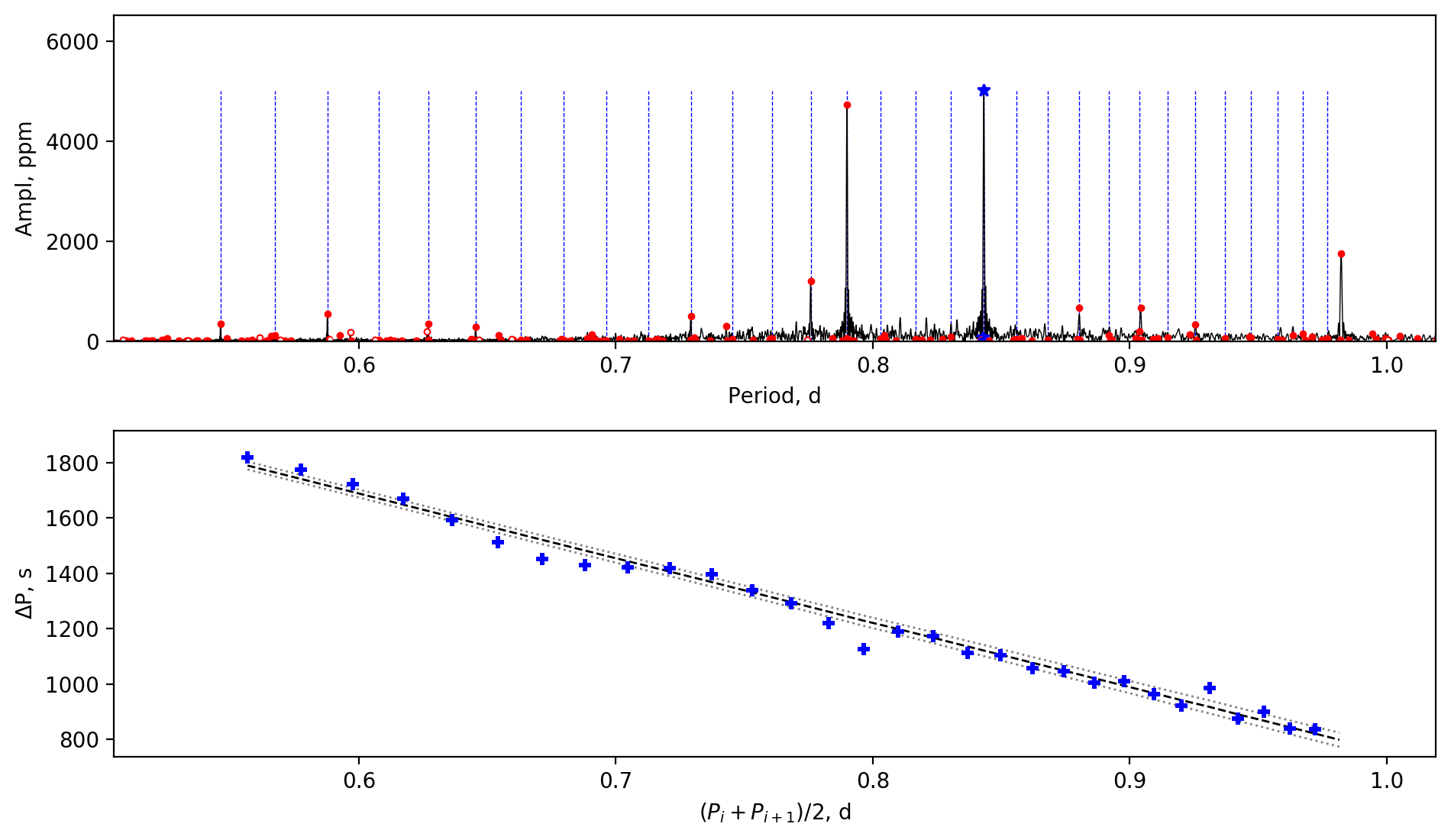}
\caption{The period spacing patterns of KIC\,1295531.}\label{fig:KIC 1295531}
\end{figure*}

\begin{figure*}
\centering
\includegraphics[width=1\textwidth]{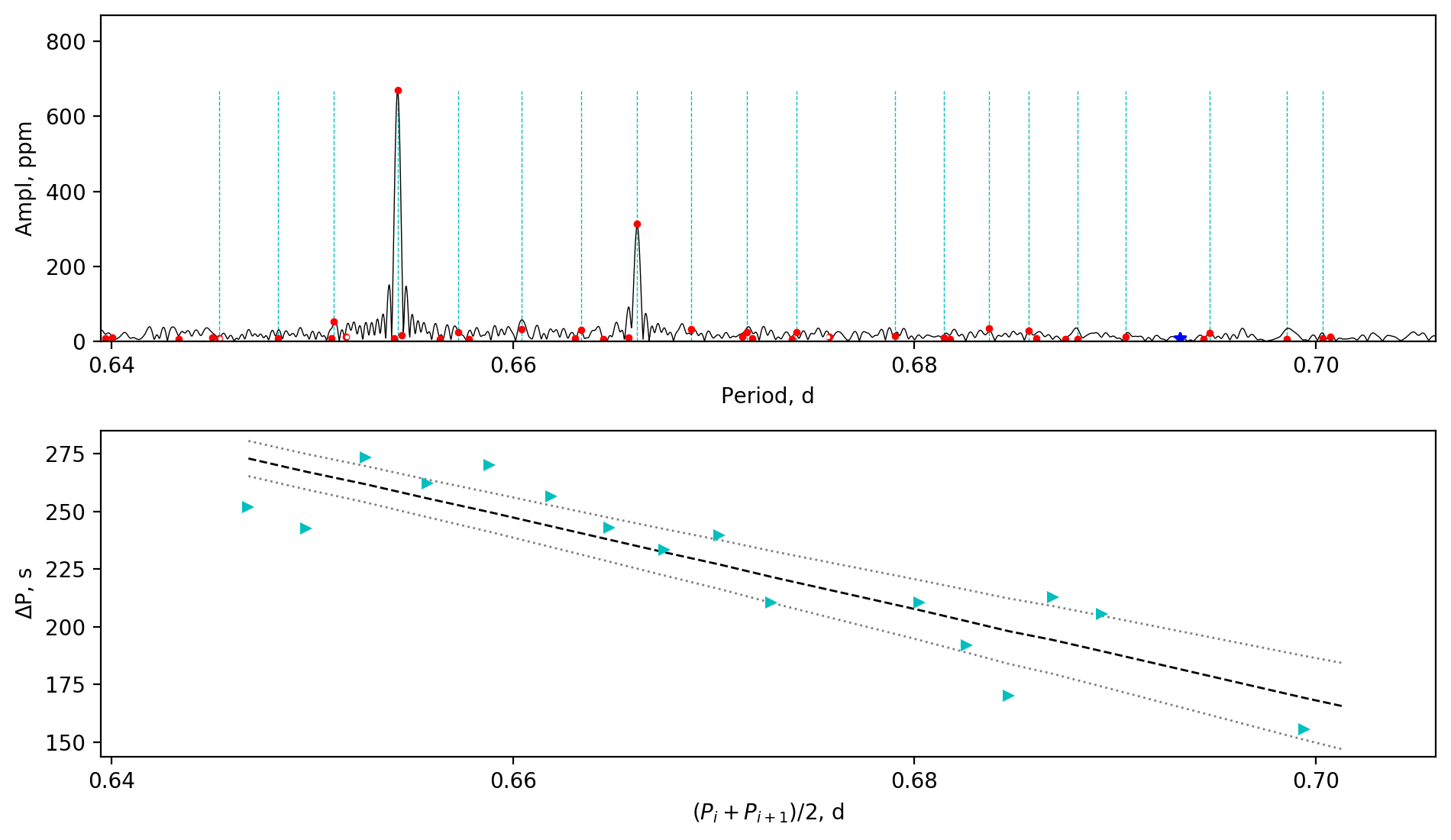}
\caption{The period spacing patterns of KIC\,7515679.}\label{fig:KIC 7515679}
\end{figure*}

\begin{figure*}
\centering
\includegraphics[width=1\textwidth]{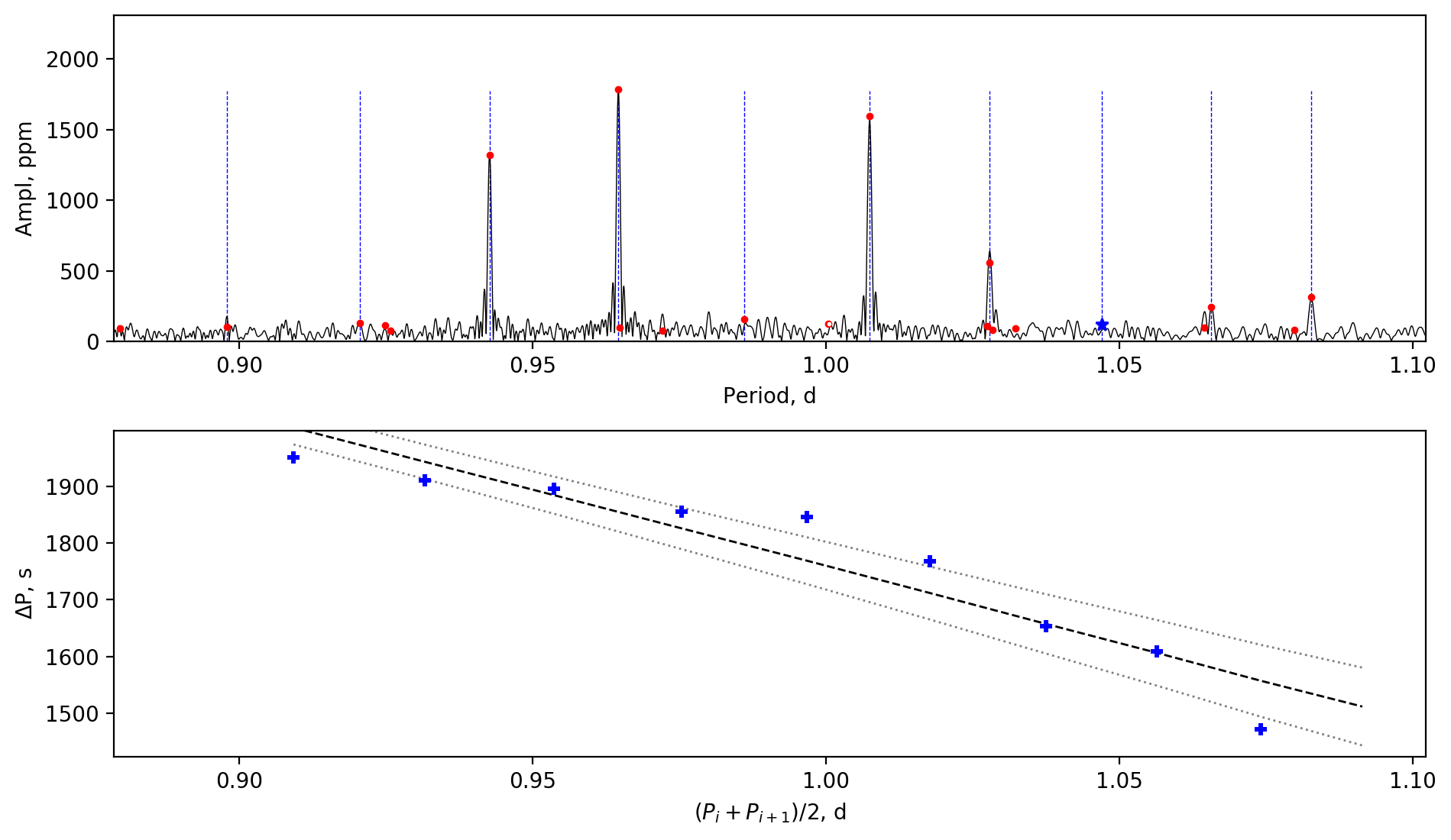}
\caption{The period spacing patterns of KIC\,12470041B.}\label{fig:KIC 12470041B}
\end{figure*}

\begin{figure*}
\centering
\includegraphics[width=1\textwidth]{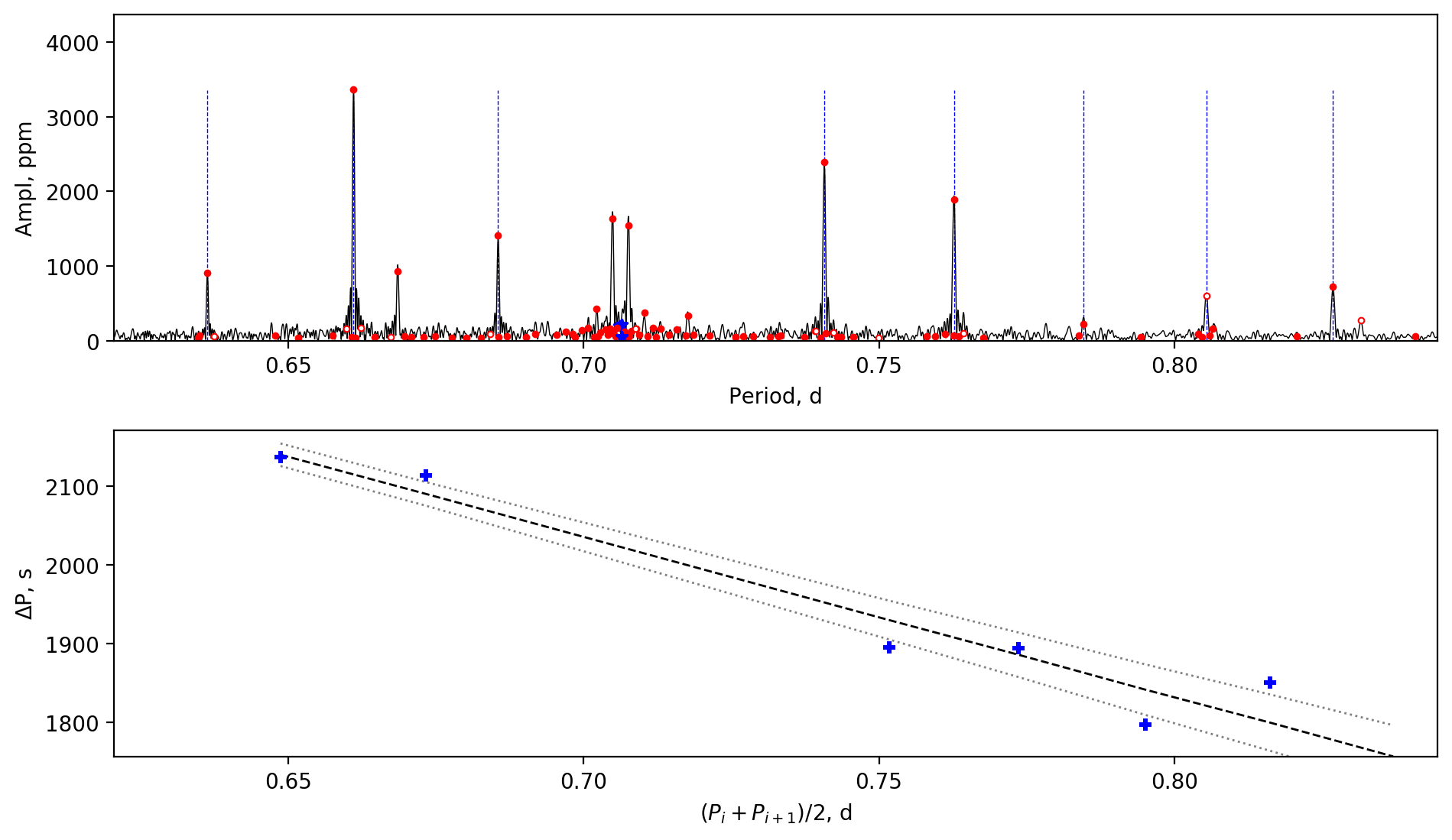}
\caption{The period spacing patterns of KIC\,5565486.}\label{fig:KIC 5565486}
\end{figure*}

\begin{figure*}
\centering
\includegraphics[width=1\textwidth]{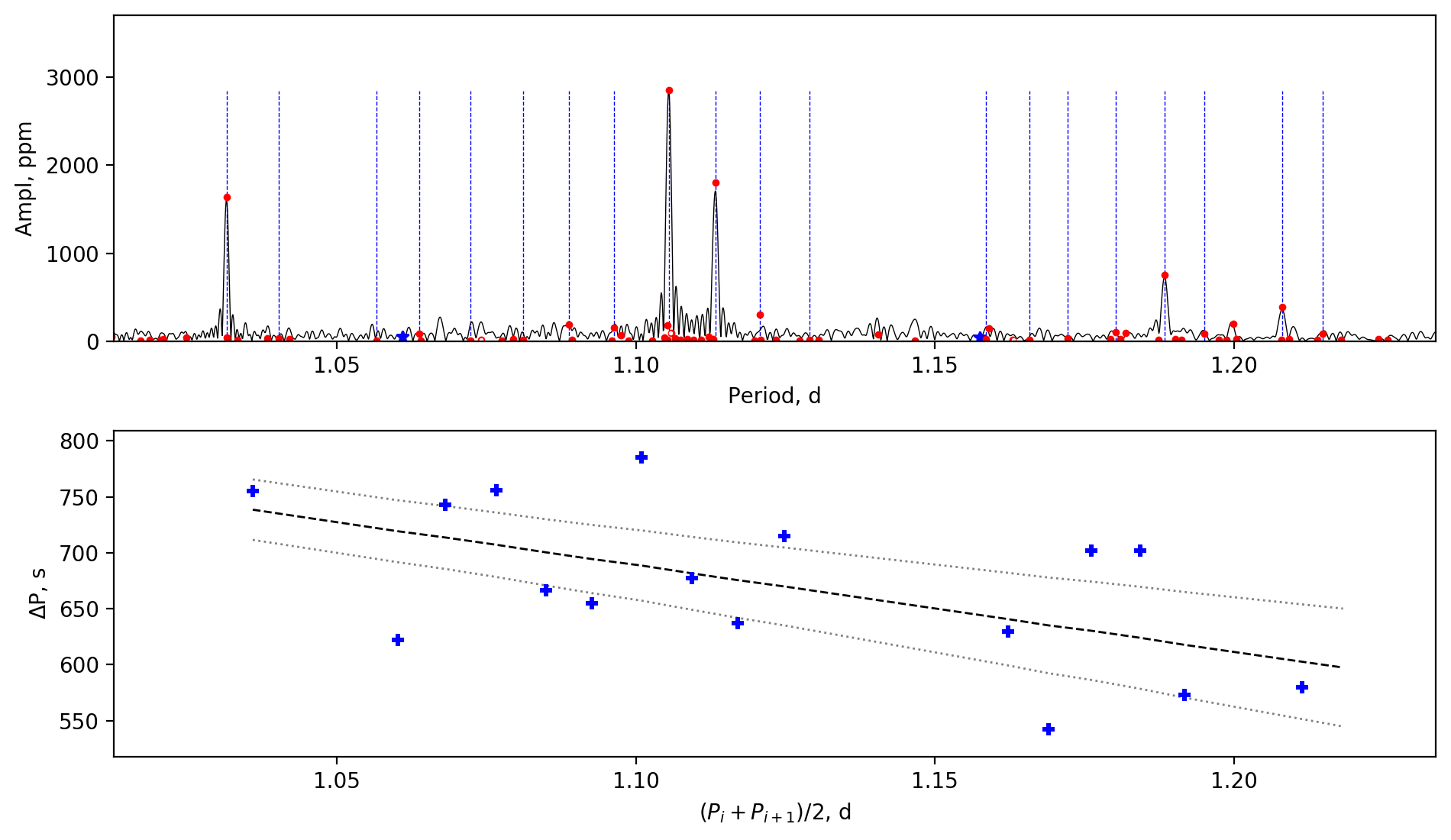}
\caption{The period spacing patterns of KIC\,11820830.}\label{fig:KIC 11820830}
\end{figure*}

\begin{figure*}
\centering
\includegraphics[width=1\textwidth]{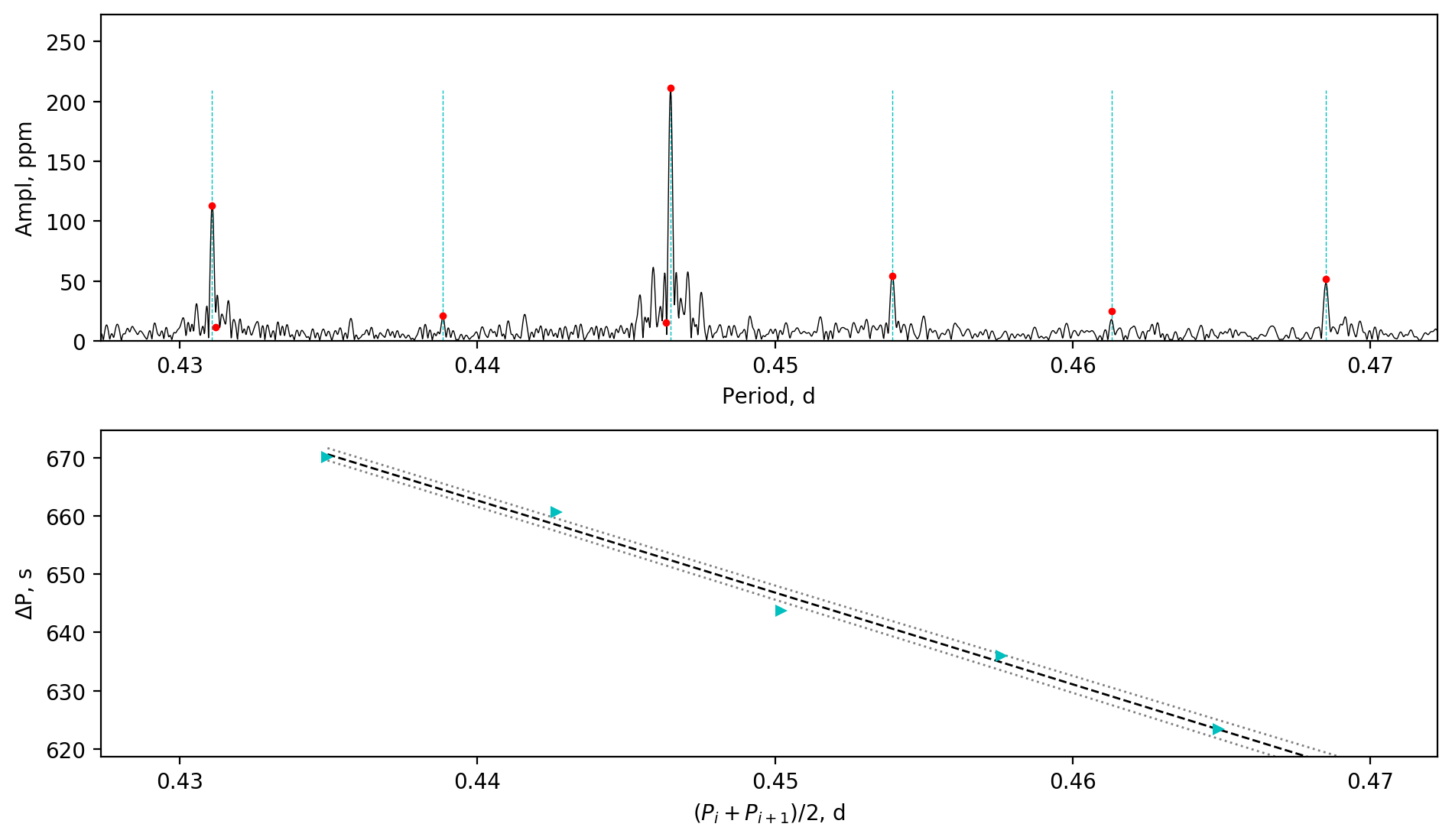}
\caption{The period spacing patterns of KIC\,9851944.}\label{fig:KIC 9851944}
\end{figure*}

\begin{figure*}
\centering
\includegraphics[width=1\textwidth]{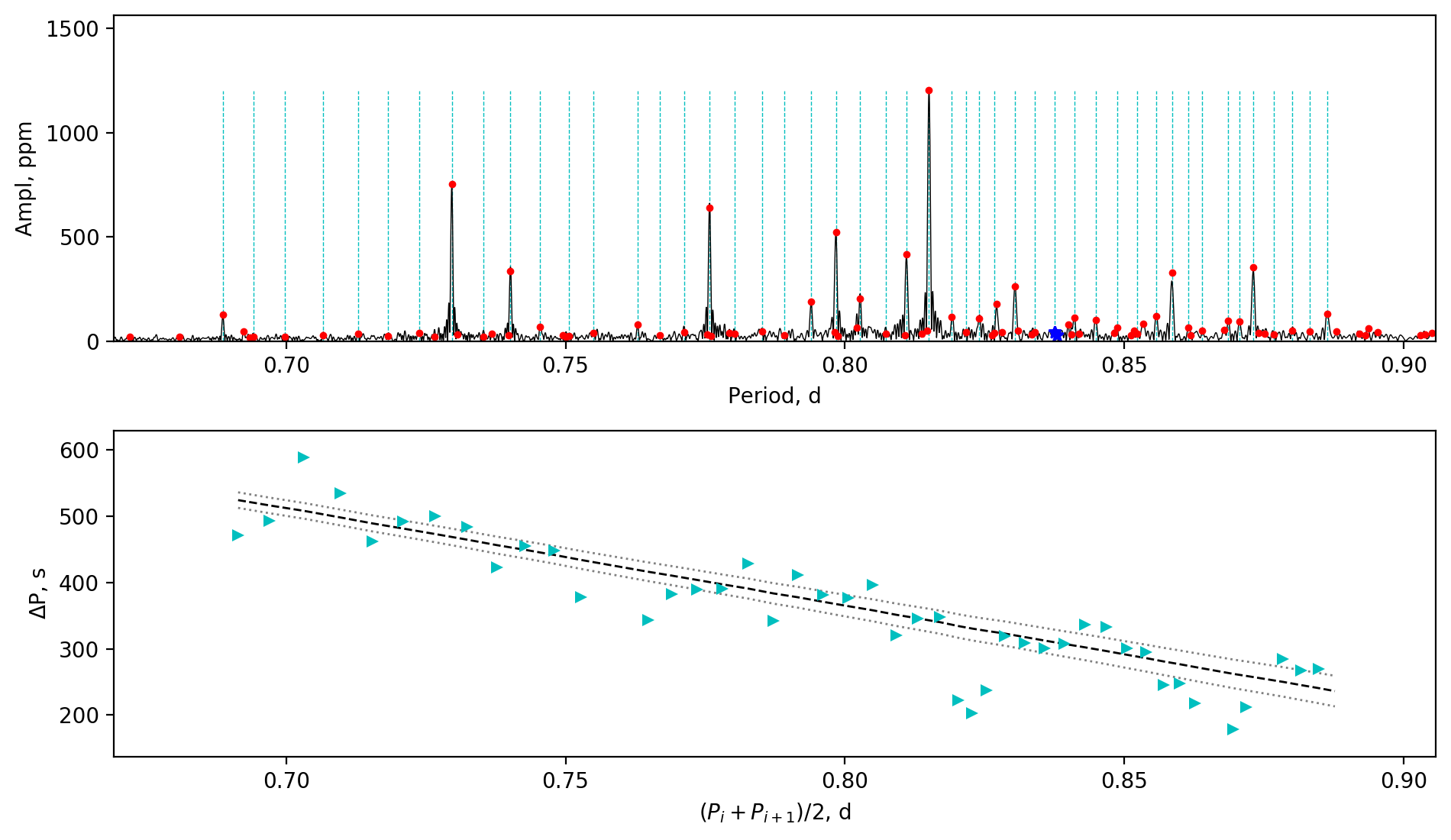}
\caption{The period spacing patterns of KIC\,8197406.}\label{fig:KIC 8197406}
\end{figure*}

\begin{figure*}
\centering
\includegraphics[width=1\textwidth]{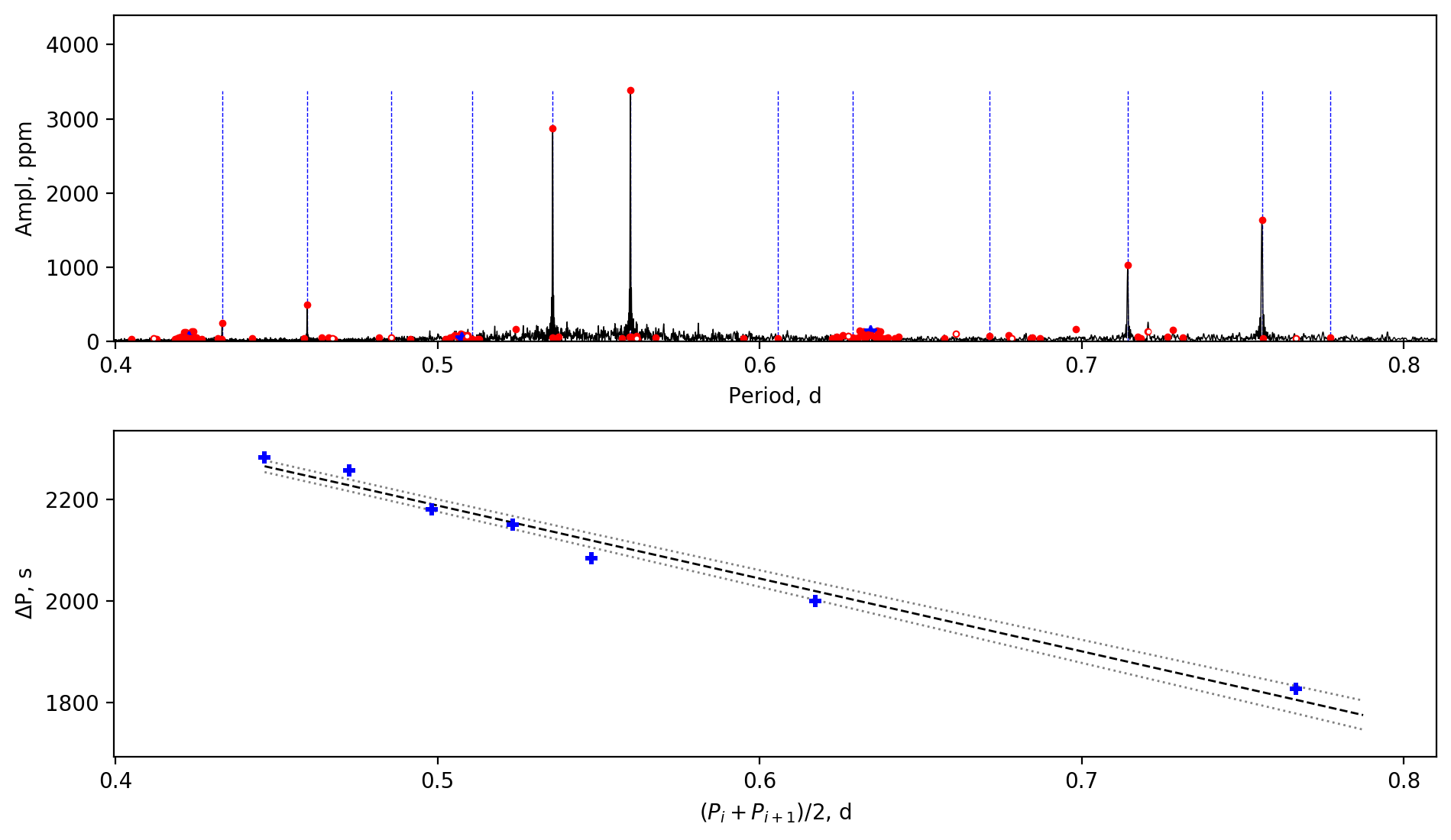}
\caption{The period spacing patterns of KIC\,9236858.}\label{fig:KIC 9236858}
\end{figure*}

\begin{figure*}
\centering
\includegraphics[width=1\textwidth]{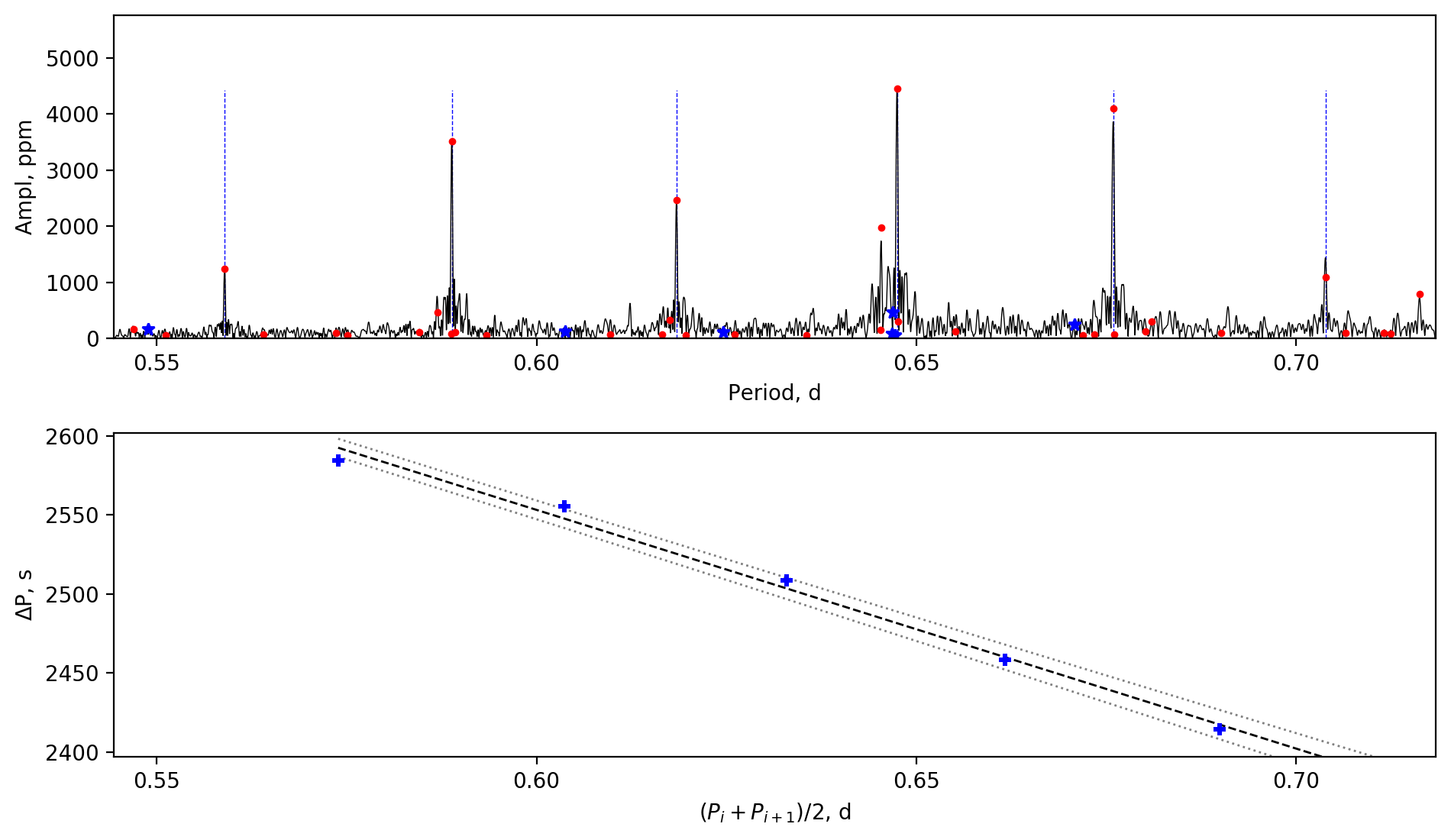}
\caption{The period spacing patterns of KIC\,4932691.}\label{fig:KIC 4932691}
\end{figure*}

\begin{figure*}
\centering
\includegraphics[width=1\textwidth]{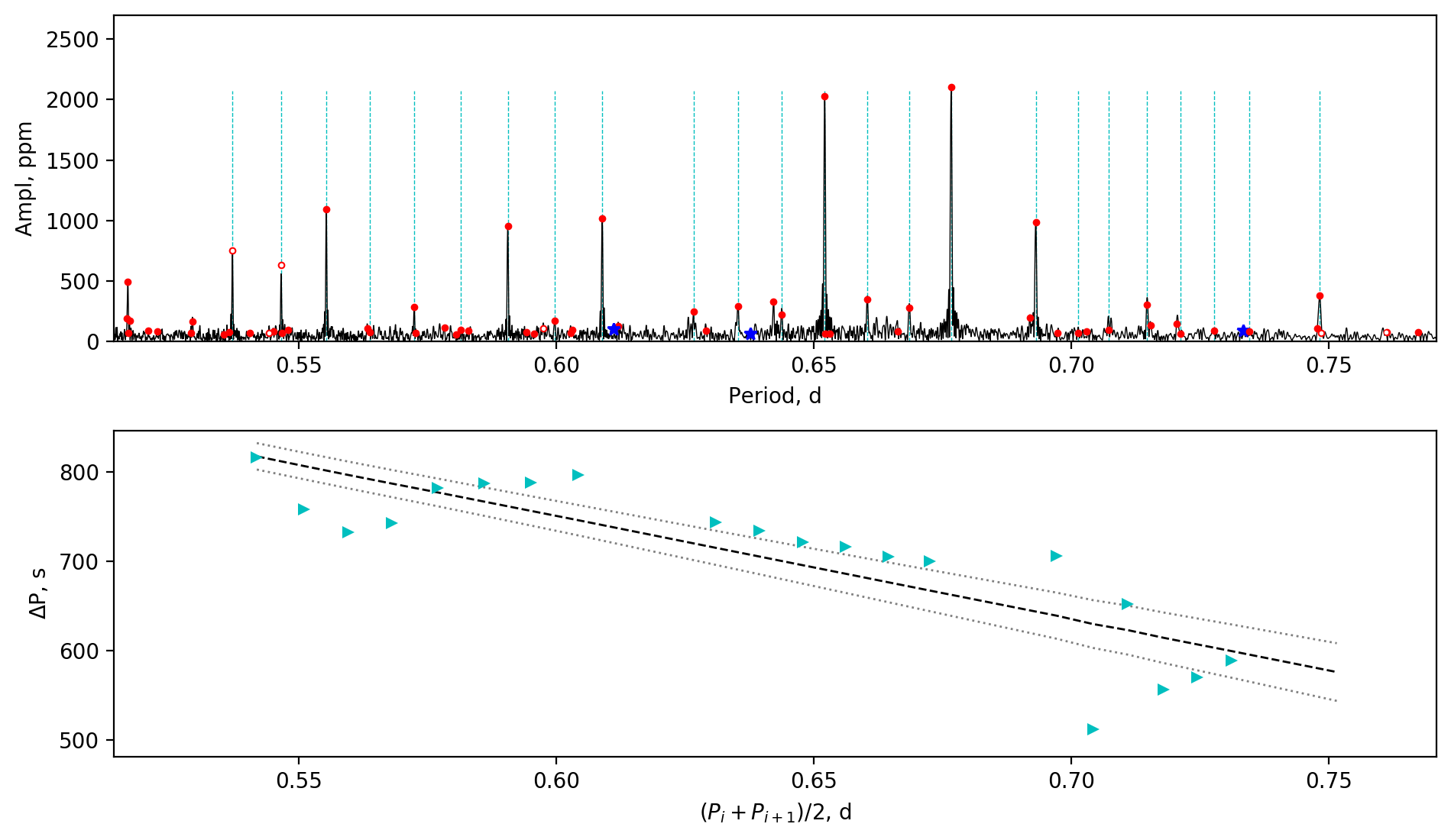}
\caption{The period spacing patterns of KIC\,12470041A.}\label{fig:KIC 12470041A}
\end{figure*}

\begin{figure*}
\centering
\includegraphics[width=1\textwidth]{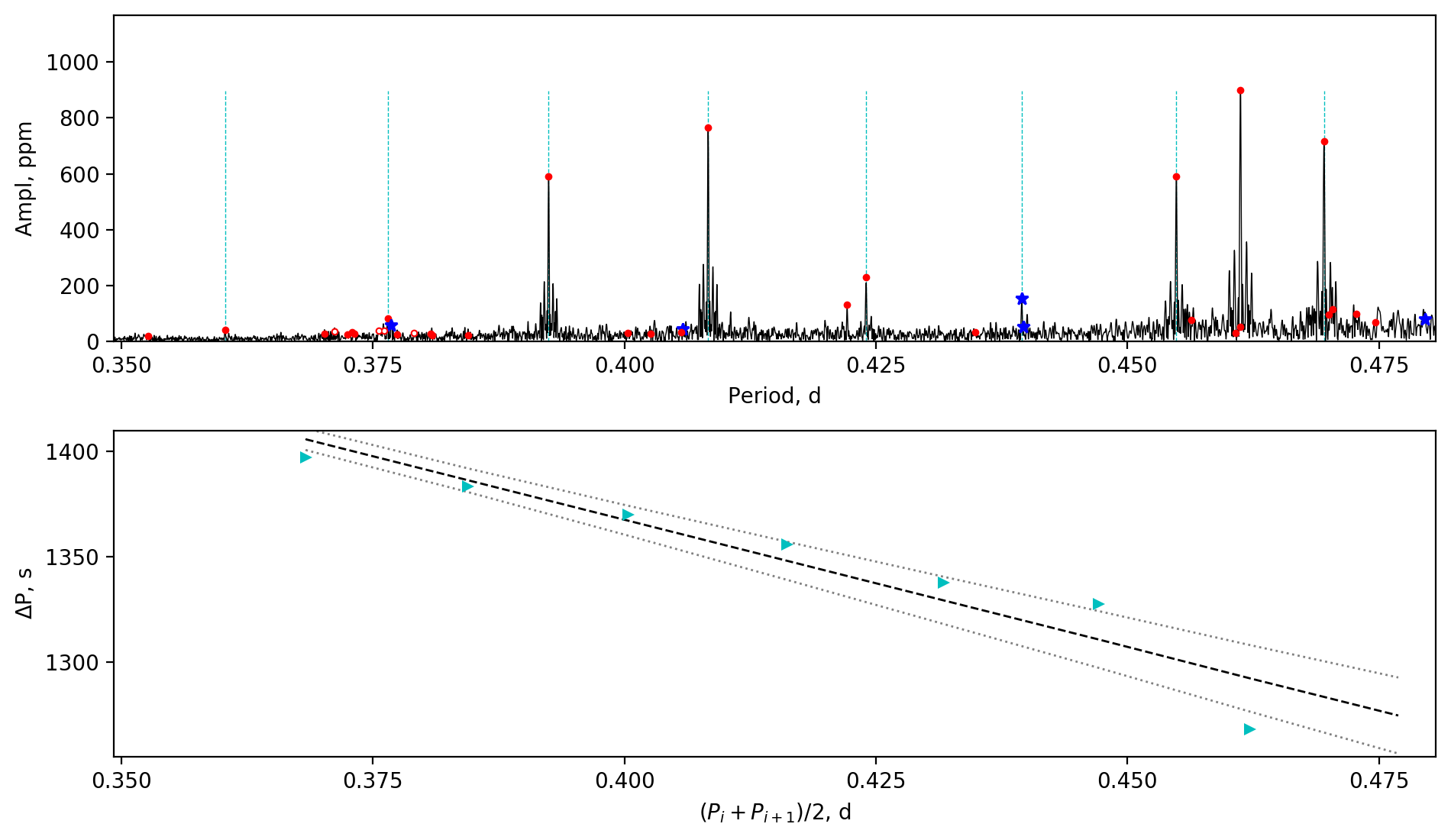}
\caption{The period spacing patterns of KIC\,10486425.}\label{fig:KIC 10486425}
\end{figure*}

\begin{figure*}
\centering
\includegraphics[width=1\textwidth]{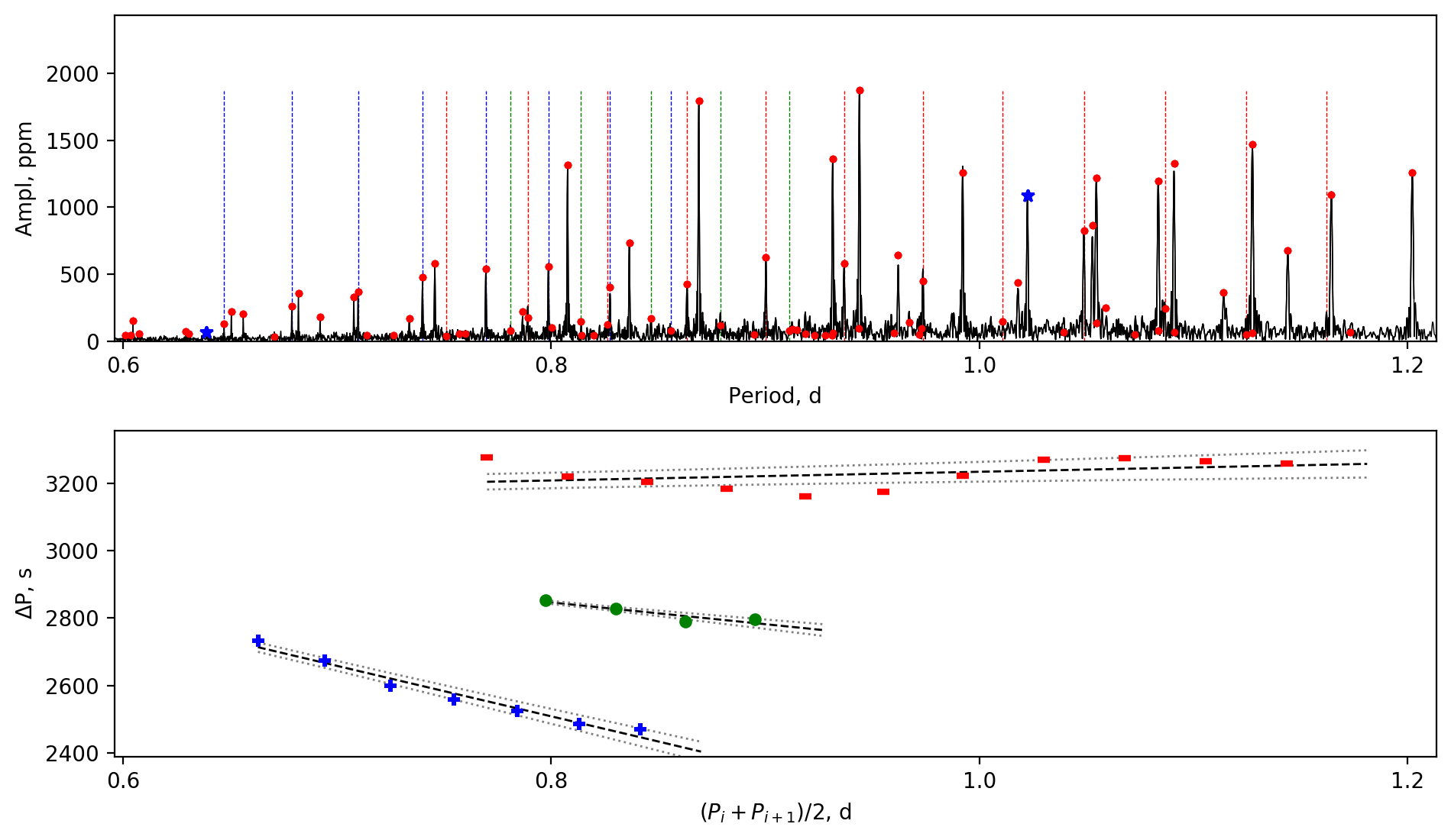}
\caption{The period spacing patterns of KIC\,10080943B.}\label{fig:KIC 10080943B}
\end{figure*}

\begin{figure*}
\centering
\includegraphics[width=1\textwidth]{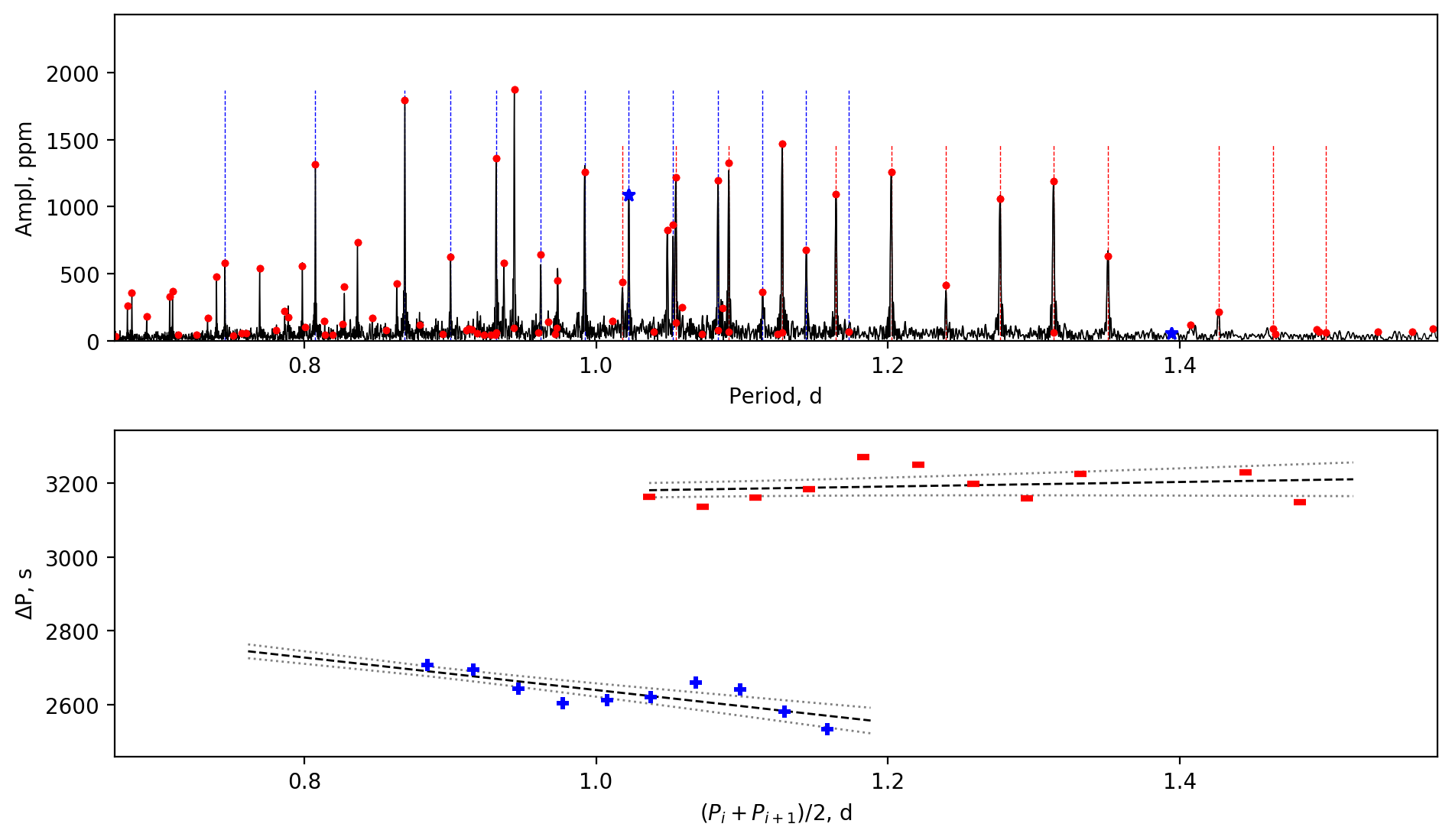}
\caption{The period spacing patterns of KIC\,10080943A.}\label{fig:KIC 10080943A}
\end{figure*}

\begin{figure*}
\centering
\includegraphics[width=1\textwidth]{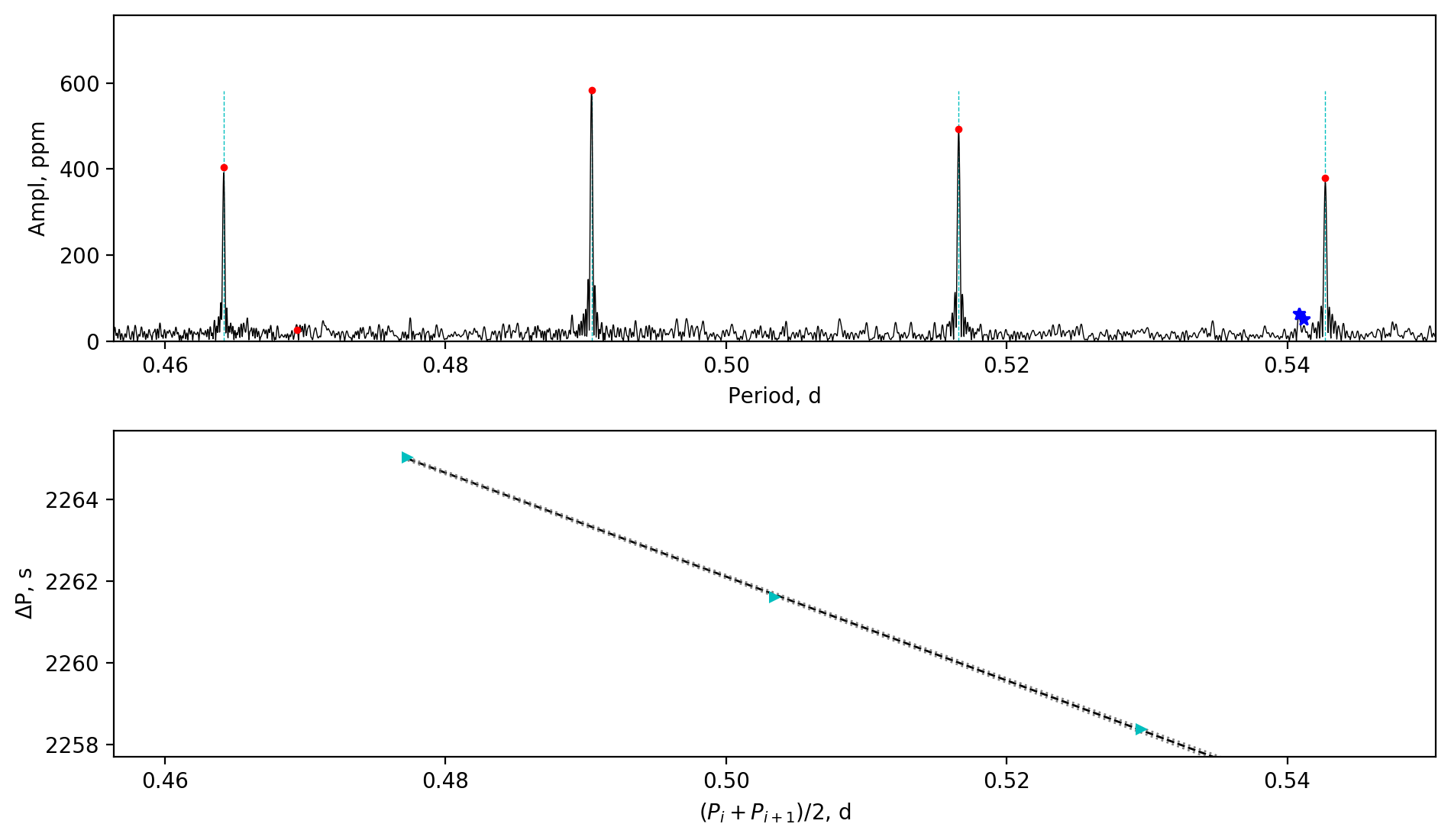}
\caption{The period spacing patterns of KIC\,8429450.}\label{fig:KIC 8429450}
\end{figure*}

\begin{figure*}
\centering
\includegraphics[width=1\textwidth]{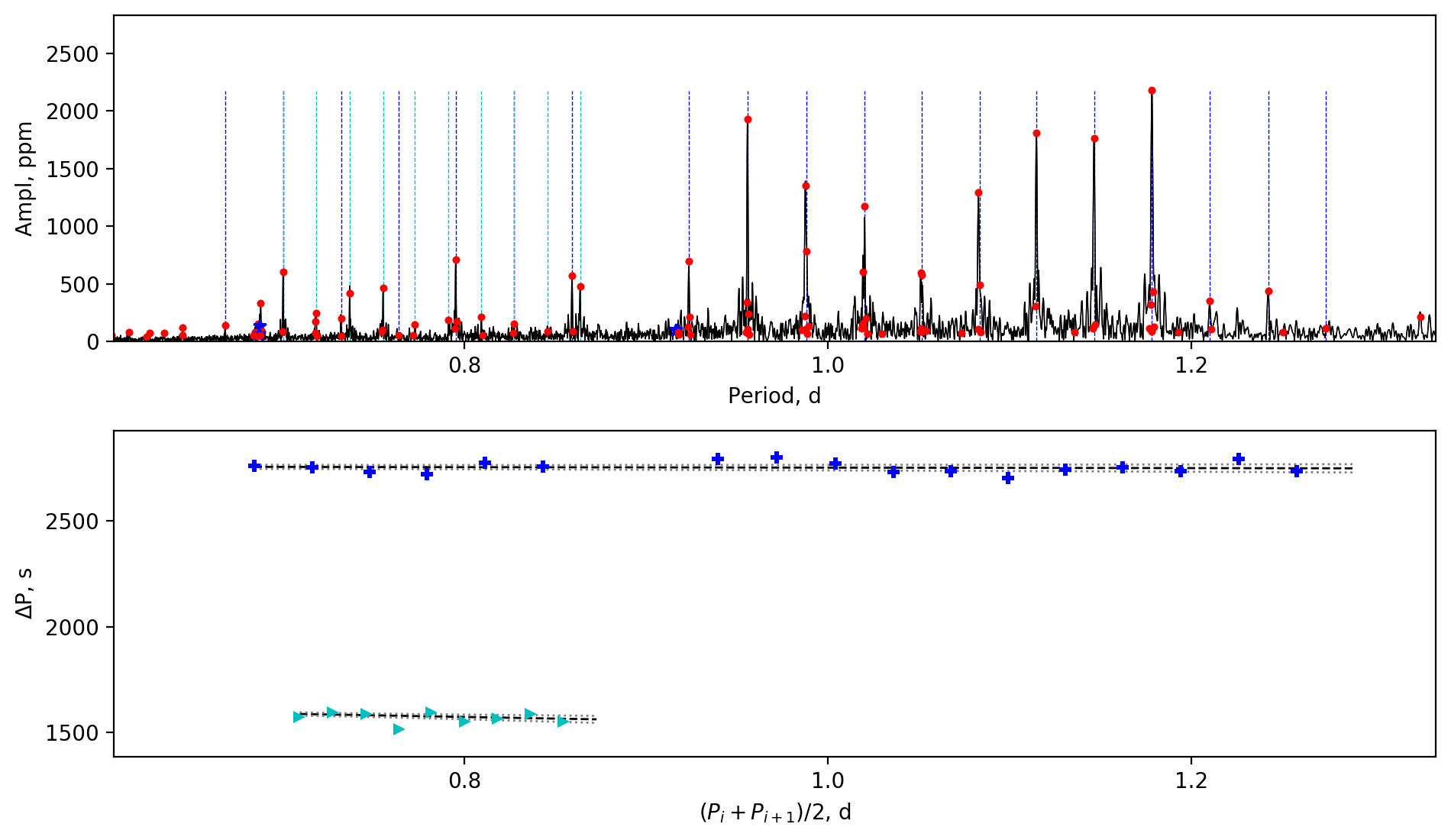}
\caption{The period spacing patterns of KIC\,9850387.}\label{fig:KIC 9850387}
\end{figure*}

\begin{figure*}
\centering
\includegraphics[width=1\textwidth]{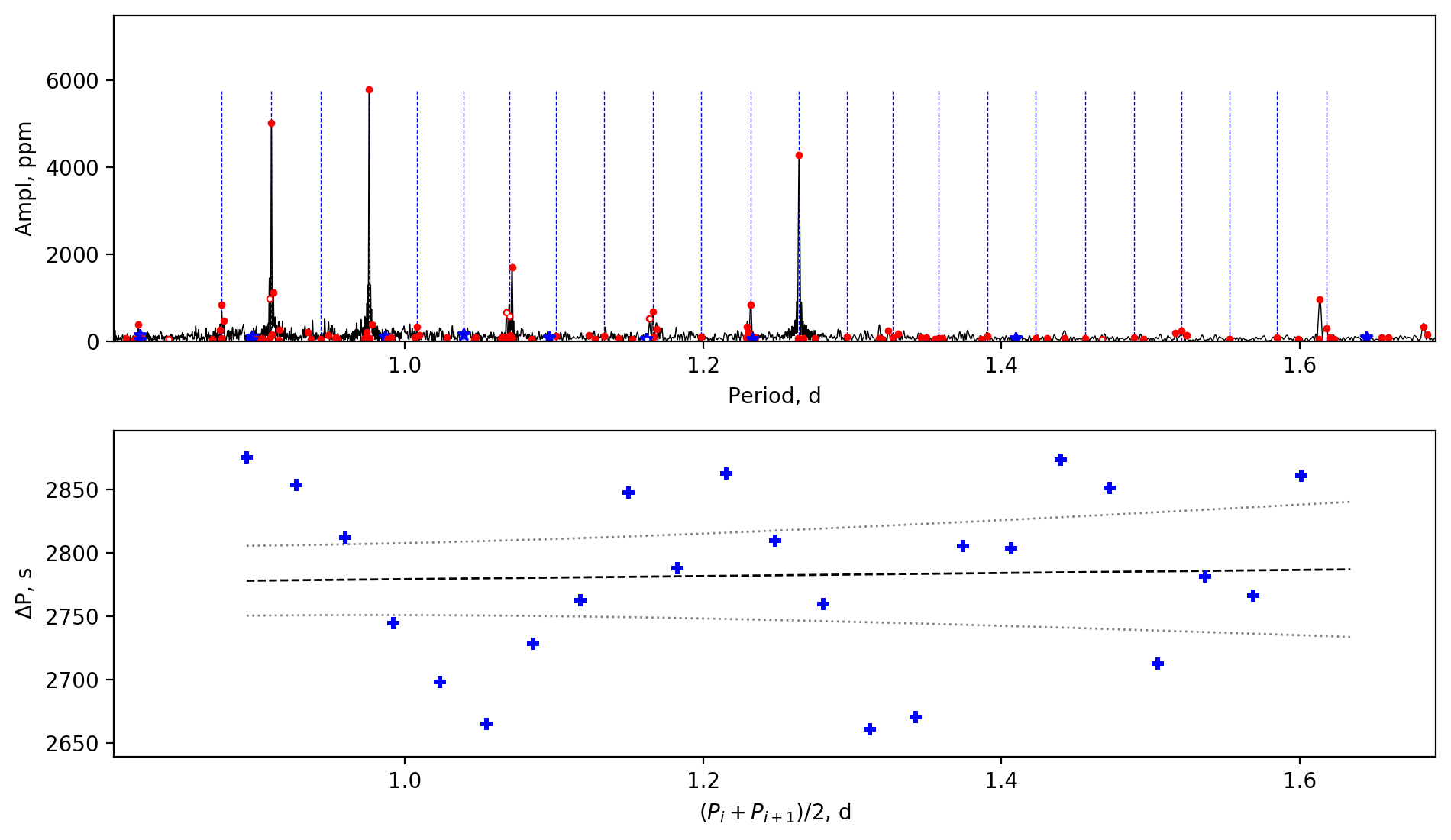}
\caption{The period spacing patterns of KIC\,8197761.}\label{fig:KIC 8197761}
\end{figure*}

\begin{figure*}
\centering
\includegraphics[width=1\textwidth]{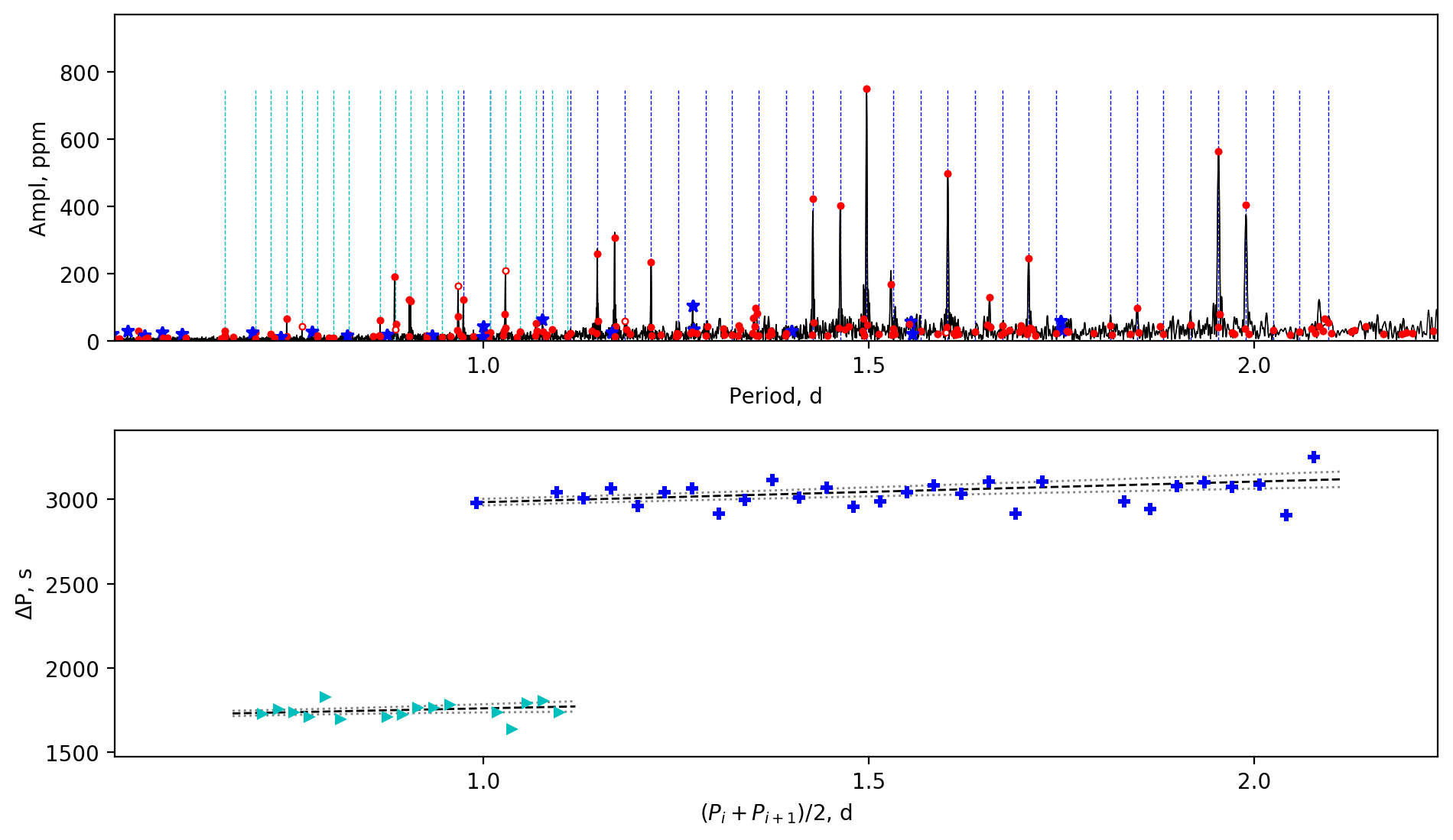}
\caption{The period spacing patterns of KIC\,4142768.}\label{fig:KIC 4142768}
\end{figure*}

% Don't change these lines
\bsp	% typesetting comment
\label{lastpage}
\end{document}